\documentclass[%
10pt,
 twocolumn,
nofootinbib,
 amsmath,amssymb,
 aps,
prb,
longbibliography
]{revtex4-2}

\usepackage{natbib,hyperref}
\usepackage{graphicx}
\usepackage{dcolumn}
\usepackage{svg}
\usepackage{bm}
\usepackage{mathtools}
\usepackage{amssymb}
\usepackage{amsmath}
\usepackage{amsthm}
\usepackage{graphicx}
\usepackage{bm}
\usepackage{listings}
\usepackage{epsfig}
\usepackage{amsfonts}
\usepackage{bbold}
\usepackage{url}
\usepackage{multirow} 
\usepackage{footnote}
\usepackage{blkarray}
\usepackage[]{appendix}
\usepackage[makeroom]{cancel}
\makeatletter
\def\cantox@vector#1#2#3#4#5#6#7#8{%
  \dimen@.5\p@
  \setbox\z@\vbox{\boxmaxdepth.5\p@
   \hbox{\kern-1.2\p@\kern#1\dimen@$#7{#8}\m@th$}}%
  \ifx\canto@fil\hidewidth  \wd\z@\z@ \else \kern-#6\unitlength \fi
  \ooalign{%
    \canto@fil$\m@th \CancelColor
    \vcenter{\hbox{\dimen@#6\unitlength \kern\dimen@
      \multiply\dimen@#4\divide\dimen@#3 \vrule\@depth\dimen@\@width\z@
      \vector(#3,-#4){#5}%
    }}_{\raise-#2\dimen@\copy\z@\kern-\scriptspace}$%
    \canto@fil \cr
    \hfil \box\@tempboxa \kern\wd\z@ \hfil \cr}}
\def\bcancelto#1#2{\let\canto@vector\cantox@vector\cancelto{#1}{#2}}
\makeatother
\usepackage[]{hyperref}
\usepackage{soul} 

\newcommand{\corr}[1] {{\color{black}{#1}}\color{black}}

\newcommand{\expSTMESR}{~\cite{Baumann_Paul_science_2015,
Natterer_Yang_nature_2017,Choi_Paul_natnano_2017,Willke_Paul_sciadv_2018,Yang_Bae_prl_2017,
Willke_Bae_science_2018,Y_Bae_advanced_science_2018,Willke_Singha_nanolett_2019,
Willke_Yang_natphys_2019,Yang_Paul_prl_2019,yang_coherent_2019,Seifert_Kovarik_pr_2020,
Seifert_Kovarik_eabc_2020,Wang2023_universal,
Weerdenburg_Steinbrecher_2020,Steinbrecher_Weerdenburg_2020,Jinkyung_Hyperfine_2022,zhang_electron_2022,kovarik_electron_2022,kot2023electric,Phark_double_resonance2023,Fe_driving_SH_2023,Wang2023AnPlatform,zhang_influence_2023,Bui2024All-electricalSpinb,kovarik_spin_2024,czap2025magneticresonanceimagingsingle, zhang2024electricfieldcontrolexchange,Huang2025,greule2025spinelectriccontrolindividualmolecules}}

\begin{document}

\title{Contrasting exchange-field and spin-transfer torque driving mechanisms in all-electric electron spin resonance} 






\author{Jose Reina-G{\'{a}}lvez}
\email{jose.reina-galvez@uni-konstanz.de}
\affiliation{Department of physics, University of Konstanz, D-78457, Konstanz, Germany}
\affiliation{Center for Quantum Nanoscience, Institute for Basic Science (IBS), Seoul 03760, Korea}
\affiliation{Ewha Womans University, Seoul 03760, Korea}

\author{Matyas Nachtigall}
\affiliation{Centro de F{\'{i}}sica de Materiales CFM/MPC (CSIC-UPV/EHU),  20018 Donostia-San Sebasti\'an, Spain}
\affiliation{Donostia International Physics Center (DIPC),  20018 Donostia-San Sebasti\'an, Spain}

\author{Nicol{\'a}s Lorente}
\affiliation{Centro de F{\'{i}}sica de Materiales
        CFM/MPC (CSIC-UPV/EHU),  20018 Donostia-San Sebasti\'an, Spain}
\affiliation{Donostia International Physics Center (DIPC),  20018 Donostia-San Sebasti\'an, Spain}

\author{Jan Martinek}
\affiliation{Institute of Molecular Physics, Polish Academy of Sciences, M. Smoluchowskiego 17, 60-179 Pozna\'n, Poland}

\author{Christoph Wolf}
\email{wolf.christoph@qns.science}
\affiliation{Center for Quantum Nanoscience, Institute for Basic Science (IBS), Seoul 03760, Korea}
\affiliation{Ewha Womans University, Seoul 03760, Korea}

\begin{abstract}
Understanding the coherent properties of electron spins driven by electric fields is crucial for their potential application in quantum-coherent nanoscience. In this work, we address two distinct driving mechanisms in electric-field driven electron-spin resonance as implemented in scanning tunneling spectroscopy. We study the origin of the driving field using a single orbital Anderson impurity, connected to polarized leads and biased by a voltage modulated on resonance with a spin transition. By mapping the quantum master equation into a system of equations for the impurity spin, we identify two distinct driving mechanisms. Below the charging thresholds of the impurity, electron spin resonance is dominated by a magnetically exchange-driven mechanism or field-like torque. Conversely, above the charging threshold spin-transfer torque caused by the spin-polarized current through the impurity drives the spin transition. Only the first mechanism enables coherent quantum spin control, while the second one leads to fast decoherence and spin accumulation towards a non-equilibrium steady-state. The electron spin resonance signals and spin dynamics vary significantly depending on which driving mechanism dominates, highlighting the potential for optimizing quantum-coherent control in electrically driven quantum systems.
\end{abstract}

\date{\today}

\maketitle

\section{Introduction}

Coherent manipulation of individual electron spins localized in single atoms or molecules on a surface using electron spin resonance (ESR) implemented in a scanning tunneling microscope (STM) is a novel platform for the study of many-body physics with potential applications for quantum simulations using artificially constructed interacting spin systems on surfaces~\expSTMESR. Such a system can be modeled as a magnetic impurity connected to two leads, which provide a harmonic alternating current (AC) drive and a static external magnetic field that Zeeman-splits the up and down spin states. When one of the leads is ferromagnetic, a polarized current is detected in the direct current (DC) signal. In this situation, adjusting the driving frequency of the AC voltage to correspond with the Zeeman splitting leads to resonance behavior in the electric current.

{\corr However, t}he mechanism mediating the coupling between the AC electric field and the spin remains a topic of debate~\cite{Ribeiro2,J_Reina_Galvez_2019,Lado_Ferron_prb_2017,Delgado_Lorente_pss_2021,J_Cuevas_C_Ast_ESR_theory_2024,PRL_Ye_ESR_theory_2024}, {\corr as selection rules indicate that such an interaction should be forbidden.}
{\corr To reconcile this,} some of us have proposed an all-electric model for ESR in the STM, which can efficiently drive the spin of an Anderson impurity model based on the principle of barrier modulation~\cite{J_Reina_Galvez_2021,J_Reina_Galvez_2023}. 
We solved the resulting quantum master equation either in the long-time limit or by simulating real-time spin dynamics via direct time propagation of the Hamiltonian~\cite{TimeESR_github}, {\corr enabling the implementation of quantum gates and highly entangled states~\cite{switzer2025unravelingspinentanglementusing}.}
At present, the model is limited to the weak coupling regime, imposing sequential tunneling processes, which implies no cotunneling effects, although the model is extended to the Coulomb Blockade regime via the tail of the resonance~\cite{J_Reina_Galvez_2023}. 

{\corr

A different line of work has proposed a time-dependent Kondo-like Hamiltonian in which the spin impurity is subjected to a driving that originates not from the Kondo coupling itself, but from the magnetic field component of the electric field, sufficiently amplified in the electromagnetic cavity formed between the tip and the spin~\cite{J_Cuevas_C_Ast_ESR_theory_2024}. This model, which represents a valuable contribution, requires the inclusion of this additional driving because a purely time-dependent Kondo scattering Hamiltonian lacks the necessary charge fluctuations to account for the experimentally observed Rabi rates, as discussed in~\cite{J_Reina_Galvez_2019}. In other words, it discards the \textit{exchange field contribution} described in~\cite{Braun_Konig_prb_2004}, which is essential in this context. However, the main concern is that t}he presence of an AC magnetic field as a product of the AC electric field contrasts with recent experimental findings~\cite{Fe_driving_SH_2023, Wang2023AnPlatform, Phark_double_resonance2023}, suggesting that a magnetic atom must be placed near the spin impurity to enable driving outside the tunnel junction but still under the influence of the electric field generated by the tip body. A long-range electromagnetic driving field appears inconsistent with the requirement for a localized magnetic field source.

{\corr

A subsequent approach, also based on a Kondo Hamiltonian,} derived an effective magnetic field proportional to the spin polarization of the tip and the AC voltage, which can drive the magnetic impurity~\cite{PRL_Ye_ESR_theory_2024}. While this proposed driving mechanism aligns to some extent with our approach, it does not fully capture all features of the experimentally observed homodyne detection.
Homodyne detection arises from the misalignment between the STM tip's spin and the magnetic impurity under AC driving.
In particular, it predicts the strongest ESR signal when the spin polarization of the tip and the magnetic impurity are mostly parallel~\cite{JK_vector_field_2021,Fe_driving_SH_2023}, whereas Ref.~\cite{PRL_Ye_ESR_theory_2024} differs from previous reports by showing the maximum ESR signal at $90$ degrees. {\corr This discrepancy is likely again related to the lack of a proper incorporation of the exchange field, an issue that we believe is also present in Ref.~\cite{Ribeiro2}.}

As has been pointed out, much of the existing literature on cotunneling (Kondo) in transport overlooks the \textit{exchange field contribution} from {\corr the first-order correction
in perturbation theory~\cite{Delgado_Rossier_prl_2011,Delgado_Rossier_prb_2011,Ribeiro2,J_Cuevas_C_Ast_ESR_theory_2024,PRL_Ye_ESR_theory_2024}. Indeed, the Schrieffer–Wolff transformation applied to the Anderson impurity model to obtain a higher-order cotunneling Hamiltonian neglects all first-order coupling terms, including the exchange field contribution. This was shown to be inaccurate when considering ferromagnetic electrodes, spin polarization,~\cite{Weymann_2006_seq_cotu,Braun_Konig_prb_2004} since the exchange field survives in the Coulomb blockade regime~\cite{Weymann_2006_seq_cotu,Braun_Konig_prb_2004,Piotr_Martinek_Hanle_effect_2023,Busz_PRB_2025}.}
A comprehensive theory that consistently incorporates both sequential and cotunneling processes, {\corr together with first-order coherent contributions such as the exchange field}, for periodically driven systems is still lacking, although recent works have made progress toward this goal~\cite{Kosov_PRB_2018_seq_cotu,A_Wacker_PE_2010_Modelling_cotu,Piotr_Martinek_Hanle_effect_2023}.

{\corr Coming back to our theoretical proposal}, a fundamental question was left unanswered: how does the barrier modulation couple to the spin, and can this mechanism be linked to any physical measurements? {\corr Does it capture the exchange field contribution?} In this work, we aim to explore these questions by establishing connections with related models for quantum dots connected to ferromagnetic leads~\cite{M_Braun_2005_hanle_seq, Konig_Jan_PRL_2003,Martinek2023,Braun_Konig_prb_2004,Piotr_Martinek_Hanle_effect_2023} where an exchange field mechanism between the tip and sample is demonstrated and applied to experimental studies~\cite{Hauptmann2008,Pasupathy_Science_kondo_effect}. This comparison is justified, as atomic or molecular spin impurities can be seen as miniaturized versions of larger lithographically defined quantum dots. We will show that the ESR mechanism is highly dependent on the applied DC bias relative to the ionization and charging energy thresholds of the impurity i.e., it depends \textit{if we are above the energy thresholds or within the Coulomb Blockade regime}. Specifically, we will contrast two driving mechanisms:
\begin{itemize}
    \item\textbf{Field-like Torque} dominates ESR for DC biases in the Coulomb blockade regime. It is mediated by the exchange magnetic field induced by the spin-polarized tip, leading to a conventional ESR mechanism, Homodyne detection and enabling coherent spin manipulation via electric pulses. 
    We refer to this driving mechanism as \textbf{FLT}. 
    
    \item\textbf{Spin-transfer Torque} drives ESR for DC biases outside the Coulomb blockade regime. It is dominated by spin accumulation on the impurity, driven by the current, which torques the impurity spin and leads to strong spin polarization of the impurity. 
    We refer to this driving mechanism as \textbf{STT}.
\end{itemize}

We emphasize that the driving mechanism is a distinct concept and exists separately from the transport regime.
The driving mechanisms (FLT, STT) are general and exist below and above the ionization thresholds. The transport regime dictates how strongly FLT and STT affect the magnetic impurity. 
{\corr It is important to mention that both driving regimes have been recently observed experimentally in a pentacene molecule~\cite{kovarik_spin_2024}, and they are expected to emerge in other molecular systems with well-defined orbital levels. The current work is inspired by the combination of key physical concepts introduced in Ref.~\cite{Braun_Konig_prb_2004} and the recent experimental realizations in the ESR-STM in Ref.~\cite{kovarik_spin_2024}}.


This paper is organized as follows. In Sec. II, we summarize the theoretical approach, drawing the comparison to quantum dot models. In Sec. III, we analyze the key physical findings from this comparison, defining the driving regimes and how they affect the quantum impurity. Finally, we perform numerical simulations to verify the key features of the theory, leading to the conclusions.


\section{Transport Model}
\subsection{Model Hamiltonian}

\begin{figure}
\centering 
\includegraphics[width=0.81945\linewidth]{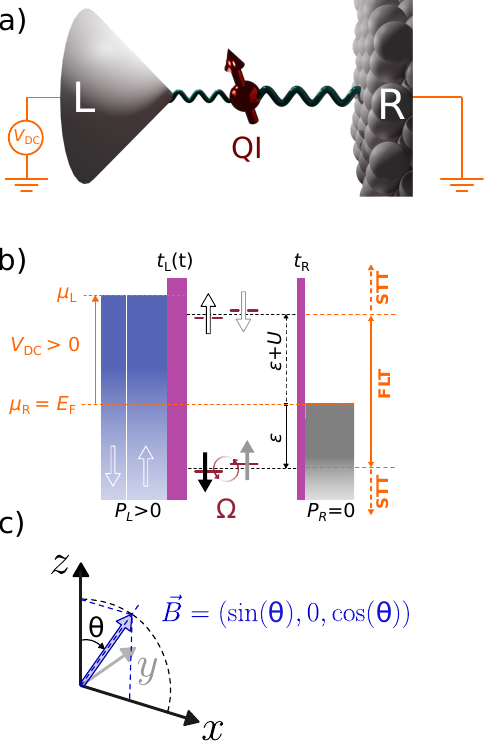}
\caption{a) Schematic of a QI connected a left (L) and right (R) lead with a bias voltage applied to the left lead. We consider the impurity to be more strongly coupled to the R lead. b)  Energy scheme of the QI. 
The hollow arrows represent unoccupied levels of the QI at $\varepsilon+U$ as a product of charging the QI with one more electron to be doubly occupied with a total energy $E_2=2\varepsilon+U$.  Meanwhile, the filled arrows, which defined the spin of the QI, are occupied single-electron states at $\varepsilon$.
Only the left electrode is polarized in this example ($P_L>0$) and modulated ($t_L=t_L(t)$) being $t_L<t_R$. The applied positive bias ($eV_{\mathrm{DC}}=\mu_L-\mu_R>\varepsilon+U>0$) implies that the Rabi process $\Omega$ is dominated by STT, as for when ($eV_{\mathrm{DC}}<\varepsilon<0$). Contrary, if $\varepsilon < eV_\mathrm{DC} <\varepsilon+U$, the Rabi process is FLT-driven, as indicated on the right side of the figure. The transport through the QI in b)is the following: 50\% of electrons up and down will tunnel from the right lead, overcoming the $\varepsilon+U$ energy, forcing the QI to be doubly populated. Since tip accepts more up electrons, the population of the up state in $\varepsilon+U$ will be depleted, leaving it down populated. Since spin-down SUMO state corresponds to a spin-up SOMO state, the QI ends up populated. The full process provides an incoherence resonance in the current when the L tunneling is modulated. c) Convention for the magnetic field direction with respect to the external coordinate system with the azimuthal angle $\phi=0$.}
\label{Energy_Sch}
\end{figure}

The model used here is a single orbital Anderson impurity model~\cite{Anderson_1961} (SAIM) of a quantum impurity (QI), $H_\mathrm{QI}$, connected to two ferromagnetic or normal electrodes (tip or substrate) that can be treated as fermionic baths by time-dependent hoppings $H_T(t)$ as introduced in previous works~\cite{J_Reina_Galvez_2021,J_Reina_Galvez_2023}, see Fig.~\ref{Energy_Sch}a.  This model is general enough to describe a wide range of physical systems, where the impurity spin can be localized in point defects, atoms, or molecules. The full system is given by $H(t)=H_{\rm L}+H_{\rm R}+H_{\rm QI}+H_{\rm T}(t),$ where the first two terms describes the spin-polarized or normal fermionic baths, the second term is the QI and the third term contains the hopping connecting baths and QI{\corr :
\begin{eqnarray}\label{eq:hres}
H_{\rm \alpha}&=&\sum_{k\sigma }\varepsilon_{\alpha k}c^{\dagger}_{\alpha k \sigma }c_{\alpha k \sigma }, 
\\
H_{\rm QI}&=& \sum_{\sigma} \varepsilon d^{\dagger}_{\sigma }d_{\sigma }
+U \hat{n}_{d  \uparrow} \hat{n}_{d \downarrow} +g\mu_B \mathbf{{B}}\cdot\mathbf{\hat{s}} \label{eq:himp0}
\\ 
H_{T}(t)&=&\sum_{\alpha k\sigma}\left(t_{\alpha }(t)c^{\dagger}_{\alpha k\sigma }d_{\sigma}+t_{\alpha  }^*(t)d^{\dagger}_\sigma c_{\alpha k\sigma }\right), \label{eq:ht0}
 \end{eqnarray}}

The bath itself can be expressed using the common form (Eq.~\eqref{eq:hres}), where each state is described by an electrode label $\alpha=L, R$, spin projection along the $z$ axis $\sigma= \uparrow, \downarrow$, and momentum $k$. 
For both ferromagnetic electrodes, we consider that their magnetizations point in the $z$ direction. In this case, the effective spin asymmetry, $D_{\alpha \uparrow} \neq D_{\alpha \downarrow}$, in the spin-dependent density of states $D_{\alpha \sigma}$ at the Fermi level can be described by the spin polarization $P_{\alpha} \equiv (D_{\alpha \uparrow}-D_{\alpha \downarrow})/(D_{\alpha \uparrow}+D_{\alpha \downarrow})$, where 
$D_{\alpha \sigma} = D_{\alpha} \left(1/2 + \sigma P_{\alpha} \right)$ with $P_{\alpha} \in [-1,1]$ and the extreme cases representing a fully spin polarized lead pointing down, $-1$, or up, $+1$. {\corr Each electrode is characterized by a electrochemical potential $\mu_{\alpha}$ and a temperature $T_\alpha$. }

For the sake of simplicity, \textit{we focus on the simplest\footnote{More complex QIs with Heisenberg exchange coupling $J$ between the transport spin $\mathbf{\hat{s}}$ and another localized spin, or between multiple localized spins, can be seamlessly incorporated into the model \cite{J_Reina_Galvez_2021}, enabling the study of arbitrary spin clusters and more intricate spin structures. Future work will explore this in depth.} QI} in Eq.~\eqref{eq:himp0}, which includes the Coulomb repulsion $U$, the onsite energy $\varepsilon$, and an external magnetic field $\mathbf{B} = B(\sin\phi \sin\theta, \cos\phi \sin\theta, \cos\theta)$ expressed in spherical coordinates, pointing in an arbitrary direction, generally noncollinear with the ferromagnetic lead magnetization.
{\corr The transport spin is defined as $\mathbf{\hat{s}} = \hbar \sum_{\sigma \sigma'} d_\sigma^\dagger \boldsymbol{\sigma}_{\sigma \sigma'} d_{\sigma'}/2$, with $\boldsymbol{\sigma}$ the vector of Pauli matrices. This definition incorporates the factor of $\hbar$ from the Bohr magneton into the spin operator}.

Our convention for the applied bias is such that $eV_{DC} = \mu_L - \mu_R$, meaning \textit{the DC bias is applied to the left electrode, which we will assign as the tip}, see Fig.~\ref{Energy_Sch}a and b. Consequently, electrons move from the sample to the tip at a positive applied bias. {\corr The onsite energy (ionization energy) $\varepsilon$ is associated with the singly occupied molecular orbital (SOMO), while $\varepsilon + U$ is linked to the singly unoccupied molecular orbital (SUMO).} Changing the sign convention to one where the DC bias is applied to the sample, $eV_{DC} = \mu_R - \mu_L$, only requires swapping the labels for $\varepsilon$ and $\varepsilon+U$ energies to keep the correct physical meaning of the orbitals. The right {\corr electro}chemical potential is assumed to be nearly aligned with the Fermi energy and grounded, which goes in line with the asymmetry tunneling conditions $t_L<t_R$ in~\cite{Tu_X_double-barrier} and follows the experimental evidence of the impurity being more strongly coupled to the substrate than the STM tip.

In this model, only the hopping terms, Eq.~\eqref{eq:ht0}, include a periodic modulation in the form:  
\begin{equation}
t_{\alpha  }(t)=t_{\alpha }^0 \left[1+A_{\alpha }\cos (\omega t) \right], \label{hopping}
\end{equation}
with $A_\alpha\propto V_{\mathrm{AC}}$. This functional {\corr is justified under appropriate parametric conditions, and} its dependence has already been discussed in previous works~\cite{J_Reina_Galvez_2019,J_Reina_Galvez_2021,J_Reina_Galvez_2023, Wolf_C_and_Delgado_F_2020}; therefore, we defer a more detailed examination to a future publication.
It is important to highlight that this approach does not explicitly include nor exclude the piezoelectric effect~\cite{Lado_2017_PRB}; rather, it accounts for all possible physical contributions to the driving parameter.

The use of a reduced density matrix formalism allows us to define any observable straightforwardly. To do so, we consider all possible charge configurations: zero, one, and two electrons in the QI (see Fig.~\ref{Energy_Sch}b). These configurations are denoted as $|p\rangle$, with $p=g,e,0,2$. The first two are the singly occupied states of the QI at $\varepsilon$ energy, corresponding to the ground and excited states, respectively, and their time evolution fully describes the spin state of the system. 
The states $0$ (empty) and $2$ (doubly occupied) are transient states necessary to facilitate transport through the impurity. {\corr The first state represents an empty quantum dot, aligned with the grounded electrochemical potential $\mu_R$ for convenience} 
and therefore chosen to be at zero energy, $E_0 = 0$. Consequently, removing one electron from the impurity requires an energy cost of $E_1 - E_0 = \varepsilon$.
Meanwhile, the doubly occupied state is formed when two electrons with opposite spins occupy the same level or orbital, with an energy given by $E_2 = 2\varepsilon + U$. To create this state, an electron from the leads must overcome an energy barrier of $E_2 - E_1 = \varepsilon + U > 0$ to be placed at the singly occupied state. From the energy perspective, the quantum impurity hosts two electrons, one at energy $\varepsilon$ and another at $\varepsilon + U$, with the latter being an effective unoccupied singly state. Thus, the relevant energies for transport are the differences between states with different numbers of electrons, i.e., $\varepsilon$ and $\varepsilon+U$, as shown in Fig.~\ref{Energy_Sch}b, rather than the absolute energy of the SIAM levels.

Since the impurity Hamiltonian does not depend on time, it is convenient to diagonalize the QI Hamiltonian and introduce the reduced density matrix. Thus, we can write $H_{\rm QI} |l\rangle = \sum_j E_j\hat{\rho}_{jj}|l \rangle = E_{l} |l \rangle.$
The Hubbard operators $\hat{\rho}_{lj}=|l\rangle\langle j|$, with the Latin characters ($l,j$ and later $u,v,\dots$) referring to eigenstates of the impurity, are closely related to the reduce density matrix~\cite{J_Reina_Galvez_2021,J_Reina_Galvez_2023,Bibek_Bhandari_2021_nonequilibrium,Hewson_book_1997}. 
{\corr These operators describe transitions between many‑body eigenstates $|l\rangle$ and $|j\rangle$ of the isolated quantum impurity, and reduce to a projector on a single state when $l=j$. If we assemble all these operators $\rho_{lj} =\mathrm{Tr} [\hat{\rho}_T(t)\hat{\rho}_{lj}]$, with $\hat{\rho}_T(t)$ being the density matrix of the full system given by the total Hamiltonian $H(t)$, then the coefficients $\rho_{l j}$ are exactly the elements of the impurity’s reduced density matrix in that eigenbasis. For the isolated impurity, subspaces with different electron numbers are decoupled, so $\rho_{l j}=0$ whenever $| l \rangle$ and $| j \rangle$ differ by one electron.
When we couple the impurity to the electrodes, these previously disconnected subspaces become linked by single‑electron tunneling, and these operators mediate the transition probability, as the transformation of the tunneling Hamiltonian in terms of Hubbard operators indicates:
\begin{equation*}
H_T(t)=\sum_{\alpha k \sigma lj} \left(t_{\alpha }(t)c^{\dagger}_{\alpha k\sigma } \lambda_{l j,\sigma} \hat{\rho}_{lj}  + h.c. \right).
\end{equation*}

The matrix element $\lambda_{lj,\sigma} = \langle l | d_\sigma |j\rangle=\langle l|0\rangle \langle \sigma |j \rangle+\langle l|\bar{\sigma}\rangle \langle 2| j \rangle$ reflects the change of the many-body configurations on the QI, they are computed in Appendix~\ref{app:lambda_coef}. Unlike ordinary fermionic creation and annihilation operators, Hubbard operators do not obey Wick’s theorem but instead a graded Lie algebra. As a result, perturbation theory is more challenging, making, for example, a cotunneling expansion with exchange‑field corrections more cumbersome.}

The specific case in Fig.~\ref{Energy_Sch}b, where the ground and excited eigenstates, $g$ and $e$, are labeled $\downarrow$ and $\uparrow$, respectively, corresponds to a positive out-of-plane magnetic field (i.e. {\corr oriented along the positive $z$ axis). For an isotropic $g$-factor and when the coupling to the electrodes is weak, the Zeeman interaction aligns a spin-$1/2$ parallel to the magnetic-field direction. Therefore, if the external field is arbitrarily oriented the spin aligns with $\mathbf{B}$ rather than with the fixed $z$ axis, and in general $H_{\mathrm{QI}}$ is no longer diagonal in the $z$-projection basis. Accordingly, by our definition, \emph{we represent the reduced density matrix in the coordinate system where $H_{\mathrm{QI}}$ is diagonal} (the \textbf{QI eigenbasis}), \emph{which here coincides with the magnetic-field direction}. In systems with strong anisotropy or for higher-spin cases, however, the spin(s) will not in general point along $\mathbf{B}$ and the eigenbasis of $H_{\mathrm{QI}}$ need not coincide with the magnetic-field axis.
Thus one must work with two coordinate systems: the fixed Cartesian frame $(x,y,z)$ in which the $z$ axis defines the computational basis (chosen parallel to the tip magnetization for convenience), and the QI eigenbasis (the axis where $H_{\mathrm{QI}}$ is diagonal). Quantities defined in the QI eigenbasis must be projected onto the fixed Cartesian frame when comparing to measurements, because the transport current probes the $z$ component of the spin since the tip magnetization is along $z$, as we will later.} Consequently, expectation values of spin components must be computed with care, see Appendix~\ref{app:spin_expectation}; one cannot simply insert Pauli matrices in the computational-$z$ basis without the appropriate basis rotation or change in the basis set.

\subsection{Quantum master equation (QME)}

At this stage, we derive a QME for the reduced density matrix by treating the coupling between the QI and the reservoirs up to the lowest order in perturbation theory in $H_T$ in the Born-Markov approximation~\cite{rammer_2007,Dorn_2021}, meaning $\hbar^{-1}\Gamma\tau_B\ll 1$ where $\Gamma$ is proportional to the rate of change of the reduced density operator due to the coupling to the bath and $\tau_B$ is the characteristic decay time of the bath~\cite{Hone_2009}. Following Refs. [\onlinecite{schoen-94,koenig-96-1,koenig-96-2,spletts-06,Esposito_Galperin_prb_2009,cava-09,Bibek_Bhandari_2021_nonequilibrium,J_Reina_Galvez_2021,J_Reina_Galvez_2023}], we write:
\begin{eqnarray}
\hbar\dot{\rho}_{lj}(t)&-& i\Delta_{lj}\rho_{lj}(t)= \sum_{vu}\left[\Gamma_{vl,ju}(t)+\Gamma^*_{uj,lv}(t)\right] \rho_{vu}(t)  \nonumber \\
&-&\sum_{vu}\left\{ \Gamma_{jv,vu}(t)\rho_{lu}(t)+\Gamma^*_{lv,vu}(t)\rho_{uj}(t)\right\},
\label{rho_master_eq_6_0}
\end{eqnarray}
where $\Delta_{lj}=E_l-E_j$ {\corr and every Latin character refers to eigenstates of the QI.}
The QME, Eq.~\eqref{rho_master_eq_6_0}, is physically meaningful in the limit of weak coupling between QI and electrodes, and, as we mentioned, it is represented in the {\corr eigenstates basis set, parallel to the external magnetic field for an isotropic spin-1/2}. In a previous work~\cite{J_Reina_Galvez_2021}, we derived a Floquet version of the QME, which has proven useful in the long-time limit for efficiently computing spectra by transforming the equations to the frequency domain. This formulation incorporates Floquet components of the density matrix, labeled by the Floquet number $n$.
The rates\footnote{While rates are typically expressed in units of $1/\text{s}$, in our formulation they are instead defined in units of energy. This reflects their interpretation as an energy width rather than a conventional rate.} in Eq.~\eqref{rho_master_eq_6_0} can be written as the sum of two contributions per electrode $\alpha$: 
\begin{equation}
\Gamma_{vl,ju}(t) =\sum_\alpha\left[ \Gamma_{vl,ju,\alpha}^{-}(t) + \Gamma_{vl,ju,\alpha}^{+}(t)\right],
\label{rate}
\end{equation}
which are
\begin{eqnarray}
	\Gamma_{vl,ju,\alpha}^{+}(t)&=&\sum_{\sigma}  \lambda_{vl,\sigma}\lambda_{uj,\sigma}^*
       \gamma_{\alpha\sigma}(t)
        {\cal I}_\alpha^+(\Delta_{ju})	\label{rateT}
\\
	\Gamma_{vl,ju,\alpha}^{-}(t)&=&\sum_{\sigma}  \lambda_{lv,\sigma}^*\lambda_{ju,\sigma}
       \gamma_{\alpha\sigma}(t)  {\cal I}_\alpha^-(\Delta_{ju}) 	.\label{rateTle}
\end{eqnarray}

{\corr Equations~\eqref{rateT} and~\eqref{rateTle} contain the coupling $\gamma_{\alpha\sigma}(t)=\left(1+A_{\alpha }\cos(\omega t) \right)^2 \gamma_{\alpha\sigma}^0$ where $\gamma_{\alpha\sigma}^0=2\pi|t^0_\alpha|^2 D_{\alpha}(1/2+\sigma P_\alpha)=\gamma_{\alpha}^0(1/2+\sigma P_\alpha)$ represents the level broadening due to the time-independent hopping, $t^0_\alpha$, for spin $\sigma$ in the electrode $\alpha$.
T}he Fermi distribution function $f_\alpha(\epsilon)=1/\left(e^{\beta_\alpha(\epsilon-\mu_{\alpha})}+1\right)$, being $\beta_\alpha=1/k_BT_\alpha$ and $\mu_\alpha$ the {\corr electro}chemical potential, is encoded in the integrals ${\cal I}^{\pm}_\alpha(\Delta_{ju})$ which are of the form:
\begin{eqnarray}
	{\cal I}_\alpha^\pm(\Delta_{ju})&=&\pm \frac{i}{2\pi}\frac{1}{1+A_\alpha\cos(\omega t)} \int^\infty_{-\infty} d \epsilon f^{\pm}_\alpha (\epsilon) \nonumber \\ &\times&
    \left(\frac{1}{\epsilon\mp\Delta_{ju}\pm i\hbar/\tau_c}  +  \frac{e^{i\omega t}A_\alpha/2}{\epsilon\mp\Delta_{ju}\mp \hbar\omega\pm i\hbar/\tau_c}  \right.\nonumber\\ &+& \left.  \frac{e^{-i\omega t}A_\alpha/2}{\epsilon\mp\Delta_{ju}\pm\hbar\omega\pm i\hbar/\tau_c}  \right).
	\label{int}
\end{eqnarray}
Notice that if $\hbar\omega\ll |\Delta_{ju}|$, Eq.~\eqref{int} simplifies to\footnote{A feasible assumption since $\hbar \omega$ will be in the 0.1 meV range or below, while $\Delta_{ju} \sim 100$ meV. This is because, \textit{in the rates only}, $\Delta_{ju}$ always represents an energy difference between charged states with different numbers of electrons; otherwise, the rate is zero due to the definition of $\lambda_{lj,\sigma}$. Conversely, $\Delta_{lj}$ in Eq.~\eqref{rho_master_eq_6_0} can be in the $\hbar \omega$ range.}:
\begin{eqnarray}
	{\cal I}_\alpha^\pm(\Delta_{ju})=\pm \frac{i}{2\pi}\int^\infty_{-\infty} d \epsilon f^{\pm}_\alpha (\epsilon) 
    \frac{1}{\epsilon\mp\Delta_{ju}\pm i\hbar/\tau_c},
	\label{int_s}
\end{eqnarray}
approximation valid for all the simulations in this work. Equation~\eqref{int_s} contains $f^+_\alpha(\epsilon)=f_\alpha(\epsilon)$ and $f^-_\alpha(\epsilon)=1-f_\alpha(\epsilon)$ following~\cite{Braun_Konig_prb_2004}.
%
%
%
Further details on the transition rates, the derivation of the quantum master equation (QME), and its solution in the long-time limit can be found in~\cite{J_Reina_Galvez_2021}{, \corr albeit with some differences, as the theory and physical interpretations have evolved over time.}

It is important to mention that $\hbar/\tau_c$, up to first order in perturbation theory, should be infinitesimal in Eqs.~\eqref{int} and~\eqref{int_s}. However, in this work, $\hbar/\tau_c \ll |\Delta_{ju}|\approx|\varepsilon|\ \mathrm{or}\ \varepsilon + U$ but nonzero, allowing us to explore two distinct regimes for a first-order perturbation theory: the one \textit{above the thresholds} $|\varepsilon|,\varepsilon + U$ where an full electron can hop in or out from the QI and the one \textit{below the thresholds} where the QI is in the so called \textit{Coulomb blockade} (CB) regime and we tunnel through the tail of the resonance~\cite{J_Reina_Galvez_2023}. Therefore, the transition between both regimes is controlled by the DC voltages and happens when it equalizes the ionization energy $\varepsilon$ or the charging one $\varepsilon+U$, which correspond to the situation of removing and adding one electron on the QI respectively. 

We are aware of the limitations a finite $\hbar/\tau_c$ introduces, as a proper characterization of the CB requires the use of cotunneling transport or a Kondo scattering Hamiltonian~\cite{Weymann_2006_seq_cotu,J_Reina_Galvez_2019}. However, as long as $\hbar/\tau_c \ll |\varepsilon|, \varepsilon + U$, we can overcome this limitation of the model and explore, to some extent, the implications of the FLT regime that governs the ESR signal in the CB regime. In this context, we expect low current at low DC biases, consistent with previous studies~\cite{Braun_Konig_prb_2004,A_P_Jauho_2006,Weymann_2006_seq_cotu,Piotr_Martinek_2021_JMMM,Piotr_Martinek_Hanle_effect_2023}. Future extensions of this model will aim to circumvent this issue for a better characterization of the system.
However, first-order perturbation theory already provides rich results that follow experimental evidence, as we will explore next.

\subsection{Expression for the current}
 
The electric current flowing out of an electrode $\alpha$ is defined following the usual formula~\cite{meir_wingreen_prl_1992} as $I_{\alpha}(t)=-e\frac{d\langle N_{\alpha}\rangle}{dt} = -i\frac{e}{\hbar} \langle [H_T(t),N_\alpha]\rangle $
which leads to:
\begin{equation}
I_{\alpha}(t)=
\frac{2e}{\hbar}\sum_{lju} \mbox{Re} \left\{\rho_{lu}(t) \left[ \Gamma_{lj,ju,\alpha}^{+}(t) - \Gamma_{lj,ju,\alpha}^{-}(t)\right] \right\}.
\label{current_t}
\end{equation}
{\corr This expression corresponds to the lowest order in the coupling (sequential tunneling) and is therefore expected to decrease rapidly in the CB regime, while remaining large and constant outside of it. Equation~\eqref{current_t}} contains both DC and AC terms as a consequence of the hopping modulation. The measurable quantity in the experiment is the electric current signal in the long-time limit, which depends on the resonance condition as well as the relative orientation of the QI spin to the direction of the tip spin polarization. We refer to this current as ESR signal in the following. The coherences (off-diagonal elements of the density matrix) are explicitly included in the expression and are crucial for the correct calculation of the ESR signal, as pointed out in a previous work~\cite{J_Reina_Galvez_2023}. The coherences in Eq.~\eqref{current_t} also lead to the so called \textit{homodyne detection} of the electric current, a product of the misalignment of the tip's spin and the spin of the impurity under it~\cite{JK_vector_field_2021,Fe_driving_SH_2023,Kubota2008_nature_phys}. {\corr Following Appendix~\ref{app:homodyne} one arrives to:}
\begin{eqnarray}
    I_L(t)&=& \frac{e}{\hbar} \gamma_L \left(1+A_{L}\cos (\omega t) \right)^2 \nonumber \\
    &\times& \left[ (\rho_1- 2P_L \langle s_z \rangle ) \mbox{Re} \left\{ {\cal I}_L^+(\Delta_{2g})\right\}\right. \nonumber \\
    &-&\left. (\rho_1+ 2P_L \langle s_z \rangle )\mbox{Re}\left\{ {\cal I}_L^-(\Delta_{0g})\right\}\right]\nonumber \\
    &=&-I_R(t)=I(t).   
    \label{homo_current}
\end{eqnarray}
$I_L=-I_R$ is due to electric current conservation {\corr and $\rho_1=\rho_{gg}+\rho_{ee}\approx 1$ if the populations of the transient states ($0, 2$) are negligible, which we assume in the following}. Equation~\eqref{homo_current}, employs again the notation $g$ to refer to the ground state for a more general treatment instead of $\downarrow$, accounting for the situation in which the spin is tilted by the magnetic field with respect to the spin polarization on $z$. We do this because the eigenstates are always defined {\corr in the axis where the QI Hamiltonian is diagonal}, see Appendix~\ref{app:spin_expectation}, {\corr parallel to the magnetic field here, as we have mentioned before.} 
As a consequence of choosing the $z$-axis as the magnetization direction of the leads, where the spin polarization lies, the electric current becomes a measurement of the expectation value $\langle s_z \rangle $ as Eq.~\eqref{homo_current} shows. In general, for any magnetization orientation of the spin-polarized lead, we expect that the current will measure the spin component parallel to it, {\corr in agreement with the physical interpretation in the experiments~\expSTMESR.}

Assuming that the magnetic field lies in the $xz$ plane, $\mathbf{B}=B(\sin\theta,0,\cos\theta)$, see Appendix~\ref{app:spin_expectation}, we obtain:
\begin{eqnarray}
    \langle s_z \rangle &=& \frac{\hbar}{2} (\rho_{ee}-\rho_{gg}) \cos\theta + \frac{\hbar}{2} (\rho_{ge}+\rho_{eg}) \sin \theta  ,
    \label{General_Sz}
\end{eqnarray}
{\corr which does not depend on $\rho_0$ and $\rho_2$, the populations of the unoccupied and doubly states, since they provide zero spin to the simplest QI.} $\theta$, also referred to as \textit{homodyne angle}, is the angle formed between the direction of the left electrode's magnetization, fixed along the $z$-axis, and the impurity spin when the $g$-factor is isotropic. We should emphasize that in Eq.~\eqref{General_Sz}, the main contribution to the populations comes from the Floquet number $n = 0$, while the contribution from the coherences arises from $n = 1$~\cite{J_Reina_Galvez_2023}. Thus, the coherences are intrinsically time-dependent in the long-time limit. For the sake of completeness, the in-plane components of the spin expectation value are:
\begin{eqnarray}
    \langle{s_x}\rangle &=&\frac{\hbar}{2}
(\rho_{ee}-\rho_{gg})\sin\theta -\frac{\hbar}{2}(\rho_{ge} + \rho_{eg})\cos\theta, \label{Sx_ave_eq}\\
\langle{s_y}\rangle &=&\frac{i\hbar}{2}
(\rho_{ge} - \rho_{eg})=-\hbar \mbox{Im}(\rho_{ge}).
\label{Sy_ave_eq}
\end{eqnarray} 

The reader might wonder why the expectation value of the spin along $z$ depends on the coherences and not only on the difference in populations or why the $x$-component depends on the population. The reason is the inclusion of mixed states as a consequence of the magnetic field being tilted from the $z$ axis. Mathematically, Eq.~\eqref{General_Sz} arises from changing the basis set from the classical up and down states of the Pauli matrix to the eigenstate basis set of the quantum impurity (QI) since ${\hat{s}}_i$ has to be expressed in the same basis as the reduced density matrix, meaning that we cannot simply use the standard Pauli matrices representation. This can be interpreted as a projection of the populations and coherences, defined in the {\corr axis where the QI Hamiltonian is diagonal}, onto the Cartesian coordinates. Equivalently, one could rotate the cartesian coordinates to the magnetic field direction, changing the basis of the reduced density matrix instead, a procedure done in the literature \cite{JK_vector_field_2021,Fe_driving_SH_2023,kovarik_spin_2024,kovarik_electron_2022,Yi_Chen_Harnessing_2023}. We emphasize that $\langle \mathbf{s}\rangle/\hbar$ describes a quantum-statistical average and every component can adopt any number between $-1/2$ and $1/2$.

Introducing Eq.~\eqref{General_Sz} into Eq.~\eqref{homo_current} allows us to obtain an electric current equation that clearly shows the dependence on $\theta$. To simplify the result, we place ourselves in the long time where the populations are constant, i.e. $\rho_{ee( g g)}=P_{e( g)}$. Additionally, we consider that the linear driving term dominates ($A_L^2\sim 0$) so that the coherences are continuously Larmor precessing only, i.e. $\rho_{ g e}(t)=\rho_{ g e,-1}e^{i \omega t}+\rho_{ g e,1}e^{-i \omega t}$, where the $\pm 1$ in the coherences refers to its Floquet number~\cite{J_Reina_Galvez_2023}. This yields:
\begin{eqnarray}
   I_L &\approx& \left(\mbox{Re} \left\{ {\cal I}_L^+(\Delta_{2 e})\right\} - \mbox{Re} \left\{ {\cal I}_L^-(\Delta_{0 e})\right\}\right) \nonumber \\  
   &-& P_L\left[(P_{ e}-P_{ g}) \cos\theta +2A_L (\rho_{ g e,1}   
   +\rho_{ e g,-1}) \sin\theta\right] \nonumber \\
   &\times&  \left(\mbox{Re} \left\{ {\cal I}_L^+(\Delta_{2 e})\right\} + \mbox{Re} \left\{ {\cal I}_L^-(\Delta_{0 e})\right\}\right) \implies \nonumber \\
      I_L &\approx& I_0 - \underbrace{P_L\left[\Delta P_{eg} \cos\theta +2A_L (\rho_{ge,1}+\rho_{eg,-1}) \sin\theta\right] I_p}_{\Delta I} \nonumber \\
   \label{homo_current_final}
\end{eqnarray}
where the new definitions $I_0,\ I_p$, containing a non-trivial DC bias dependence, see Eqs.~\eqref{int} and~\eqref{int_s}, and $\Delta P_{eg}$ can be deduced by comparing terms. {\corr Later on, we conveniently write $\Delta I = I_L - I_\mathrm{BG}$, where the background current (i.e., baseline or off-resonant current) is defined as $I_\mathrm{BG} = I_L - \Delta I = I_0 - 2\Delta I$, which depends on the spin polarization $P_L$ and  the homodyne angle $\theta$.
}

Equation~\eqref{homo_current_final} highlights the role of spin polarization. For $P_L=0$, the electric current simply has a constant unpolarized value, originating from electron or hole hopping from the left to right (or vice versa) through the QI. In this case, the current is independent of the QI populations and coherences, rendering it insensitive to changes in the QI state. Consequently, it cannot induce spin flips or resonant behavior, even under AC driving. A finite spin polarization $P_L\neq 0$ breaks this situation, producing an imbalance between up and down states that modifies the tunneling magnetoresistance.

In this regime, a polarized electric current emerges and $\langle s_z\rangle$ becomes relevant in Eq.~\eqref{homo_current_final}. On resonance and in the long time limit in the CB regime, the two singly occupied levels are equally populated ($1/2$ each) and $\langle s_z\rangle\rightarrow 0$, assuming finite but small coherences. By contrast, for $P_L\neq 0$ and $A_L=0$, the QI retains a net spin polarization $\langle s_z\rangle\neq 0$, but the absence of coherences prevents population exchange, and no resonance occurs at the Larmor frequency.
Equation~\eqref{homo_current_final} also reveals the dependence on the homodyne angle $\theta$. For $\theta=0$, $\sin\theta=0$, no population exchange takes place, mirroring the undriven case. At $\theta=\pi/2$, the current becomes independent of the populations and the ESR signal is governed solely by the coherences. Thus, Eq.~\eqref{homo_current_final} demonstrates that homodyne detection arises from both the misalignment between the tip and impurity spin and the presence of AC {\corr response being measured into the DC current.}

\subsection{Mapping onto a spin problem}

Following Ref.~\cite{Braun_Konig_prb_2004}, we now aim to map the QME, Eq.~\eqref{rho_master_eq_6_0}, onto a system of equations for the spin. This will become particularly useful for building some physical intuition on the definitions of the Rabi rate and the renormalization of the resonance energy. Using Eqs.~\eqref{General_Sz},~\eqref{Sx_ave_eq} and~\eqref{Sy_ave_eq}, the system of equations for the spin components and the population of the transient states is\footnote{Equations~\eqref{spin_equations} and~\eqref{spin_equations_b} assume a negligible Zeeman energy in the rates compared to the ionization and charge energy, $\varepsilon$ and $\varepsilon + U$ {\corr in order to simplify the equations. This is valid s}ince these energy thresholds are $\sim 100$ meV or more while $\Delta_{eg}=g\mu_B B < 0.1$ meV in the STM-ESR context \expSTMESR. This allows us to write $\sum_\alpha P_\alpha \left[\mbox{Im}(\Gamma_{ g 0,0  g,\alpha}^{P_\alpha=0}- \Gamma_{ g 2,2  g,\alpha}^{P_\alpha=0}) \right] \sin\theta = \sum_{v=0,2} \mbox{Im} (\Gamma_{ e v,v  g}(t)) $, which further simplifies the spin system of equations; see Appendix~\ref{app:mapping_onto_spin}{\corr.~\textit{However, in the numerical simulations, the Zeeman energy is not neglected anywhere.} } }:
\begin{subequations}\label{eq:spinrho}
\begin{equation}
  \begin{aligned}
   \hbar\langle \dot{s}_z \rangle &=\Delta_{ g e}\langle {s}_y \rangle\sin\theta  - \hbar\langle {s}_z \rangle /\tau_{\mathrm{rel}}  +\hbar \langle \dot{s}_z \rangle_{\mathrm{acc}} \\
    \hbar\langle \dot{s}_x \rangle &=\left[ 2\sum_{v=0,2} \mbox{Im} (\Gamma_{ e v,v  g}(t))/\sin\theta - \Delta_{ g e}\cos\theta\right]\langle {s}_y \rangle \\
    &- \hbar\langle {s}_x \rangle /\tau_{\mathrm{rel}}    \\
    \hbar\langle \dot{s}_y \rangle &=
   -2\sum_{v=0,2}\mbox{Im} (\Gamma_{ e v,v  g}(t)) \langle s_x\rangle/\sin\theta 
   -  \hbar\langle {s}_y \rangle /\tau_{\mathrm{rel}}  \\ 
   &- \Delta_{ g e} \left(\langle s_z\rangle \sin\theta -\langle s_x\rangle \cos\theta \right)
   \label{spin_equations}
  \end{aligned}
\end{equation}
\begin{equation}
 \begin{aligned}
  \hbar\dot{\rho}_0 &= 2\mbox{Re}(\Gamma_{ g 0,0 g}^{P_\alpha=0}(t))(1-\rho_0-\rho_2) -4\mbox{Re}(\Gamma_{ g 0,0 g}^{P_\alpha=0}(t))\rho_0 \\
  &+4\sum_\alpha \mbox{Re}(\Gamma_{ g 0,0 g,\alpha}^{P_\alpha=0}(t)) P_\alpha \langle s_z\rangle /\hbar   \\
  \hbar\dot{\rho}_2 &= 2\mbox{Re}(\Gamma_{ g 2,2 g}^{P_\alpha=0}(t))(1-\rho_0-\rho_2)-4\mbox{Re}(\Gamma_{ g 2,2 g}^{P_\alpha=0}(t))\rho_2 \\
  &-4\sum_\alpha \mbox{Re}(\Gamma_{ g 2,2 g,\alpha}^{P_\alpha=0}(t)) P_\alpha \langle s_z\rangle /\hbar .
  \label{spin_equations_b}
 \end{aligned}
\end{equation}
\end{subequations}

The notation $\Gamma_{lj,uv}^{P_\alpha=0}(t)$ indicates zero spin polarization for the rates in Eq.~\eqref{rate}. 
It is important to emphasize that the rates $ \Gamma_{ e v,v g}(t)$, that “flip'' the excited state to the ground one or vice versa, are proportional to the spin polarization and $\sin\theta$, making them a consequence of both the lead spin polarization and the homodyne angle, i.e. those rates are exactly zero when $P_L=0$ and $\theta=0$. Thus $2\sum_{v=0,2} \mbox{Im} (\Gamma_{ e v,v  g}(t))/\sin\theta$ \textit{does not depend on} $\theta$.
 
Equations~\eqref{spin_equations} introduces the spin relaxation rate $1 /\tau_{\mathrm{rel}}$ and the non-equilibrium spin accumulation $\langle \dot{s}_z \rangle_{\mathrm{acc}}$, parallel to the spin polarization as defined in Ref.~\cite{Braun_Konig_prb_2004},
\begin{equation}
    \begin{aligned}
      1/\tau_{\mathrm{rel}} &= \sum_{v=0,2}\mbox{Re}(\Gamma_{ g v,v g}^{P_\alpha=0}(t)+\Gamma_{ e v,v e}^{P_\alpha=0}(t))/\hbar  
       \label{spin_rel}
       \end{aligned}
\end{equation}
\begin{equation}
    \begin{aligned}
     \langle \dot{s}_z \rangle_{\mathrm{acc}} &=2\sum_\alpha P_\alpha \Bigg[\mbox{Re}(\Gamma_{0 g, g 0,\alpha}^{P_\alpha=0}(t))\rho_0 -\mbox{Re}(\Gamma_{2 g, g 2,\alpha}^{P_\alpha=0}(t))\rho_2 \\
     &+ (\mbox{Re}(\Gamma_{ g 2,2 g,\alpha}^{P_\alpha=0}(t)-\Gamma_{ g 0,0 g,\alpha}^{P_\alpha=0}(t))) (1-\rho_0-\rho_2)/2\Bigg],
       \label{spin_acc}
       \end{aligned}
\end{equation}

The non-equilibrium \textbf{spin accumulation} term, Eq.~\eqref{spin_acc}, depends on the populations of transient and single-occupied states of the QI, and describes the spin accumulation via tunneling to and from spin-polarized leads, which renders it proportional to the current. In short, the spin accumulation term is a \textit{source term that builds up the spin polarization on the QI}. {\corr The time-independent spin accumulation defines a characteristic timescale, $\hbar / \langle \dot{s}_z \rangle_\mathrm{acc}$, over which it acts on the spin. For DC voltages within the CB regime, the effect of the spin accumulation can be partly neglected, since spin relaxation time is much shorter than the spin accumulation time and dominates the spin dynamics.}

The \textbf{spin relaxation} rate Eq.~\eqref{spin_rel} describes the \textit{decay of every spin component towards the steady state} by electrons with a specific spin tunneling out of the QI, a process with $v=0$, or by tunneling in of a second electron with an opposite spin state, forming a doubly occupied state ($v=2$). The spin accumulation and relaxation govern the spin dynamics of the QI if the spin polarization and magnetic field are parallel, $\theta=0$. However, this is not the complete picture of the dynamics since the angle $\theta$ enters Eqs.~\eqref{spin_equations}. This orientation dependence can be encompassed in the components of the following rotational term:
\begin{equation}
    \begin{aligned}
      \langle \mathbf{\dot{s}} \rangle_{\mathrm{rot}} &= \langle\mathbf{s}\rangle \times \sum_\alpha g\mu_B \mathbf{B_\alpha}(t)/\hbar 
       \label{spin_rot}
       \end{aligned}
\end{equation}
A direct comparison with Eq.~\eqref{spin_equations} shows that:
\begin{equation}
    \mathbf{B_\alpha}(t)=\left( B\sin\theta,0,\frac{2 \sum_{v}\mbox{Im} (\Gamma_{ e v,v  g,\alpha}(t))}{g\mu_B\sin\theta}+  B\cos\theta\right).
    \label{Btotal}
\end{equation}
In this expression, we identify the so-called exchange magnetic field as 
\begin{eqnarray}
    &&g\mu_B \mathbf{B_{exch}}(t)=2 \sum_{v=0,2}\left[\mbox{Im} (\Gamma_{ e v,v  g}(t))/\sin\theta\right] \mathbf{z}\nonumber \\ &=&
-\sum_\alpha\frac{ \gamma_{\alpha}(t)}{2\pi}P_{\alpha}\int^{\infty}_{-\infty}  d \epsilon \left(\frac{\epsilon-\varepsilon}{(\epsilon-\varepsilon)^2+(\hbar/\tau_c)^2}f_\alpha^- (\epsilon)\right.  \nonumber \\
		&+& \left.  \frac{\epsilon-\varepsilon-U}{(\epsilon-\varepsilon-U)^2+(\hbar/\tau_c)^2}f_\alpha^+ (\epsilon) \right)\ \mathbf{z},
\label{exchange_field}
\end{eqnarray}
where we have inserted the definition of the imaginary part of the rate. The name of this effective field originates from the fact that it arises due to virtual particle exchange between the electrode and the QI. This field is directly proportional to the electrode's spin polarization and vanishes in the electron-hole symmetry point $eV_\mathrm{DC}=\varepsilon+U/2$, where we have assumed that the voltage is applied to the left {\corr electro}chemical potential while the right one is grounded so $eV_\mathrm{DC}=\mu_L-\cancel{\mu_R}^{0}$. 
Furthermore, the exchange field vanishes in the absence of Coulomb interaction ($U = 0$), demonstrating that its origin is a quantum many-body effect. It is important to notice that Eq.~\eqref{exchange_field} does not change the charge of the QI, contrary to the spin accumulation term. Therefore, it can be finite even at zero voltage, indicating its independence from the DC electric current. However, Eq.~\eqref{exchange_field} does depend on the charging energies $\varepsilon,\ \varepsilon+U$ and therefore can be understood as an exchange field due to virtual particle exchange with a spin-polarized lead.

In the CB regime, where the spin accumulation is small, the dynamics of the spin are dominated by Eq.~\eqref{spin_rot} as long as $\theta\neq 0,\pi$. These two extreme cases are interesting though since they imply a simple relaxation of the $z$ component of the spin to the steady without any evolution of the $x$ and $y$ components, see Eqs.~\eqref{spin_equations}. In the absence of driving, the exchange field will generate a spin splitting of the QI, a Zeeman splitting, even at $B=0$. When the driving is turned on, both the rotation and accumulation term will exhibit time-dependent behavior that will yield FLT and STT resonance signals in the frequency domain respectively.

\section{Hopping-driven SAIM: Rabi rate, \texorpdfstring{$T_2$}{} times and energy dressing}

In the following, we will discuss features of our model, continuing to use a spin-1/2 impurity. This approach results in shorter and more manageable expressions, leading to a more intuitive understanding of the underlying physics. We emphasize that while higher-spin systems can significantly alter the equations presented here by introducing additional singly and transient states, the central conclusions regarding the driving mechanism (FLT and STT) and the transport (above or below the energy thresholds) will remain valid.

\subsection{Rabi rate}

In previous studies, the Rabi frequency was introduced as a mathematical consequence of the model, conveying that the modulation of the tunneling barrier generates a driving of reasonable amplitude~\cite{J_Reina_Galvez_2019,J_Reina_Galvez_2021,J_Reina_Galvez_2023}. {\corr While formally correct, that presentation does not give an immediate, physical intuition for how the tunneling-modulation acts on the spin. It is therefore instructive to recast the driving mechanism into a more intuitive picture and to examine explicitly the equations of motion for the spin components in the two driving regimes considered here: FLT and STT.

For the FLT-driven regime, which is predominant in the CB regime, the equations reproduce the typical behavior of a spin under a time-dependent magnetic field, the exchange field. Thus the impurity undergoes Rabi oscillations governed by the torque term (the cross product in Eq.~\eqref{spin_rot}), and the observed ESR signal is strongly shaped by homodyne readout. In this regime the Rabi frequency is set by the component of the time-dependent exchange field that acts perpendicular to the instantaneous spin (i.e., the torque projection).
By contrast, in the STT-driven regime, dominant above the energy thresholds that delimit the CB regime, the time-dependent spin accumulation governs the ESR signal. Physically, spin-transfer torque originates from the frequent repeated transfer of tunneling electrons with a specific polarization (orientation), generating a non-equilibrium, dissipative torque that drives the impurity incoherently toward a new polarized steady state. Since the populations are not being affected, the ESR signal is fully controlled by the coherences and proportional to $\sin\theta$; thus, it presents neither asymmetric contribution nor homodyne dependence.}

\subsubsection{Rabi rate in the FLT-driven regime}

There are two essential ingredients for the Rabi rate to emerge naturally in the QME: the spin polarization of one of the leads and the driving amplitude. We already showed the consequence of introducing polarization on the rates by mapping the QME onto a spin problem in Eqs.~\eqref{spin_equations} and~\eqref{spin_equations_b}: the spin accumulation and rotation terms emerge as convenient descriptors for the spin dynamics. In the CB regime, the spin accumulation is rather small and can be partly discarded. Thus, only the rotation and relaxation terms strongly influence the time evolution of the spin. 

When the exchange field and the spin form an angle $\theta \neq 0,\pi$, the former exerts a torque-like force on the initial spin, while spin relaxation ensures convergence to the steady state. If this FLT generated by the exchange field is time-dependent, it follows that Rabi oscillations will be induced during the relaxation process, $\pi/2$ radians out of phase with the driving term and the STT-driven mechanism~\cite{kovarik_spin_2024,Kubota2008_nature_phys}. The corresponding Rabi rate is determined by the component of the exchange field acting as the torque on the spin, namely:
    $\Omega_{\mathrm{FLT}}\propto B_{\mathrm{exch}} \cos (\pi/2-\theta). $

Looking at Eq.~\eqref{exchange_field} we can easily identify
\begin{equation}
    2\hbar\Omega_{\mathrm{FLT}}= g\mu_B B_{\mathrm{exch}} \sin (\theta)=2\sum_{v=0,2}\mbox{Im}(\Gamma_{ g v,v  e}).
    \label{Rabi_simple}
\end{equation}
This simple expression for the Rabi rate highlights its physical meaning: it corresponds to the projection of the exchange field onto the axis perpendicular to the spin alignment (the {\corr eigenbasis} axis) within the plane defined by the spin and the exchange field. A comparison between the QME and the optical Bloch equations leads to the same identification of the Rabi rate as the imaginary part of the spin-flip rate, since only the imaginary part generates oscillatory dynamics. As a direct consequence, the Rabi rate is maximized at $\pi/2 + m\pi$, with $m$ an integer. However, this maximum does not necessarily translate into a stronger ESR signal amplitude because of the homodyne detection mechanism: the ESR signal measured through the tunneling current reflects only the $z$ component of the spin, as will be demonstrated in the simulation results.

Note that the exchange field includes up to second harmonic generation since 
\begin{equation}
    \gamma_{\alpha}(t)=\gamma_{\alpha}^0\left[1+\frac{A_\alpha^2}{2} +2A_\alpha \cos(\omega t) + \frac{A_\alpha^2}{2} \cos(2\omega t) \right],
    \label{coupling_full}
\end{equation}
but it could be expanded to higher harmonics if one considers the full Bessel function expression~\cite{jauho_wingreen_prb_1994,arrachea-moskalets-2006} or a more complex oscillating behavior in~\ref{coupling_full}. Following this, we can define the Rabi rates per harmonic $n=0,1,2$ as:
\begin{eqnarray*}
    \hbar\Omega^{0\omega}_{\mathrm{FLT}}&=& \frac{1}{2} g\mu_B B_{\mathrm{exch}}^z(A_L=0) \left(1+A_L^2/2 \right) \sin\theta \\
    \hbar\Omega^{1\omega}_{\mathrm{FLT}}&=& g\mu_B B_{\mathrm{exch}}^z(A_L=0) A_L \sin\theta \\
    \hbar\Omega^{2\omega}_{\mathrm{FLT}}&=& \frac{1}{4}g\mu_B B_{\mathrm{exch}}^z(A_L=0) A_L^2 \sin\theta
\end{eqnarray*}
Therefore, the exchange field provides zero-order and higher harmonic signals directly, with the latter in agreement with recent experiments on a strongly driven pentacene molecule spin-1/2 system~\cite{2ndharmonic_paper}
The $n$-th harmonic signal appears at a frequency of $f_{1\omega}/n$, with $f_{1\omega}$ being the first harmonic resonance frequency and $n>0$ the harmonic number. In this way, the second harmonic signal provides an additional resonance at $f_{1\omega}/2$ in the frequency spectrum. This can be easily verified since this 2th harmonic signal will emerge when $ 2\hbar\omega=4\pi \hbar f_{2\omega} = \Delta_{eg}\implies f_{2\omega}=\Delta_{eg}/2\hbar=f_{1\omega}/2$. If the frequency is fixed whilst sweeping the magnetic field, the $n$-th resonance magnetic field will roughly follow the relation $nB_{1\omega}$ with $B_{1\omega}$ the magnetic field of the first resonance. The zero-harmonic feature is a special case, and it is found at zero frequency in the frequency domain and near zero in the magnetic field domain. 

One can show that the resonance for the zero-harmonic signal occurs when $B=-B_{\mathrm{exch}}\cos\theta$ when the frequency is fixed. This suggests that this signal position can be used as a measurement of the exchange field. 
From our model's perspective, the origin of this zero-field feature arises from spin-polarized electron precession in a magnetic field generated by the exchange interaction with the spin polarized tip. {\corr Nonetheless, experimental observation of this ``resonance'' is still lacking in the CB regime,
but they could offer coherent control of a single spin at zero AC driving, using only DC voltages.} In the absence of driving, no ESR signal is expected for any harmonic when sweeping over the frequency for a fixed magnetic field, even at zero frequency, since the $\cos(\omega t)$ term vanishes in the rates.

\subsubsection{Rabi rate in the STT-driven regime}

When increasing the DC bias $eV_\mathrm{DC}$ beyond the energy thresholds, $eV_\mathrm{DC}< \varepsilon$ and $eV_\mathrm{DC}>\varepsilon+U$, we \textit{are outside the CB regime} and the electric current through the system drastically increases at constant tip-atom distance. In this regime, the role of the exchange field is suppressed by the dominance of the spin accumulation term and the rotation term starts to decrease logarithmically. This does not imply the lack of an ESR signal when the QI is driven in the STT regime. In fact, we can define the Rabi rates in this regime since the rates in the spin accumulation term (Eq.~\eqref{spin_acc}) can be written as the real part of a rate linked to a Rabi process by applying the following relations:
\begin{eqnarray*}
        \hbar\Omega_{\mathrm{STT}}(t) &=& \sum_{v=0,2}\mbox{Re}(\Gamma_{ g v,v e}(t))\\
        &=&\sum_\alpha P_\alpha\left[\mbox{Re}(\Gamma_{ g 0,0 g,\alpha}^{P_\alpha=0}(t) - \Gamma_{ g 2,2 g,\alpha}^{P_\alpha=0}(t))\right]\sin\theta \nonumber\\
        \hbar\Omega_{\mathrm{STT}}^{v=0,2}(t) &=& \mbox{Re}(\Gamma_{v  g,  e v}(t))\\
        &=&\pm\sum_\alpha\mbox{Re}(\Gamma_{v g, g v,\alpha}^{P_\alpha=0}(t))P_\alpha\sin\theta ,
\end{eqnarray*}
where the $+$ sign applies to $v=0$ while the $-$ sign is for $v=2$. The definitions of the Rabi rates above are easily confirmed by comparing our QME with the optical Bloch equation and considering that the Rabi rate is a complex number with a real and an imaginary part. If driving is applied, the Rabi rates above will modulate the spin accumulation, providing an ESR purely governed by the coherences~\cite{J_Reina_Galvez_2023}. On resonance, this forces the spin to undergo Larmor precession, in phase with the driving term, while it rapidly evolves to the steady state.  
{\corr The precession occurs in the plane perpendicular to the {\corr eigenbasis axis}, parallel to the magnetic field, so the populations, always defined {\corr with respect to that axis for us}, remain unaltered even on resonance, a result that persists in the long time limit situation. 
}

As a consequence of the populations remaining unchanged even on resonance,  the spin will not perform Rabi oscillations since, essentially, it is being forcedly locked (accumulated) in the direction set by the external magnetic field. Moreover, the ESR signal does not follow the homodyne detection~\cite{JK_vector_field_2021,Fe_driving_SH_2023}, leading to a fully symmetric resonance {\corr in the absence of exchange field}~\cite{J_Reina_Galvez_2023}. Another key feature of this regime is that Rabi rates are proportional to the electric current, exhibiting anharmonic behavior, since both the current and the Rabi rate depend on the coupling to the tip. As a result, the STT-driven regime naturally gives rise to 0-th and higher harmonic resonances~\cite{2ndharmonic_paper}, similar to what the FLT-driven regime does. In Ref.~\cite{kovarik_spin_2024}, the authors observed a peak near zero magnetic field with an intensity comparable to or slightly higher than that of the first harmonic peak. Notably, this zero-field signal appears even in the absence of driving. 
Since these measurements were obtained in the STT-driven regime, explanations based on spin bottlenecking~\cite{Stephen_Flatte_B0_PRL_2020} or the Hanle effect~\cite{M_Braun_2005_hanle_seq,Piotr_Martinek_Hanle_effect_2023} are quite suitable.  {\corr Physically, our 0-th harmonic is consistent with these interpretations.}

It is important to emphasize the inherent competition between the external magnetic field and the coupling to the spin-polarized electrode, which determines where the spin will accumulate, as discussed in~\cite{Braun_Konig_prb_2004}. {\corr In the undriven regime, i}f $g\mu_B B \gg \gamma_L^0$, the spin-accumulation term is too weak to align the spin with the $z$-axis, and instead the spin accumulates {\corr approximately parallel to the magnetic field orientation. This alignment is not perfect: the residual influence of spin accumulation together with the torque exerted by the exchange field prevent the spin from pointing fully along $\mathbf{B}$.} Conversely, if $g\mu_B B \ll \gamma_L^0$, the spin-accumulation term dominates, forcing the spin to align with the $z$-axis. This mechanism underlies the Hanle effect outside the CB regime~\cite{M_Braun_2005_hanle_seq,Piotr_Martinek_Hanle_effect_2023}.

In both cases, the sign of the spin accumulation determines whether the QI is spin-polarized positively or negatively, independent of the magnetic-field orientation. In this work, we choose coupling and magnetic field values that realistically reflect experimental conditions, reproducing FLT Rabi frequencies consistent with experiment and placing us in the $g\mu_B B \gg \gamma_L^0$ regime. {\corr When AC driving is applied, however, the situation changes in the rotating frame and on resonance as $\theta$ varies from out-of-plane to in-plane orientations. In the latter case, the spin accumulates along the $z$-axis in the rotating frame, while for intermediate angles there is competition between accumulation along $z$ and parallel to the magnetic field.}

\subsubsection{Analytical form of the Rabi rate}

We could encompass both the FLT and STT Rabi rates in a complex rate term as $\hbar \Omega(t) =\hbar (\Omega_{\mathrm{STT}}(t)+i\Omega_{\mathrm{FLT}}(t))=\sum_{v=0,2}\Gamma_{ g v,v  e}(t)$. Every rate is calculated numerically in the wide band limit,
which mandates a cutoff in the integral.
If the cutoff energy, $E_c$, is much larger than any other energy in the QI, i.e. $E_c\gg |\varepsilon|,\ |\varepsilon+U|$, the rates will converge to a value independent of the cutoff~\cite{Braun_Konig_prb_2004,Piotr_Martinek_2021_JMMM,Piotr_Martinek_Hanle_effect_2023,Busz_PRB_2025}, which is the situation in all our calculations. In this case, at extremely low temperatures, $k_B T_{\alpha}\ll |\varepsilon-\mu_\alpha|,|\varepsilon+U-\mu_\alpha|$, we can assume that the Fermi distribution function behaves as a step function\footnote{An analytical expression for the imaginary part of the Rabi rate at finite temperatures can be expressed using digamma functions \cite{Busz_PRB_2025}, but we are interested in the low temperature scenario.}:$
f_\alpha^+(\epsilon)\approx 1-\Theta(\epsilon-\mu_\alpha),\  
 f_\alpha^-(\epsilon) \approx \Theta(\epsilon-\mu_\alpha),$
with $\Theta(x)$ being the Heaviside step function.  This approximation allows us to write the simplest analytical form of the Rabi rate for a spin-$1/2$ system as:
\begin{eqnarray}
    \hbar\Omega(t)&=& -\sum_\alpha\frac{1}{4\pi}\gamma_{\alpha}^0 (1+A_\alpha\cos(\omega t))^2 P_{\alpha}\sin\theta  \nonumber \\
    &\times&\left[ \arctan\left(\frac{\mu_\alpha-\varepsilon-U}{\hbar/\tau_c} \right)+ \arctan\left(\frac{\mu_\alpha-\varepsilon}{\hbar/\tau_c}\right) \right. \nonumber \\
    &+&\left. \frac{i}{2}\ln\left(\frac{(\mu_\alpha-\varepsilon-U)^2+(\hbar/\tau_c)^2}{(\mu_\alpha-\varepsilon)^2+(\hbar/\tau_c)^2}\right)\right].
  \label{Rabi_eq_analytical}
\end{eqnarray}

In the limit $\hbar/\tau_c \ll |\varepsilon|, \varepsilon + U$, the real part of the Rabi frequency becomes close to zero in the CB regime, which results in nearly zero current as well. In contrast, outside of the CB regime, the sum of the $\arctan$ terms gives $\pi\mbox{sign}(\mu_L)=\pi\mbox{sign}(V_\mathrm{DC})$, which depends on the sign of the {\corr electro}chemical potential of the polarized electrode. On the other side, the imaginary part of Eq.~\eqref{Rabi_eq_analytical} does not vanish in the CB regime, indicating its independence from the current, as we mentioned before. 

\subsection{Spin relaxation time and quality factor \texorpdfstring{$\Omega T_2$}{}}
Let us move on to the spin relaxation Eq.~\eqref{spin_rel}. Since our model does not consider any pure dephasing mechanism and we have not included cotunneling processes, the time independent part of $\tau_{\mathrm{rel}}$ is actually the coherence time $T_2$, which is purely lifetime-limited so that $T_2=2T_1$. Assuming low temperature again, one arrives at an analytical expression for $T_2$:
\begin{eqnarray}
	\frac{1}{T_2}&=&\frac{1}{h}\sum_\alpha \gamma_{\alpha}^0\left(1+\frac{A_\alpha^2}{2} \right)\nonumber \\ & \times &\int^\infty_{-\infty} d \epsilon \left[f_\alpha^- (\epsilon)\frac{\hbar/\tau_c}{(\epsilon-\varepsilon)^2+(\hbar/\tau_c)^2} \right. \nonumber \\
  &+& \left. f_\alpha^+ (\epsilon) \frac{\hbar/\tau_c}{(\epsilon-\varepsilon-U)^2+(\hbar/\tau_c)^2} \right] \nonumber \\
  &\approx&\frac{1}{h}\sum_\alpha \gamma_{\alpha}^0\left(1+\frac{A_\alpha^2}{2} \right)\left[\pi-\arctan\left(\frac{\mu_\alpha-\varepsilon}{\hbar/\tau_c}\right)\right.\nonumber \\
  &+& \left. \arctan\left(\frac{\mu_\alpha-\varepsilon-U}{\hbar/\tau_c} \right)\right],
\label{T2_spin_one_half}
\end{eqnarray}
An important feature of Eq.~\eqref{T2_spin_one_half} is that it saturates rapidly as we increase $|V_\mathrm{DC}|$ outside of the CB regime to
\begin{equation}
    \frac{1}{T_2}=\frac{1}{2\hbar} \gamma_{L}^0\left(1+\frac{A_L^2}{2} \right),
    \label{T2_high_Vdc}
\end{equation}
where the right {\corr electro}chemical potential is grounded and its contribution can be discarded compared to the left one for $eV_\mathrm{DC}=\mu_L<\varepsilon$ and $eV_\mathrm{DC}>\varepsilon+U$. Equation~\eqref{T2_high_Vdc} does not depend on the induced broadening $\hbar/\tau_c$, since the DC bias is above the energy thresholds and is essentially controlled by the time independent left coupling $\gamma^0_L$, which is where the driving is applied and the tip-atom distance is encoded.

In the CB regime, however, the coherence times depend on the induced broadening, and a finite value of $\hbar/\tau_c$ prevents $T_2$ from diverging. This feature is a clear limitation of the model, as it was mentioned before in the introduction, and going beyond it requires extending to higher orders in perturbation theory of the tunneling amplitude $t_\alpha$, which will lead to cotunneling~\cite{Weymann_2006_seq_cotu,J_Reina_Galvez_2019}. Future extensions of this model will aim to circumvent such limitation, nonetheless, we set $\hbar/\tau_c\ll |\varepsilon|,\varepsilon+U$ so its influence on Eq.~\eqref{T2_spin_one_half} is minimized. In this way, we can explore the behavior of $T_2$ in the CB regime, although, at low DC voltage, we expect low current in line with previous studies~\cite{Braun_Konig_prb_2004,A_P_Jauho_2006,Weymann_2006_seq_cotu,Piotr_Martinek_2021_JMMM,Piotr_Martinek_Hanle_effect_2023,Busz_PRB_2025} as the cotunneling regime becomes more significant in the deep CB regime. 

The product $\Omega T_2$ per harmonic is the quality factor of a qubit, giving the number of Rabi oscillation the spin performs on resonance before reaching the steady state. We focus on the Rabi frequency related to $1\omega$, the first harmonic, with $eV_\mathrm{DC}=\mu_L$, and the spin polarization and driving only applied to the left lead. 
When the QI is driven under STT, the real part of the Rabi rate dominates, and we obtain:
\begin{equation}
    \Omega_{\mathrm{STT}}^{1\omega}T_2 \approx -  \frac{\mathrm{sign}(V_\mathrm{DC})\sin\theta}{2}  \frac{A_L P_L}{\left(1+\frac{A_L^2}{2} \right)}.
    \label{RabiT2_SEQ}
\end{equation}

Equation~\eqref{RabiT2_SEQ} is proportional to the ESR amplitude and already suggests an optimal value for the driving. Additionally, it indicates that the ESR signal is independent of the couplings to the leads. If the driving is increased substantially, Eq.~\eqref{RabiT2_SEQ} will tend toward zero, and the ESR signal will diminish as well. Specifically, the driving value that maximizes the first harmonic signal is $\sqrt{2}$, yielding $\Omega_{\mathrm{STT}}^{1\omega} T_2 = -\mathrm{sign}(V_\mathrm{DC}) \sin \theta P_L \sqrt{2}/4$. Even in the ideal scenario of a fully polarized electrode and a magnetic field perpendicular to the exchange field, $|\Omega_{\mathrm{STT}}^{1\omega} T_2| =\sqrt{2}/4 \approx 0.35$, meaning that in the STT-driven regime, we are unable to create a robust spin-qubit, as it will complete barely 1/3 of a Rabi oscillation before fully decaying to the steady state. This does not improve further if we consider the surviving Rabi rate of the FLT regime instead. Therefore, the spin is always overdamped in this scenario, and the ESR driven by STT is mainly useful for polarizing the QI as an initialization step, although the ESR signal can still be reasonably strong under suitable conditions~\cite{J_Reina_Galvez_2021,J_Reina_Galvez_2023}. 

On the other hand, the quality factor in the FLT-driven regime, $\Omega_{\mathrm{FLT}}^{1\omega} T_2$, offers more flexibility to achieve larger values because, first, the Rabi rate does not vanish in the CB regime, and second, a reasonable $T_2$ can be maintained in this regime while keeping the induced broadening $\hbar/\tau_c$ low. Considering again that the DC bias applied to the tip and the sample is grounded, we arrive at: 

\begin{widetext}
   \begin{eqnarray}
    \Omega_{\mathrm{FLT}}^{1\omega} T_2\approx \frac{-i A_L \gamma_{L}^0 P_{L}\sin\theta\ln\left(\frac{(eV_\mathrm{DC}+\varepsilon+U)^2+(\hbar/\tau_c)^2}{(eV_\mathrm{DC}+\varepsilon)^2+(\hbar/\tau_c)^2}\right)/2}{\gamma_{L}^0\left(1+\frac{A_L^2}{2} \right)\left[\pi-\arctan\left(\frac{eV_\mathrm{DC}-\varepsilon}{\hbar/\tau_c}\right)+\arctan\left(\frac{eV_\mathrm{DC}-\varepsilon-U}{\hbar/\tau_c} \right)\right]+ \gamma_{R}^0\left[\pi+\arctan\left(\frac{\varepsilon}{\hbar/\tau_c}\right)-\arctan\left(\frac{\varepsilon+U}{\hbar/\tau_c} \right)\right]}.\nonumber \\
    \label{Rabi_T2_CB}
\end{eqnarray} 
\end{widetext}

Equation~\eqref{Rabi_T2_CB} depends on both left and right couplings, unlike Eq.~\eqref{T2_high_Vdc}, since now $\varepsilon<eV_\mathrm{DC}<\varepsilon+U$ and the terms left and right are both of a similar order of magnitude. It is also required to retain the broadening terms to avoid divergences as we are considering low temperatures. Increasing the coupling of the unpolarized electrode to values where $\gamma_R^0 \gg \gamma_L^0$ suppresses the ESR signal since $\Omega_{\mathrm{FLT}}^{1\omega} T_2 \rightarrow 0$, while increasing $\gamma_L$ causes the ESR amplitude to reach a saturation point that becomes independent of the couplings. \textit{The former implies influence from the substrate}, indicating that extremely strong coupling to it \textit{leads to no ESR}. Equation~\eqref{Rabi_T2_CB} depends on the DC voltage and enables further optimization of the qubit's quality factor beyond the limit of 1/3 of the STT-driven regime, as we will detail in the results section. Both extreme situations between the couplings yield an electric current independent of the spin polarization, suggesting an ideal factor between $\gamma_R^0$ and $\gamma_L^0$ that maximizes the ESR signal in the FLT-driven regime. 

It is worth noting that \textit{cotunneling influences the ESR signal through the decoherence time $T_2$}, which yields, for example, unexpected DC bias behavior with the ESR signal that our model cannot reproduce, as shown in~\cite{J_Cuevas_C_Ast_ESR_theory_2024}. Nonetheless, the strength of our model lies in the correct characterization of the Rabi rates through the exchange field and the spin accumulation, meaning that the driving is correctly accounted for and naturally given by the model itself once an AC electric field is considered. 

Outside of the CB regime, Eq.~\eqref{Rabi_simple} does not fully vanish, but it logarithmically tends toward zero. However, the $T_2$ drops much faster, making Eq.~\eqref{Rabi_T2_CB} extremely small for any driven regime, preventing any quantum control. Thus, both FLT and STT regimes will give a measurable ESR signal in the electric current, though their origins and utilities differ significantly, as we will show in the results section: STT-ESR provides a strong spin polarization of the QI, whereas for the FLT-ESR, the QI can be reliably used as a spin-qubit.

\subsection{Energy dressing}

Returning to the exchange field, we observed that the FLT Rabi rate can be interpreted as the projection of the exchange field onto the axis perpendicular to the spin in the plane containing both the exchange field and the spin itself. However, we initially ignored the projection of this field along the {\corr eigenbasis} axis. 
On it, the Hamiltonian is diagonal, and the projection of the exchange on this axis will change the energy of the states, usually referred to as \textit{dressing or Lamb shift}. This projection provides a Zeeman-like splitting even in the absence of an external magnetic field, as long as the exchange field and spin are not perpendicular. When an external magnetic field is applied, the energy shift will modify the resonance frequency of the spectrum, shifting it relative to the bare resonance frequency given by $f_0=\Delta_{ e g}/h=g\mu_B B/h$. Following this geometric interpretation, we define the energy (or frequency) shift as:
\begin{eqnarray}
    \delta E (t)&=& h\delta f (t)=h(f_0^\prime-f_0)= g\mu_B B_\mathrm{exch}(t) \cos \theta \nonumber \\
    &=& 2\sum_\alpha P_\alpha \left[\mbox{Im}(\Gamma_{ g 0,0  g,\alpha}^{P_\alpha=0}(t)- \Gamma_{ g 2,2  g,\alpha}^{P_\alpha=0}(t)) \right]\cos\theta, \nonumber \\ 
    \label{energy_dress}
\end{eqnarray}
{\corr where $f_0^\prime$ is the dressed resonance frequency.}
The time-dependent component in Eq.~\eqref{energy_dress} does not affect the spin dynamics, as it aligns with the {\corr eigenbasis} axis, though we include it here for completeness.

By neglecting the Zeeman energy relative to the charge energies, we obtain a useful relation between the Rabi rate and the energy shift:
\begin{equation}
    h\delta f (t) \sin\theta = 2\hbar \Omega_{\mathrm{FLT}}(t)\cos\theta
    \label{dress_rabi}
\end{equation}
This relation was used to derive Eqs.~\eqref{spin_equations} and~\eqref{spin_equations_b}. Equation~\eqref{dress_rabi} provides a straightforward method to effectively measure the Rabi frequency, given the knowledge of the frequency shift and the angle between the tip and the external magnetic field. Another important detail about the energy dressing is how it affects zero-harmonic and higher-harmonic signals. Writing Eq.~\eqref{rho_master_eq_6_0} in Floquet components, as in~\cite{J_Reina_Galvez_2021,J_Reina_Galvez_2023}, shows that in general the resonance frequency per harmonic is $hf_{n\omega}=(\Delta_{ e g}+h\delta f)/n$, which makes the energy shift dependent on the harmonic $n$ (or Floquet number). Interestingly, if we move to the perspective of the magnetic field being swept at a fixed frequency, the resonance magnetic field per harmonic will be $B_{n\omega} =h (nf-\delta f) /g\mu_B$. This implies that the shifting in the magnetic field domain is the same per harmonic since $\delta f$ is not altered by $n$. Following the equations above, the zero-harmonic feature in the frequency domain appears at zero frequency, while in the magnetic field domain, it does at $-h\delta f/g\mu_B=-B_\mathrm{exch}\cos\theta$. As a consequence, this offset can be used to measure the exchange field indirectly, a proposal in line with other studies~\cite{Piotr_Martinek_Hanle_effect_2023}. 

\section{Results}

In this section, we examine key points from the previous discussion of the theory through numerical analysis, highlighting the differences between ESR signals of the \textit{first harmonic only} driven by STT and FLT.
Model parameters were chosen to ensure a strong ESR signal from an $S=1/2$ system in the FLT-driven regime while keeping the system weakly coupled to the leads under driving, as per similar requirements in~\cite{J_Reina_Galvez_2023}. We summarize \textit{the parameters employed in our simulations next}. 

The left lead represents the spin-polarized tip with a spin polarization of $P_L = 0.8$ in agreement with the findings in~\cite{J_Cuevas_C_Ast_ESR_theory_2024}. The coupling strengths are set to $\gamma_{R}^0 = 3 \times \gamma_{L}^0 = 15\ \mu$eV for all simulations, and the temperature is equal to $T=1$ K for both the left and right leads. In the context of ESR-STM experiments, this implies that the tip-sample distance is constant and the background current will change accordingly with the DC bias. The DC bias drop is applied to the tip so $eV_{DC} = \mu_L$ since we align the {\corr electro}chemical potential of the right lead with the Fermi energy, which defines the zero energy. The singly occupied energy is $\varepsilon = -100$ meV, and the Coulomb repulsion is $U = 250$ meV. These values offer a realistic enough exploration of both tunneling regimes, not so far off from DFT and other predictions~\cite{zhang2024electricfieldcontrolexchange,Corina_TMPc_paper_2025}. 

We have increased the intrinsic width of the impurity states to $\hbar / \tau_c = 0.25 \ $meV, 
compared to the one used in~\cite{J_Reina_Galvez_2023}, allowing exploration of the CB regime
while maintaining $\hbar / \tau_c \ll |\varepsilon|, \ \varepsilon + U$. The driving amplitude is set to $A_L = 50\%$, applied only to the polarized lead, consistent with the experimental setup. Finally, the Zeeman splitting provides a bare resonance frequency of approximately $f_0 = 16.8$ GHz, with an isotropic $g$-factor of 2 {\corr and $B = 0.6$ T. The Zeeman energy that appears in the rates is not neglected in the numerical simulations, even though its contribution is extremely small}.. The energy dressing will modify the resonance position to a value $f_0'=\delta f+f_0$, see Eq.~\eqref{energy_dress}, which is taken into account to perform on resonance simulations. Additional parameters are specified in figure captions.

\subsection{Spin accumulation, {\corr current and populations} vs. DC bias \corr{under no } AC voltage}

\begin{figure}
    \centering   
    \hspace*{-0.6cm}
\includegraphics[width=0.52895\linewidth]{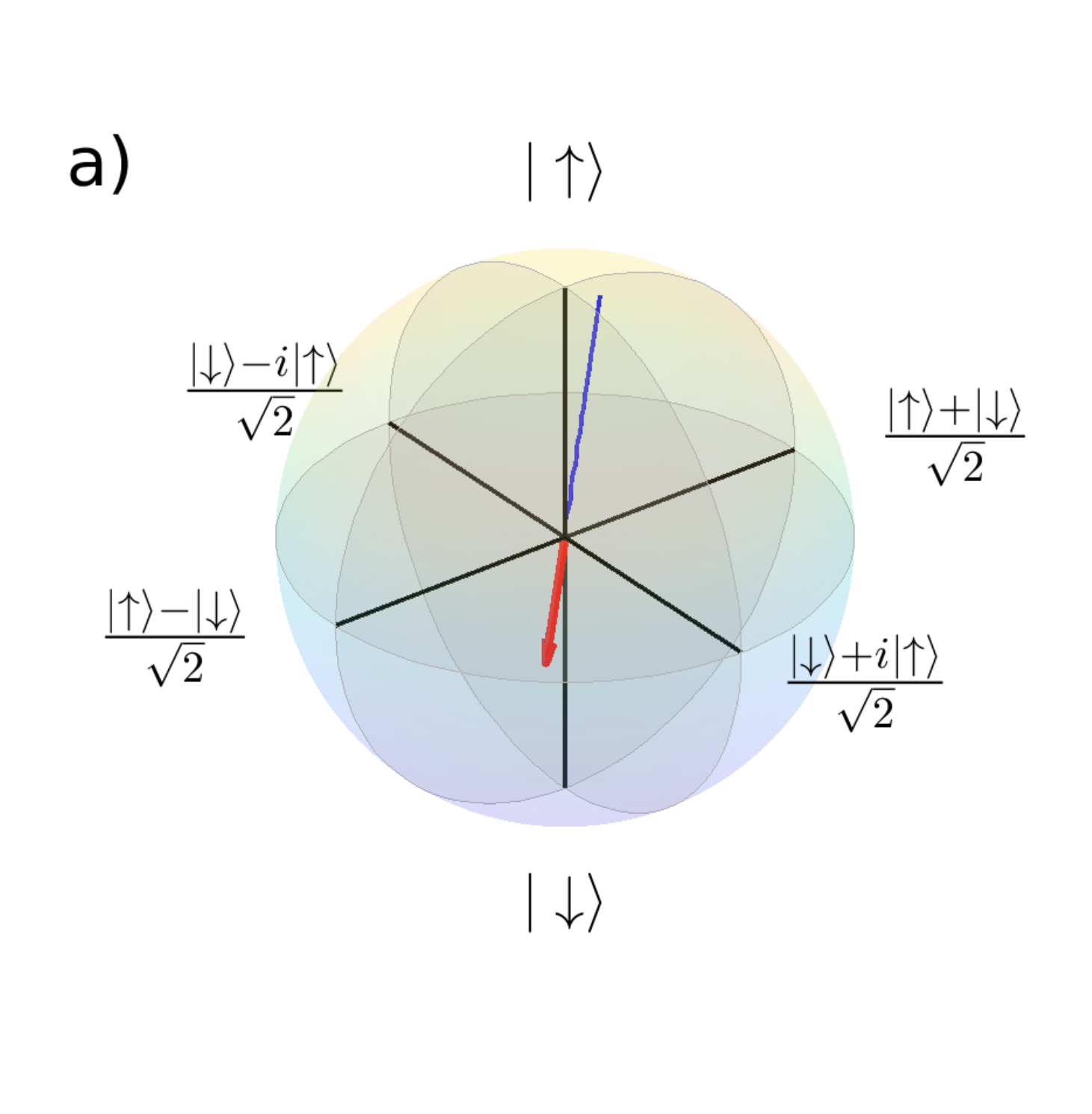}
\includegraphics[width=0.5182\linewidth]{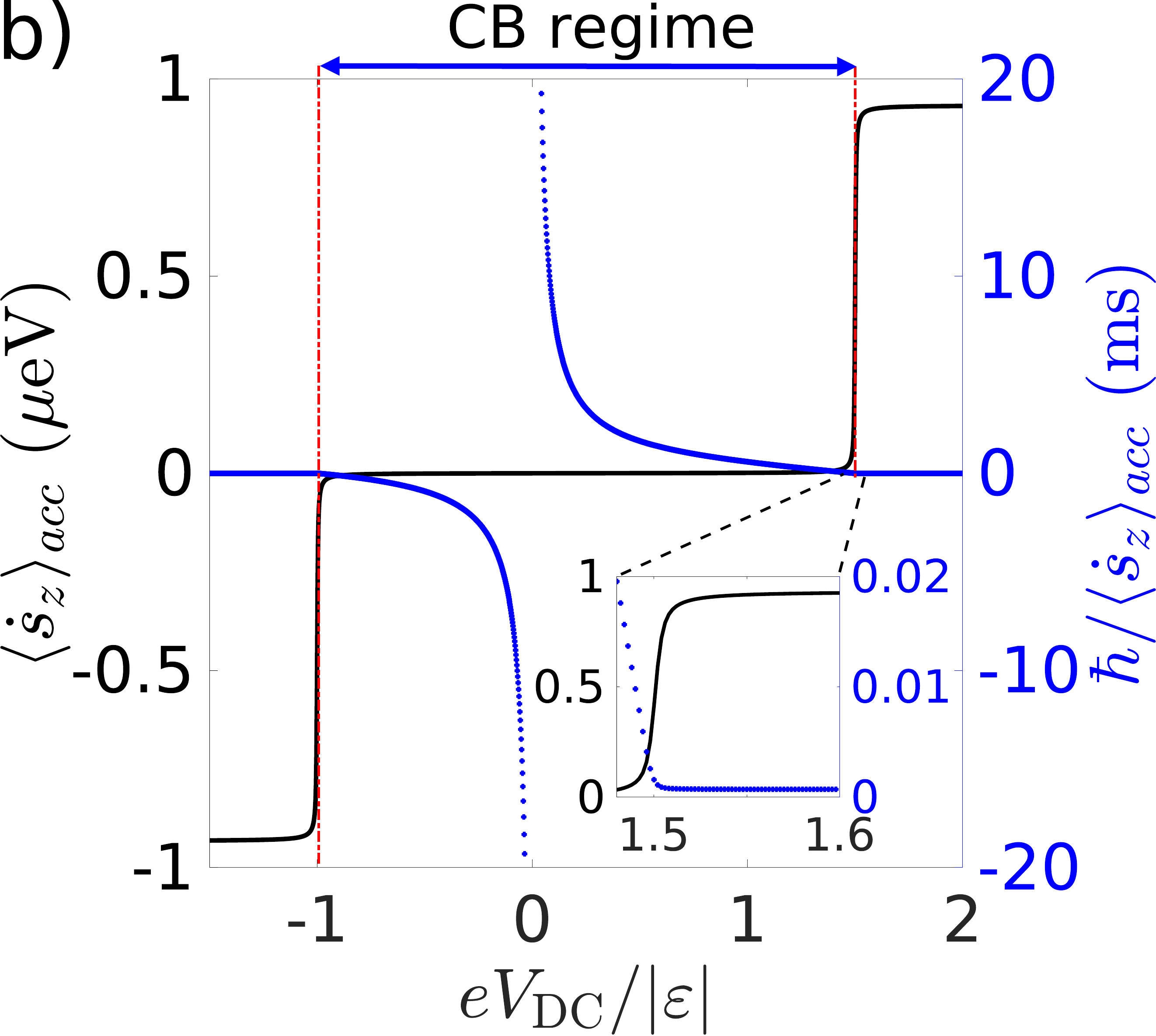}
\includegraphics[width=1.\linewidth]{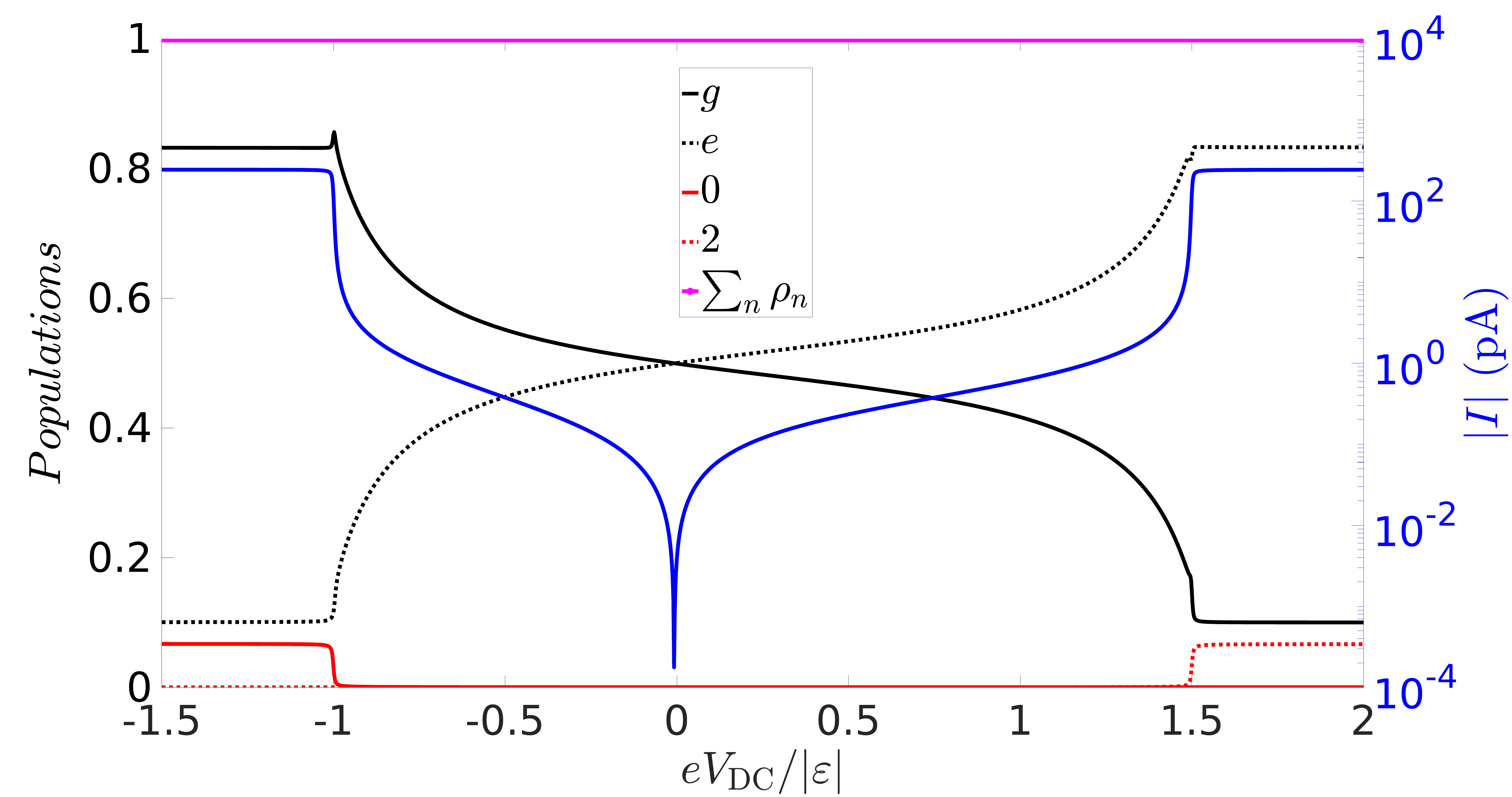}
\caption{\textbf{Influence of the transport regime on the spin dynamics without AC driving.}
a) Time evolution of the impurity spin under a positive DC bias above $\varepsilon+U$ with a value $eV_\mathrm{DC}=-2\varepsilon>0$. In the steady state, the initial state is polarized towards the $ |\mkern-5mu \uparrow\rangle$ state from the thermal initialization $\langle \mathbf{s} \rangle_\mathrm{thermal} = (-0.033, 0, -0.188)$ (red arrow), which mostly points at $ |\mkern-5mu \downarrow\rangle$, since $\theta=\pi/18$. This means the quantum impurity rapidly \textit{torques} when the DC bias is on and positive spin-polarized electrons are injected from the tip. 
b) Spin accumulation rate (Eq.~\eqref{spin_acc}) and its inverse as a function of the DC bias at constant tip-sample distance. The spin accumulation term dominates above the energy thresholds $\varepsilon$ and $\varepsilon+U$ but drastically drops off in the CB regime. The opposite behavior is observed when looking at its inverse (in blue), which diverges at zero voltage. The inset shows the transition region at positive voltages, where $\hbar/\langle\dot{s}_z\rangle_\mathrm{acc}$ saturates to 0.72 ns, indicating fast dynamics of the spin accumulation acting on the QI. {\corr c) Populations (left axis) and absolute value of the current (in blue, right axis, log scale) as functions of the normalized DC bias. Outside the CB regime, the current is large and constant ($\sim \mathrm{sign}(V_\mathrm{DC})\times 240$ pA), and higher‐energy states (doubly and empty QI states) can be slightly populated. Inside the CB regime, the population is fully governed by the singly occupied states, and the current rapidly decreases, exhibiting an $\arctan$ behavior, see Eq.~\eqref{eq:current_analytical}. The QI satisfies detailed balance ($\sum_l \rho_{ll}=1$) at all DC bias values.}}
    \label{fig:Sacc_no_driving}
\end{figure}

As a preamble, we first look at the behavior of the impurity spin in the absence of an AC driving field to separate the dynamics induced by the electric current from those of the ESR processes. As a consequence, only the spin accumulation and relaxation terms will influence the dynamics since the time independent exchange is much smaller than the external magnetic field (teslas compared to milliteslas for the parameters used). Fig.~\ref{fig:Sacc_no_driving}a shows the effect of spin accumulation on the spin dynamics on the Bloch sphere at $eV_\mathrm{DC}>\varepsilon+U$, torquing the thermally initialized QI towards a positive spin polarization, mostly aligning with $ |\mkern-5mu \uparrow\rangle$, following the sign of $\langle\dot{s}_z\rangle_\mathrm{acc}$ in Fig.~\ref{fig:Sacc_no_driving}b at large positive DC bias. 
{\corr The spin does not accumulate along the magnetization direction of the tip (z-axis) because the strong magnetic field, with $\theta=\pi/18$ or $10$ degrees,  dominates and aligns the spin approximately parallel to it, pointing towards the excited or ground states} depending on whether the sign of $\langle\dot{s}_z\rangle_\mathrm{acc}$ is positive or negative, respectively. If the Zeeman splitting is much smaller than the coupling to the spin polarized electrode, the spin impurity will be spin-accumulated out of plane, aligned with the magnetization of lead. 

The effect of spin accumulation persists even in the CB regime, albeit weaker, vanishing completely at zero voltage. There is, however, a competition between the relaxation time, which ranges from 0.2 to 50 ns (Fig.~\ref{fig:Rabi_T2}a), and $\hbar/\langle\dot{s}_z\rangle_\mathrm{acc}$, leading to different results depending on the DC bias. In the CB regime, $\hbar/\langle\dot{s}_z\rangle_\mathrm{acc}\gg \tau_\mathrm{rel}$, so the spin evolution is dominated by the spin relaxation time, as the accumulation term acts extremely slowly on the QI. This opens the possibility of an FLT-ESR signal since the time dependent exchange field will have ``time" to torque the spin coherently, producing Rabi oscillations. On the other hand, above the energy thresholds outside of the CB regime, both the relaxation and spin accumulation times are of the same order of magnitude, forcing the QI to be spin-accumulated in the {\corr eigenbasis} axis before any coherent evolution occurs. Therefore, AC driving that leads to a time-dependent spin accumulation will provide an incoherent STT-ESR signal, as we will see in the following sections.

{\corr As a final remark, Fig.~\ref{fig:Sacc_no_driving}c shows the populations as a function of the DC bias, revealing a QI that is singly occupied throughout the CB regime; the empty and doubly occupied states become populated once the DC bias exceeds the thresholds at $\varepsilon$ and $\varepsilon+U$, respectively. This occurs when enough energy is available to remove or add an electron to the QI. The small overshoot in the ground-state population at negative bias (also present, but smaller, at positive bias) results from the positive spin polarization together with the excited state leaving the CB regime slightly before the ground state owing to the small Zeeman splitting. Consequently, the left lead, which is positively polarized, reduces the population of the up-like (excited) state, an effect difficult to visualize since this population already begins to decrease as $eV_{\mathrm{DC}}\to\varepsilon$, producing a sudden increase in the ground-state population (down-like) at $eV_{\mathrm{DC}}=\varepsilon+g\mu_B B/2$. At the positive threshold energy, the overshooting is less pronounced due to the inversion of populations at the $\varepsilon+U$ level. If the spin polarization is negative, the behavior of Fig.~\ref{fig:Sacc_no_driving}c is essentially reversed with respect to the DC bias, being the strongest sudden overshooting at $eV_{\mathrm{DC}}=\varepsilon+U-g\mu_B B/2$. Figure~\ref{fig:Sacc_no_driving}c also displays the absolute value of the current, Eq.~\eqref{homo_current}, which remains constant outside the CB regime and decays rapidly within it, in agreement with the analytical expression in Appendix~\ref{app:homodyne}, Eq.~\eqref{eq:current_analytical}. Due to numerical limitations, the current at $V_{\mathrm{DC}}=0$ is quite small but non-zero.}

\subsection{ESR in the STT-driven regime}

\begin{figure}
    \centering    
\includegraphics[width=0.52895\linewidth]{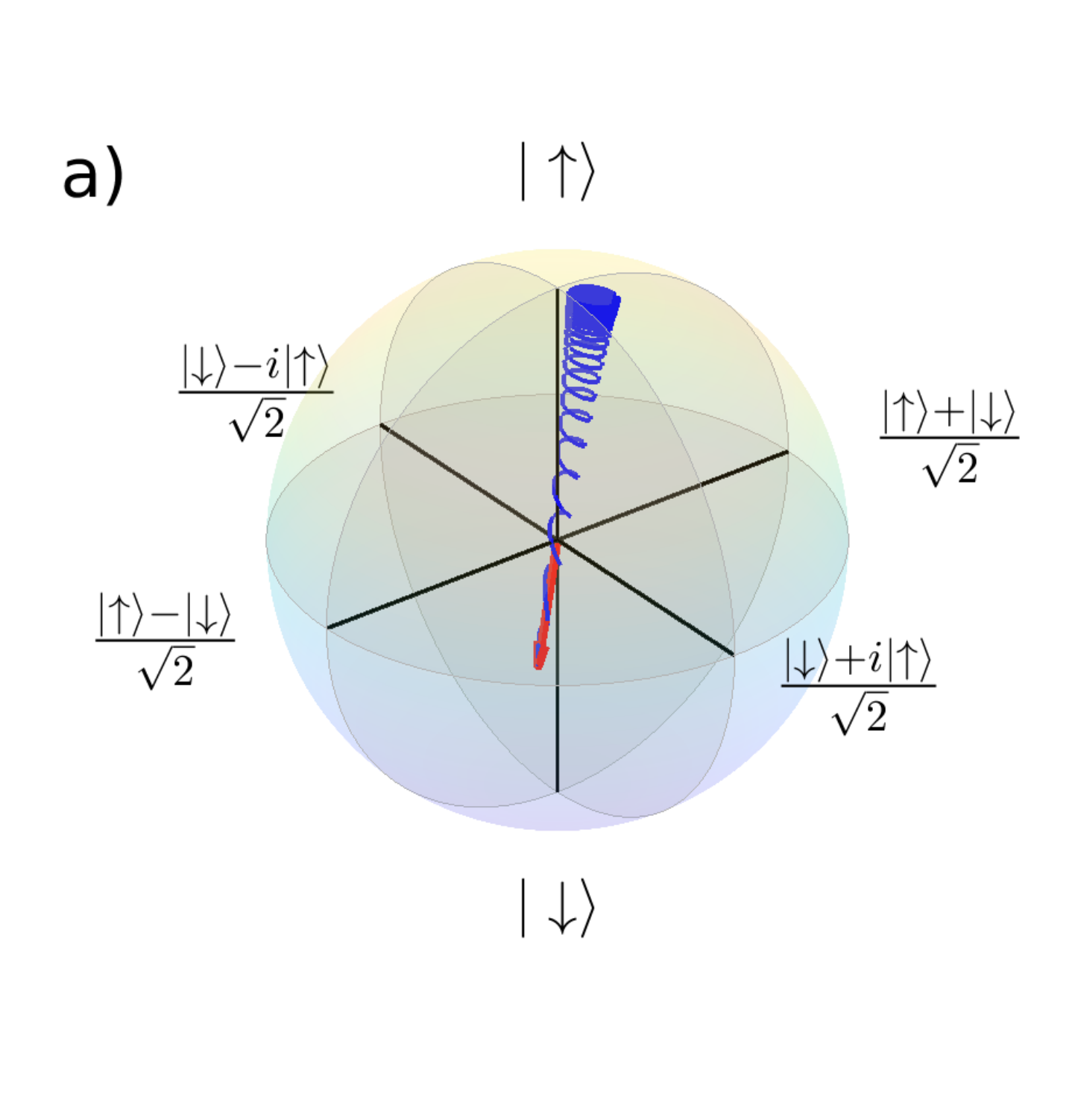}
    \includegraphics[width=0.46\linewidth]{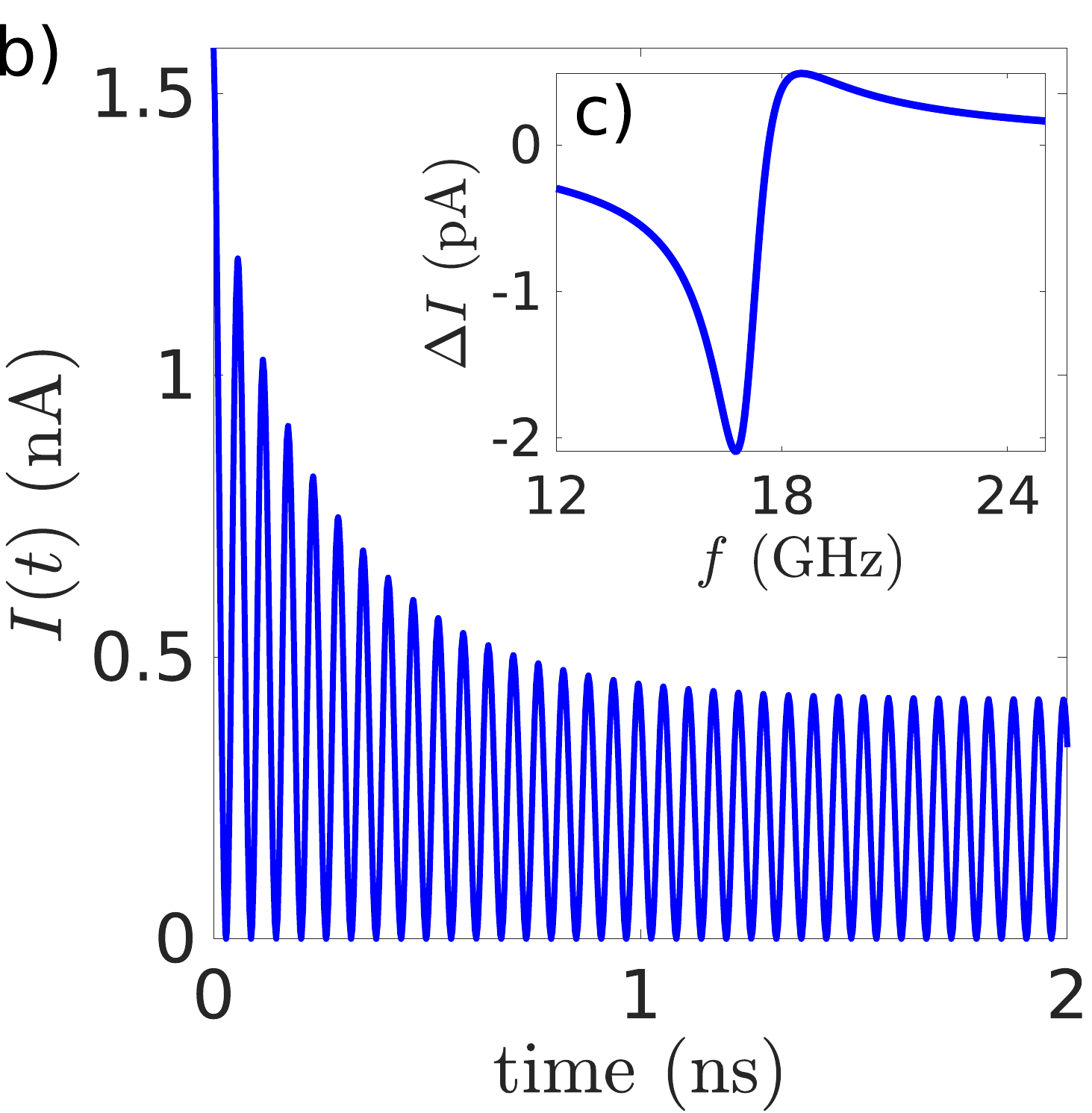}
    \caption{\textbf{ESR outside of the CB regime; out of plane magnetic field.} a) Bloch sphere representation of the spin components, on resonance with the AC field, in the \textit{lab frame} for $\theta = \pi/18$ and positive $eV_\mathrm{DC}/|\varepsilon| = 2$. The spin is initialized thermally at $\langle \mathbf{s} \rangle_\mathrm{thermal}/{\corr \hbar } = (-0.033, 0, -0.188)$ (red arrow). The positive DC bias causes the spin to flip to its spin polarization in the steady state as a consequence of the spin accumulation term. At the same time, the spin continuously performs Larmor oscillations even in the steady state. The Bloch sphere axes are reduced for clarity. b) Time-evolution of the current, Eq.~\eqref{homo_current}, from the thermal state, reaching the steady state after 2 ns. c) Corresponding ESR electric current signal, $\Delta I=I_L-I_\mathrm{BG}$, in the STT-driven regime in the frequency domain. The resonance frequency is slightly different to the bare one at 16.8 GHz, being now $f_0'=\Delta f+f_0=17.07$ GHz due to the energy dressing. }
    \label{fig:SEQ_out_of_plane}
\end{figure}

\begin{figure}
    \centering    
\includegraphics[width=0.52895\linewidth]{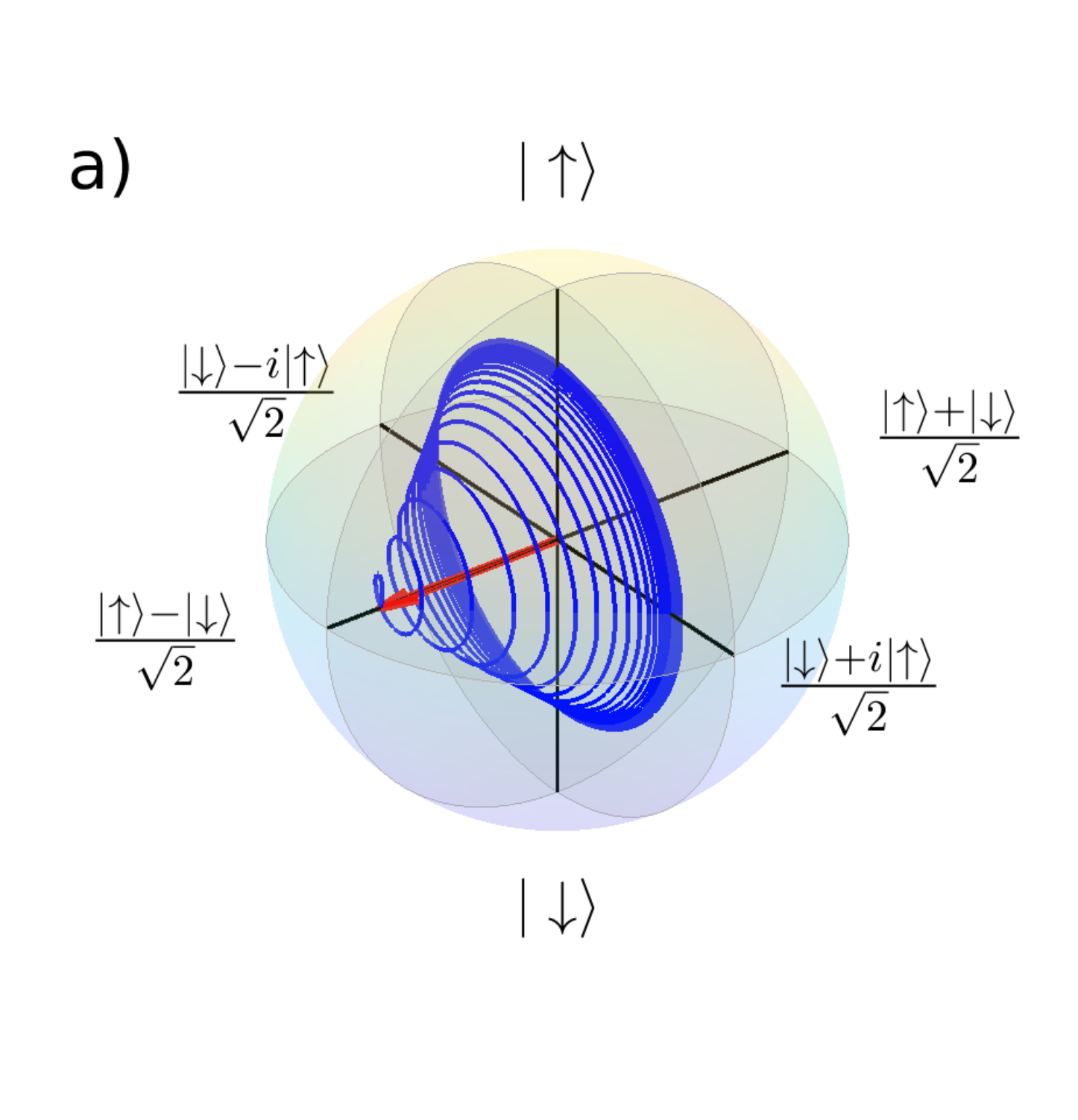}
    \includegraphics[width=0.46\linewidth]{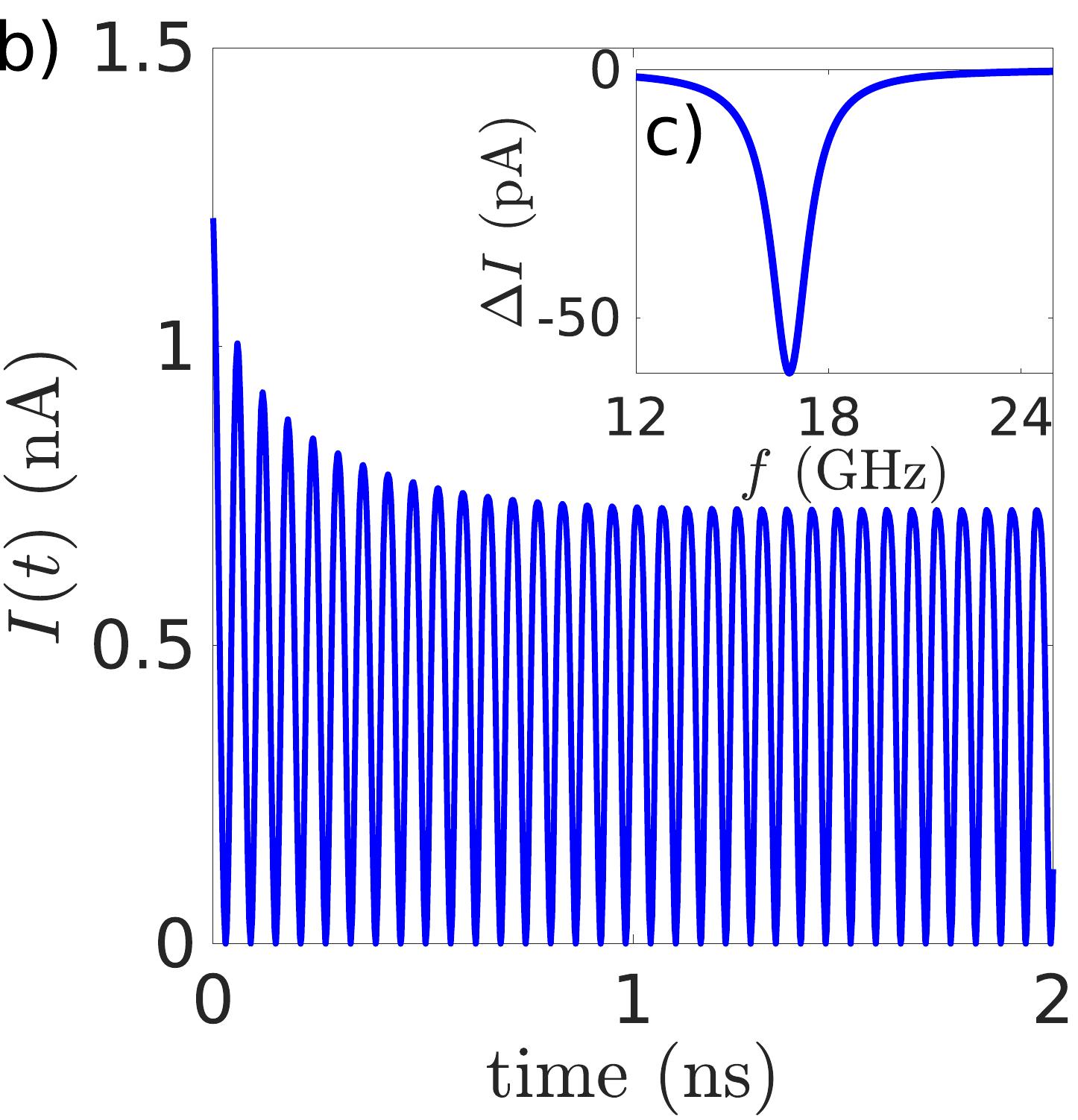}
    \caption{\textbf{ESR outside of the CB regime; in-plane magnetic field.} a) Bloch sphere representation of the spin components, on resonance with the AC field, in the \textit{lab frame} for $\theta = \pi/2$ and $eV_\mathrm{DC}/|\varepsilon| = 2$. The spin is thermally initialized at $\langle \mathbf{s} \rangle_\mathrm{thermal}/{\corr \hbar } = (-0.19, 0, 0)$ (red arrow). Here, the spin accumulation forces alignment with the {\corr eigenbasis} axis of the spin in the steady state, leading to continuous Larmor oscillations in the $zy$-plane. b) Time-evolution of the current, Eq.~\eqref{homo_current}, from the thermal situation. c) ESR electric current signal in the STT regime for the case of a) in the frequency domain, $\Delta I=I_L-I_\mathrm{BG}$. Since in-plane initialization leads to not dressing, the resonance occurs at the bare one $f_0=16.8$ GHz.}
    \label{fig:SEQ_in_plane}
\end{figure}

We now examine the effect of adding an AC component to the bias, which will lead to an ESR driven by current, an effect also referred to as spin-transfer torque (STT) in~\cite{kovarik_spin_2024}. Specifically, we analyze two scenarios {\corr when the AC field is on resonance with the Zeeman splitting}: an almost out-of-plane magnetic field (Fig.~\ref{fig:SEQ_out_of_plane}) and an in-plane magnetic field (Fig.~\ref{fig:SEQ_in_plane}). The current outside of the CB region reaches amplitudes of nA with the chosen coupling values. A completely out-of-plane magnetic field will give no ESR signal for a spin-1/2 system, as the spin and the driving field (from spin accumulation or exchange field) are aligned. In Fig.~\ref{fig:SEQ_out_of_plane}a, we choose the magnetic field to be mostly out of plane ($\theta = \pi/18$ or 10 degrees) as in Fig.~\ref{fig:Sacc_no_driving} and show the time evolution of the spin vector starting from the thermal state on the Bloch sphere. The applied voltage reverses the populations of the singly occupied states, similarly to the non-driven case, which was initially mostly $ |\mkern-5mu \downarrow\rangle$, resulting in the spin of the quantum impurity being polarized towards $ |\mkern-5mu \uparrow\rangle$ for positive bias. Contrary, for a negative DC bias, the spin would polarize towards $ |\mkern-5mu \downarrow\rangle$. As it was mentioned in the previous section, this behavior is dictated by the sign of the spin accumulation, which is determined by the DC voltage and spin polarization.

Fig.~\ref{fig:SEQ_out_of_plane}b illustrates the spin dynamics, characterized by fast Larmor oscillations that persist even in the long-time limit. These oscillations occur in a plane perpendicular to the {\corr eigenbasis} axis, {\corr parallel to the external magnetic field,} see Fig.~\ref{fig:SEQ_out_of_plane}a, and arise from the time-dependent spin accumulation. This accumulation, proportional to the polarized current, attempts to torque the accumulated spin toward the spin polarization axis, which in result makes the crossed product $\mathbf{s} \times\mathbf{B} $ non-zero, forcing the spin to torque with an intensity proportional to $\sin\theta $.
In this regime, the populations remain practically frozen, and the Larmor oscillations only modify the coherences, making the ESR current fully governed by them~\cite{J_Reina_Galvez_2023}.

Fig.~\ref{fig:SEQ_out_of_plane}c shows the corresponding frequency behavior. Although the ESR signal is significant, the background current is three orders of magnitude higher, resulting in an extremely low ESR signal relative to the background. This is expected in the STT-driven regime for the parameters used, as reflected in the low-quality factor $\Omega_{\mathrm{STT}} T_2$. The signal exhibits an asymmetric Fano profile due to the exchange field and the residual presence of homodyne detection. However, it becomes fully symmetric when $eV_{\mathrm{DC}} \gg \varepsilon + U$, where the exchange field vanishes (see Eq.~\eqref{Rabi_eq_analytical}).

For the in-plane configuration, Fig.~\ref{fig:SEQ_in_plane}a, the strongest Larmor oscillations and ESR signals (Figs.~\ref{fig:SEQ_in_plane}b and~\ref{fig:SEQ_in_plane}c)) are achieved. The Larmor precession now occurs in the $zy$-plane, again perpendicular to the magnetic field and the {\corr eigenbasis} axis. If the STT-driven regime would follow the homodyne detection previously reported~\cite{JK_vector_field_2021,Fe_driving_SH_2023}, the ESR signal would be minimal or zero since the impurity spin and spin polarization od the tip are perpendicular, and the ESR signal is proportional to their scalar product. However, in this regime, this limitation does not hold as the spin will be driven towards the $z$ axis by the spin accumulation. After this initial push, the spin feels the torque from the external magnetic field, from the cross product $\langle \mathbf{s} \rangle \times \mathbf{B}$. This torque is maximized for in plane configurations, producing a stronger ESR signal compared to the out of plane situation. This interaction allows for the initial oscillatory evolution of the spin, observed in both Fig.~\ref{fig:SEQ_in_plane}a and Fig.~\ref{fig:SEQ_out_of_plane}a. Shortly thereafter, the spin settles into oscillations within the $zy$ plane in Fig.~\ref{fig:SEQ_in_plane}a, driven by the time-dependent spin accumulation and the torque exerted by the external magnetic field.

The role of the exchange field in this scenario is analogous to the out-of-plane case; however, the ESR signal in Fig.~\ref{fig:SEQ_in_plane}c is nearly symmetrical. This symmetry arises because, when the spin is initialized along a Cartesian axis, $\langle s_z \rangle$ in Eq.\eqref{General_Sz} depends solely on either the populations or the coherences, but not on a combination of both. Consequently, no Fano profile emerges. Specifically, for an in-plane magnetic field, the $z$ component of the spin depends only on the real part of the coherences, while the $x$ component reflects the populations. {\corr Another way to view this is that the FLT-driven ESR \textit{is not detected/observed} at $90$ degrees because of the Homodyne detection, and the measured response is due to the STT-driven ESR alone, which provides a fully symmetric ESR.}

Importantly, in the lab frame, with an initial in-plane configuration, the spin cannot polarize because the spin accumulation term is fully perpendicular to the external magnetic field. Note that in the in-plane configuration, the populations, measured in the $x$ axis, also remain unperturbed in the long-time limit as observed in the out-of-plane situation Fig.~\ref{fig:SEQ_out_of_plane}. 
{\corr Interestingly, in the rotating frame, the torque exerted by the spin accumulation becomes increasingly influential as the angle changes from the out-of-plane to the in-plane configuration, with the latter causing the spin to align with the $z$-axis in the rotating frame.}


The reader may notice that the ESR signal in both cases (Figs.~\ref{fig:SEQ_out_of_plane}c and~\ref{fig:SEQ_in_plane}c) are dips (a reduction of the current on-resonance) which is a direct result of the STT-driven regime at positive bias. In this regime, the coherences peak on resonance while remaining zero off resonance. According to Eq.~\eqref{homo_current}, this behavior produces a dip in the signal. Conversely, under a negative DC bias ($eV_\mathrm{DC} < \varepsilon$), the current becomes negative and the ESR signal will flip from a dip to a peak. However, the ESR signal cannot flip by merely changing the sign of the spin polarization because an additional sign change occurs in the $z$ component of the spin, effectively preserving the overall response. Higher order tunneling processes could induce additional flipping of the signal though.

\subsection{ESR in the FLT-driven regime}

We now shift our focus to the CB regime, where FLT-driven resonance dominates. This regime favors quantum-coherent manipulation of the spin over the spin polarization of the impurity spin since the spin accumulation term is significantly minimized. This reduces the impact of the completely decoherent polarized current on the impurity, enabling the spin to be used as a qubit. FLT driven regime is basically equivalent to conventional ESR, where a time dependent magnetic field induces a resonant transition. As highlighted in the theoretical discussion, this allows for quality factors $\Omega_{\mathrm{FLT}} T_2$ to reach their upper limits, constrained only by the ability to enhance $T_2$ for a fixed Rabi rate. Notably, this enhancement is strongly dependent on the DC bias, making it the critical factor in achieving high coherence for spin manipulation.

\begin{figure}
	\centering    
	\includegraphics[width=0.52895\linewidth]{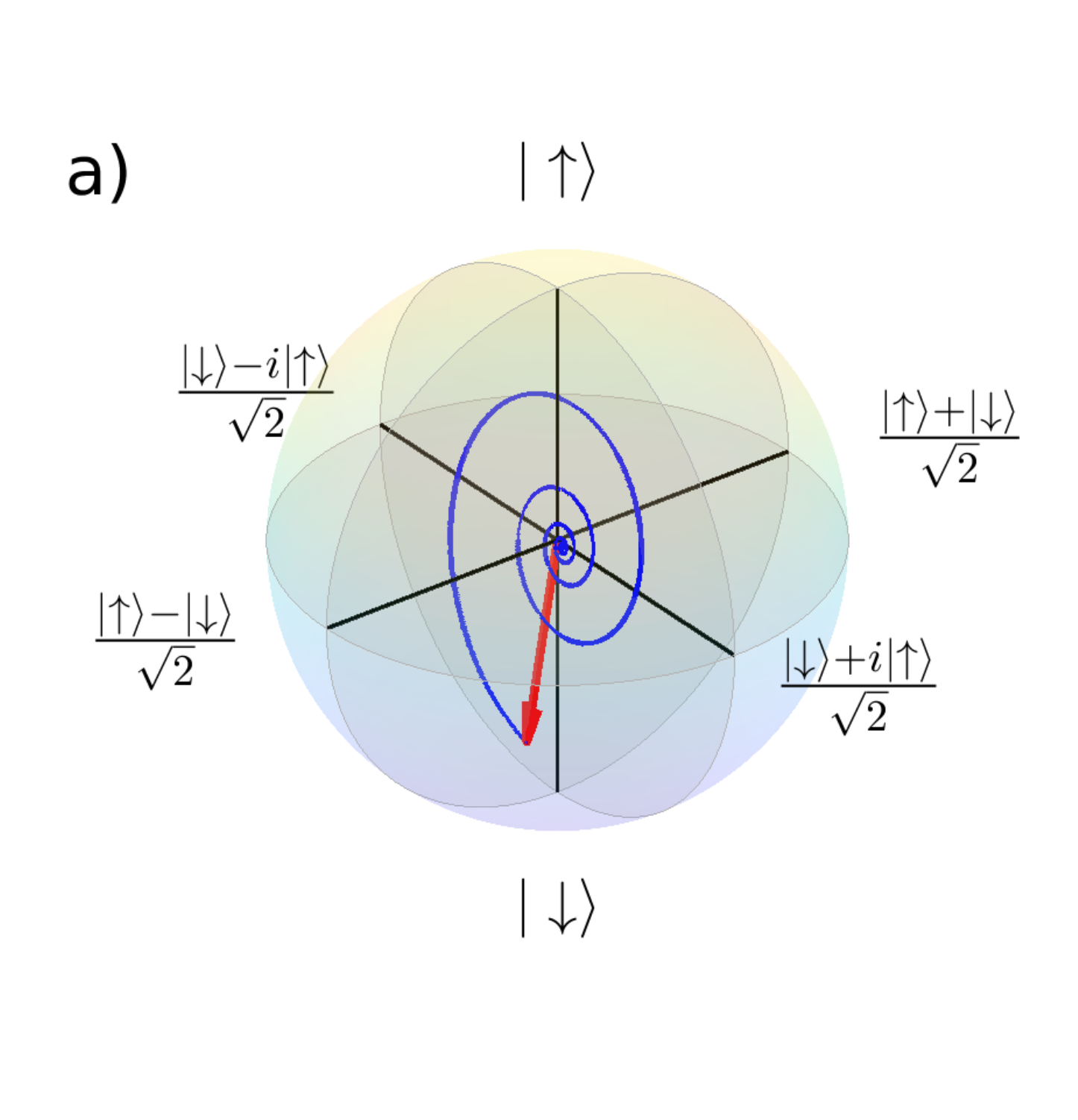}
    \includegraphics[width=0.46\linewidth]{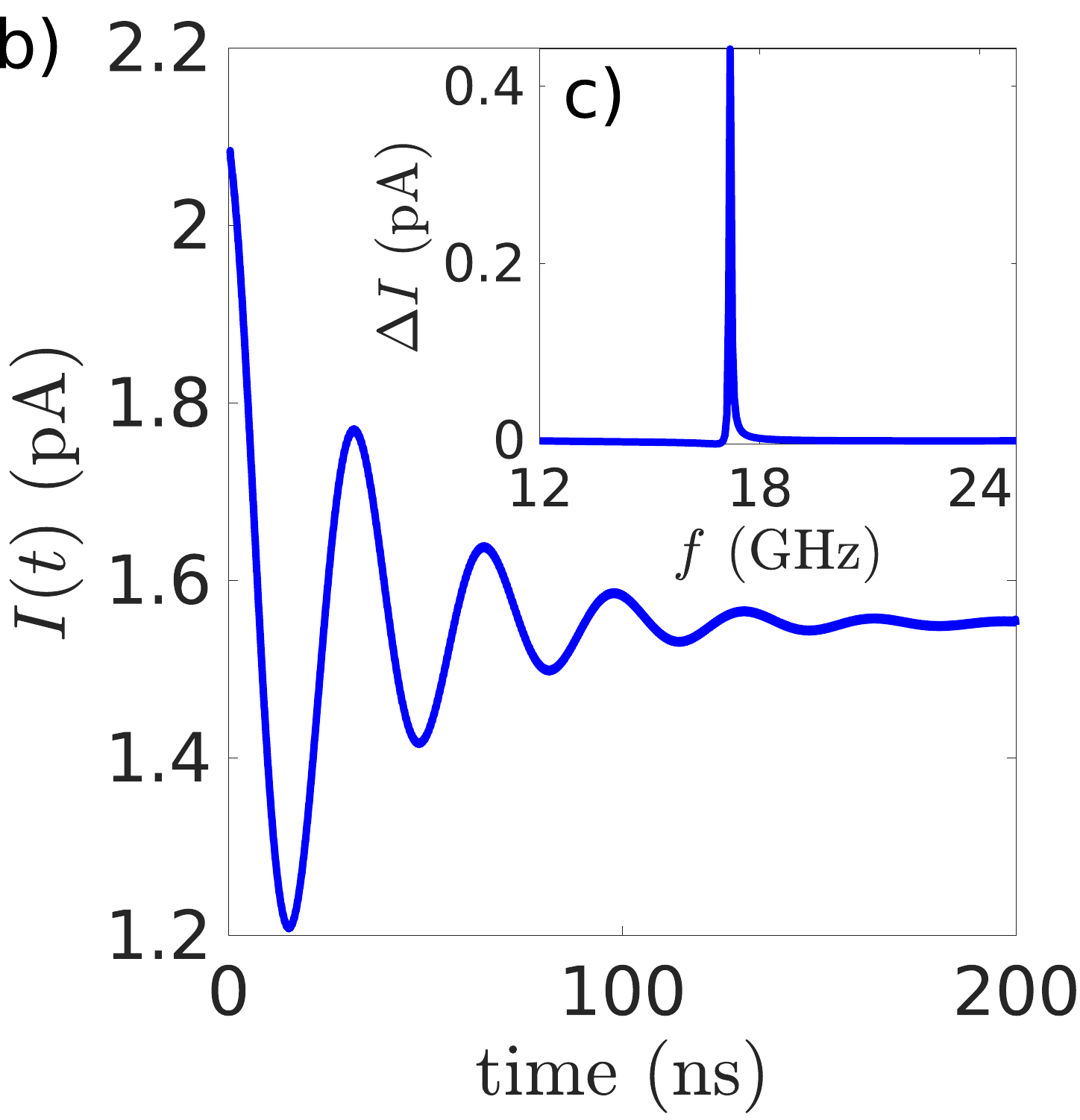}
	\caption{\textbf{ESR in the Coulomb Blockade regime; out of plane magnetic field.} a) Bloch sphere representation of the spin components in the \textit{rotating frame} for a magnetic field tilted by $\theta=\pi/18$ and a bias voltage of $eV_\mathrm{DC}/|\varepsilon|=1.275$, where the maximum quality factor $\Omega T_2 \approx 6$ is achieved. The spin is thermally initialized at $\langle \mathbf{s} \rangle_{\text{thermal}} = (-0.033, 0, -0.188)$ and experiences a torque due to the time dependent exchange field. This torque causes the spin to precess in a plane with the normal vector $\mathbf{n} = (-\sin(4\pi/9), 0, \cos(4\pi/9))$. b) Time evolution of the current, Eq.~\eqref{homo_current}, showing up to six Rabi oscillation cycles before decaying to the steady state in the rotating frame. c) ESR signal, $\Delta I=I_L-I_\mathrm{BG}$, DC contribution, featuring a peak with a subtle Fano profile. The resonance frequency is slightly tuned with respect to the bare one to be $f_0'=\Delta f+f_0=17.146$ GHz due to the energy dressing. }
	\label{fig:CB_out_of_plane}
\end{figure}

In Fig.~\ref{fig:CB_out_of_plane}a, we present the first example of Rabi oscillations in the spin system under a magnetic field tilted by $\pi/18$ radians relative to the spin polarization axis. This is shown at the optimal voltage, $eV_\mathrm{DC}/|\varepsilon| = 1.275< eV_\mathrm{DC}/(\varepsilon+U)=1.5$, which maximizes the quality factor $\Omega_{\mathrm{FLT}} T_2$, as will be discussed in the next section. Specifically, Fig.~\ref{fig:CB_out_of_plane}a illustrates the coherent exchange of populations between the mostly $ |\mkern-5mu \uparrow\rangle$ and $ |\mkern-5mu \downarrow\rangle$ states. Importantly, during these oscillations, the transient states $|0\rangle$ and $|2\rangle$ remain nearly unpopulated, in stark contrast to the STT-driven regime, where these states significantly contribute to the spin accumulation term. This highlights the pivotal role of the DC bias value, as sweeping it enables a transition from the STT regime to the FLT regime. Such a transition allows the STT regime to serve as a spin initialization and the FLT regime to facilitate precise spin control, aligning with the suggestions made in~\cite{kovarik_spin_2024}.

The current in the rotating frame, as shown in Fig.~\ref{fig:CB_out_of_plane}b, directly reflects the Rabi oscillations of the spin, as it corresponds to the $z$-component of the spin. Notably, the DC contribution here is significantly smaller than that in Fig.~\ref{fig:SEQ_out_of_plane}b, as we are in the CB regime. This reduced DC contribution allows for a relatively long coherence time $T_2$ of approximately 35 ns without diminishing the contribution of the exchange field. An interesting aspect is observed in the ESR signal (Fig.~\ref{fig:CB_out_of_plane}c), which now exhibits a sharp peak instead of the dips seen earlier. This occurs because the ESR signal depends not only on the coherences but also on the populations of the spin states when the spin Rabi oscillates. On resonance, the populations tend toward a 50/50 distribution if the driving force is sufficiently strong, as expected in classical ESR simulations. At positive bias and spin polarization, the off-resonance population in the long-time limit leads to $\langle s_z \rangle > 0$, a product of, albeit small, spin accumulation, and consequently, the ESR signal on resonance manifests as a peak, see Eq.~\eqref{homo_current}. Conversely, at negative bias, the signal flips to a dip, with an inverted Fano profile.

\begin{figure}
	\centering    	\includegraphics[width=0.52895\linewidth]{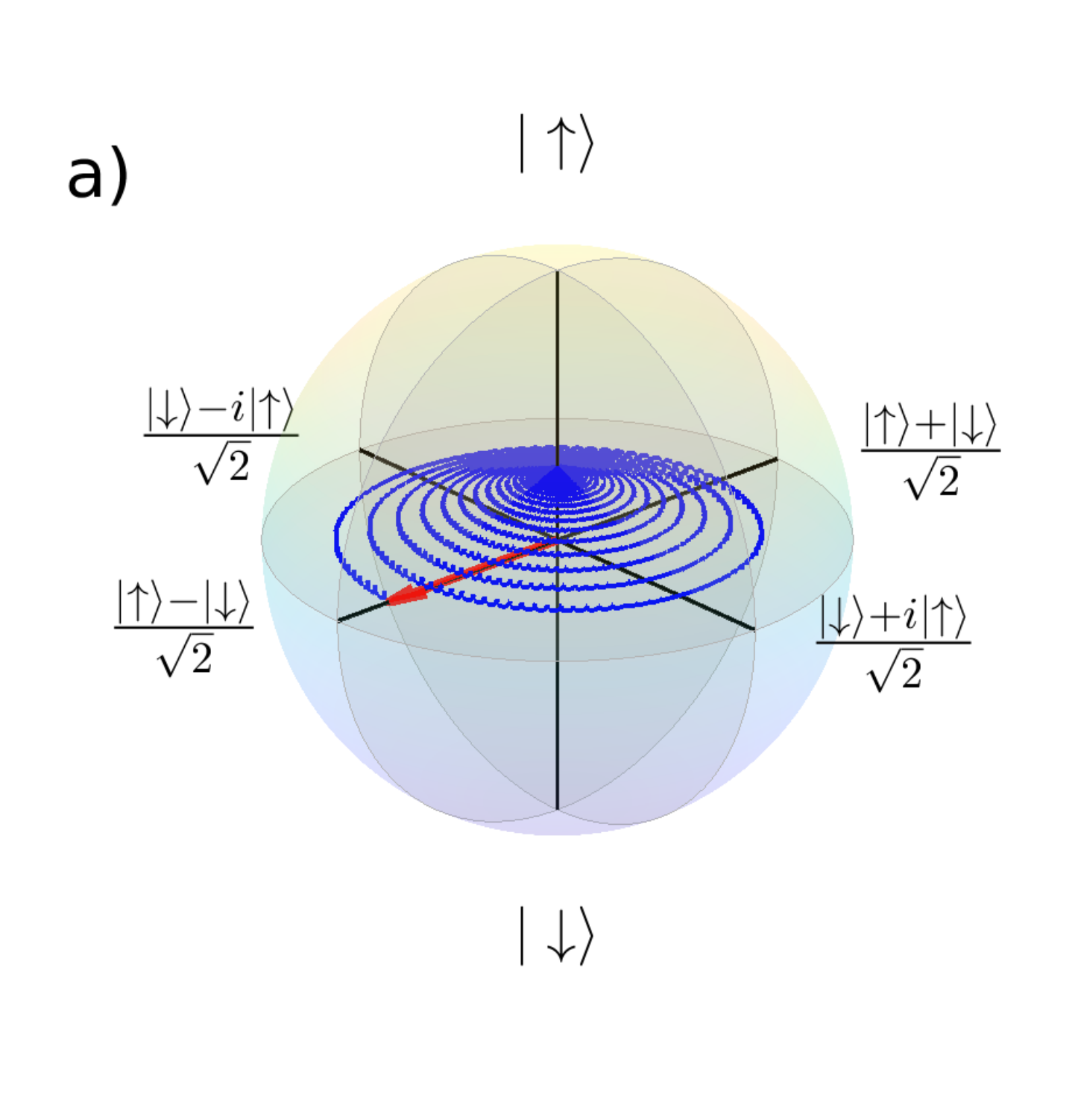}
\includegraphics[width=0.46\linewidth]{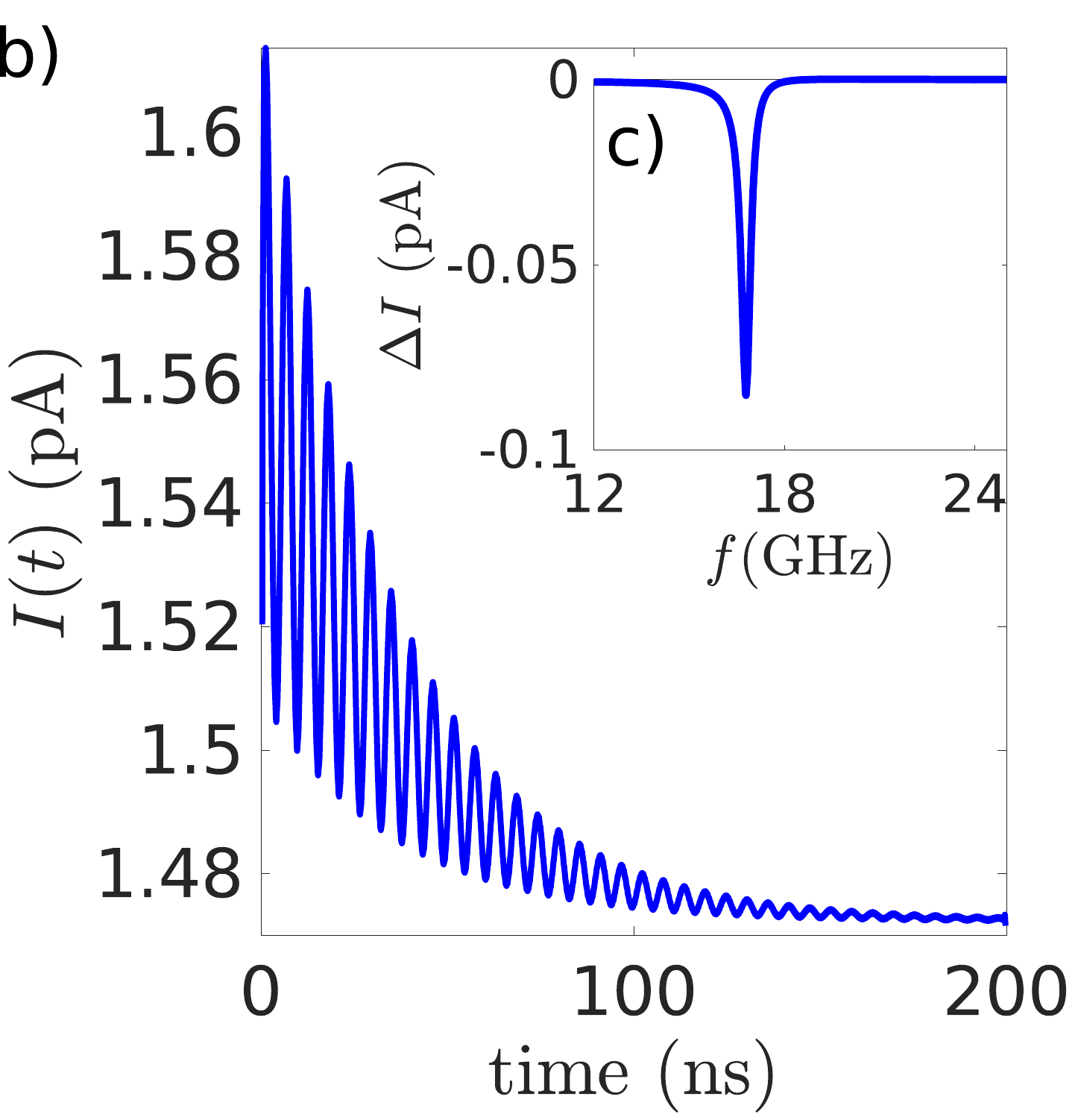}
	\caption{\textbf{ESR in the Coulomb Blockade regime; in plane magnetic field.} a) Bloch sphere representation of the spin components in the \textit{rotating frame} for a magnetic field aligned with the plane ($\theta = \pi/2$) and a bias voltage of $eV_\mathrm{DC}/|\varepsilon|=1.275$, where the maximum quality factor $\Omega T_2 \approx 36$ is achieved. The spin is thermally initialized at $\langle \mathbf{s} \rangle_{\text{thermal}} = (-0.19, 0, 0)$ and undergoes torque to the exchange field, leading to precession in a plane with the normal vector $\mathbf{n} = (0, 0, 1)$. b) Time evolution of the current, Eq.~\eqref{homo_current}, showing approximately 36 Rabi oscillations before decaying to the steady state in the rotating frame, closely matching the z-component of the spin, not the $x$-component. c) ESR signal, $\Delta I=I_L-I_\mathrm{BG}$, DC contribution, displaying a dip without a Fano profile in the frequency domain. Since in-plane initialization leads to not dressing, the resonance frequency is $16.8$ GHz.}
	\label{fig:CB_in_plane}
\end{figure}

We now turn to the time evolution of an in-plane initialized spin, Fig.~\ref{fig:CB_in_plane}a. The Rabi rate $\Omega_\mathrm{FLT}$ is maximized at $\theta = \pi/2$ radians, where the projection of the exchange field perpendicular to the spin is greatest, effectively maximizing the torque exerted by the exchange field. A notable feature in this simulation is the spin being pulled toward the positive $z$-axis of the Bloch sphere. This phenomenon is attributed to a non-negligible spin accumulation, which is expected given the DC bias's proximity to the threshold $\varepsilon + U$. This residual spin accumulation explains a non-zero ESR signal even in the perpendicular configuration between the tip's magnetization and the impurity spin, where typically no signal is found due to the homodyne detection rendering the FLT-driven component unobservable, even though the Rabi rate is maximized~\cite{JK_vector_field_2021,Fe_driving_SH_2023}. {\corr Since this measured response is due to the remanent contribution from the STT-driven regime alone, a fully symmetric ESR is obtained, see Fig.~\ref{fig:CB_in_plane}c, although without the strong damping effect of the STT regime since $\hbar/\langle \dot{s}_z\rangle_\mathrm{acc}\gg T_2$.}

Since the current measures the $z$-component of the spin, it does not reflect the Rabi oscillations as clearly as in the out-of-plane case, as confirmed in Fig.~\ref{fig:CB_in_plane}b. However, the current’s maxima and minima in the rotating frame still correspond to the spin being in the ground and excited states, respectively, demonstrating controllability of the spin{\corr , visible only due to the small driving contribution from STT. Further reducing this contribution by lowering the DC bias would completely suppress the ESR signal.} It is worth noting that while our theory lacks a fully developed cotunneling framework, it captures key aspects of the detection mechanism, particularly homodyne detection. This is evident from the intensity minimum of the ESR signal in Fig.~\ref{fig:CB_in_plane}c, despite the higher quality factor $\Omega_{\mathrm{FLT}} T_2$ in the in-plane configuration compared to the out-of-plane case Fig.~\ref{fig:CB_out_of_plane}c. Thus, we conclude that while our model has limitations, it successfully incorporates the homodyne detection, and it aligns with experimental results, indicating that out-of-plane magnetic fields yield stronger ESR signals than in-plane configurations.

If we examine the spin dynamics on the Bloch sphere in the lab frame instead of the rotating frame, the spin exhibits an additional Larmor oscillation perpendicular to the plane in which the spin undergoes Rabi oscillations. This arises from the cross product between the spin and the external magnetic field.

\subsection{On the measurability of QI spin dynamics via the current}

{\corr We wish to clarify the physical interpretation of the charge current as a probe of the quantum impurity (QI) spin state, particularly in the context of coherent dynamics. While the current results from tunneling processes that inevitably affect the occupancy of the QI, it nonetheless acts as a form of \emph{weak measurement} of the QI spin. Specifically, in the CB regime, the impurity remains singly occupied for long durations, and tunneling events are rare and weakly perturbative. In this regime, the spin-dependent tunneling rates provide a continuous readout of the QI spin projection along the axis defined by the spin polarization of the lead. The average current is thus sensitive to the expectation value of the spin component parallel to this axis, a principle that underlies the interpretation of ESR-STM magneto-resistant contrast and resonance features in numerous experimental works~\cite{Baumann_Paul_science_2015, Yang_Bae_prl_2017, Willke_Bae_science_2018, Wang2023AnPlatform,Bui2024All-electricalSpinb}. Since in the CB regime, where the QI remains mostly singly occupied, tunneling-induced decoherence is rare and minimal, allowing for coherent spin evolution over timescales much longer than the average time between tunneling events. 

This interpretation is analogous to weak projective measurements in quantum computing, where the measurement of a qubit's state is performed via its coupling to a pointer observable, such as a dispersive resonator or a charge detector~\cite{Kastner2017}. There, as in our case, the observable read out (current or voltage) does not project onto a pure eigenstate in a single shot, but rather accumulates statistically meaningful information over many repetitions or over long-time averages. Our theoretical simulations incorporate this principle: we compute the time-dependent current arising from the QI spin dynamics and extract measurable features (resonance position, linewidth, contrast) by ensemble averaging encoded in the populations, coherences or expectation values of the spin components.

The situation is essentially similar in the STT regime, but transport is now highly incoherent. Frequent tunneling events rapidly degrade spin coherence, as reflected in the short $T_2$ times obtained in our simulations. Although $T_2$ can be partially enhanced, the incoherent nature of STT-driven ESR persists. Nonetheless, the current remains sensitive to the quantum evolution of the spin, particularly under resonant driving. We therefore interpret the current not as a fully projective measurement of the spin state at each instant, but as a time-resolved, spin-selective probe whose averaged behavior reflects the underlying quantum dynamics. This perspective is fully consistent with experimental ESR-STM protocols~\expSTMESR and provides a natural basis for using the current as both a readout and a feedback channel for coherent and incoherent spin manipulation.}

{\corr Let us now take a closer look at the experimental measurements and what setup enables precise spin readout. For concreteness, we refer to Ref.~\cite{yang_coherent_2019}. In STM, Rabi oscillations and other coherent control experiments are implemented using pulsed-ESR, where the spin state is driven by nanoscale RF pulses at the resonance frequency.  The readout is based on lock-in detection of the tunneling current, which alternates between two measurement cycles. In \textit{cycle A}, the RF drive is applied for a finite duration before the spin is allowed to relax, while in \textit{cycle B} only the DC current is measured, RF is off. Subtracting the signals from these two cycles isolates the contribution of the driven spin dynamics.
Rabi oscillations are then reconstructed by systematically varying the RF pulse length in \textit{cycle A}. Longer pulses rotate the spin further on the Bloch sphere before relaxation, producing a distinct modulation of the tunneling current. Each data point corresponds to the average over hundreds of repeated A–B cycles and sweeping the RF pulse length in the  \textit{cycle A}, reconstructs the full Rabi oscillation. 

Homodyne detection ensures that the measured current reflects the spin component parallel to the tip's magnetization axis. In our model, this detection operates only in the FLT regime. In contrast, STT relies on a different detection mechanism but it also encodes the spin state under an incoherent evolution~\cite{kovarik_spin_2024}. In all cases, the extremely fast Larmor oscillations are not resolved due to the limited temporal resolution of the instruments. Consequently, FLT measurements capture only the slower Rabi oscillations, as shown in Figs.~\ref{fig:CB_out_of_plane}b and~\ref{fig:CB_in_plane}b, whereas STT measurements exhibit an exponential decay, i.e., without the fast precession seen in Figs.~\ref{fig:SEQ_out_of_plane}b and~\ref{fig:SEQ_in_plane}b. These results are in agreement with Ref.~\cite{kovarik_spin_2024}.
}

\subsection{Behavior of \texorpdfstring{$\Omega$}{} and \texorpdfstring{$T_2$}{} under DC bias}

\begin{figure}
    \centering
\includegraphics[width=0.925\linewidth]{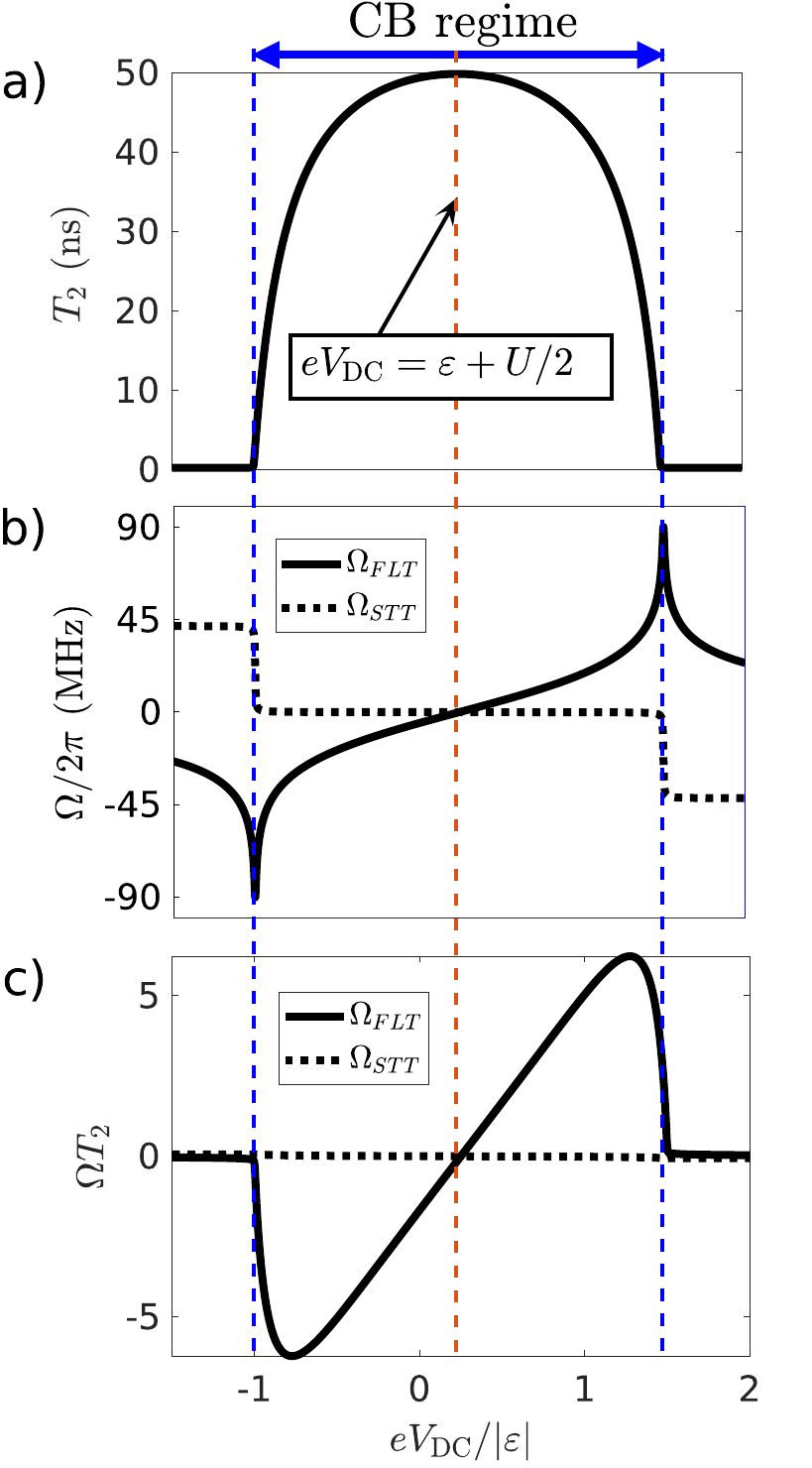}
    \caption{\textbf{Rabi rate and coherence time behavior under DC bias} a Coherence time $T_2$ as a function of the DC bias. A maximum of roughly 50 ns is achieved at the electron-hole symmetry point (vertical red dash-dotted line). The thresholds at $\varepsilon$ and $\varepsilon+U$ are marked as vertical blue dashed lines, after which the $T_2$ time drops to 0.23 ns. b) Rabi frequency behavior in both the FLT (solid line) and STT (dotted line) driven regimes as a function of DC bias. Both Rabi rates follow the equations from the theory section. (c) Figure of merit $\Omega T_2$, which determines the number of Rabi cycles and the efficiency of the ESR signal if the homodyne detection is not playing a role. Even in the absence of clear Rabi oscillations due to a small $T_2$, the product $\Omega T_2$ can still be large enough to produce a significant ESR signal.}
    \label{fig:Rabi_T2}
\end{figure}

In the following, we explore the behavior of the Rabi frequency $\Omega$ and coherence time $T_2$ as functions of the easily adjustable experimental parameter $eV_\mathrm{DC}$ for the out-of-plane situation studied in Figs.~\ref{fig:SEQ_out_of_plane} and~\ref{fig:CB_out_of_plane}. The result for $T_2$ is shown in Fig.~\ref{fig:Rabi_T2}a while the one for the Rabi rates is depicted in Fig.~\ref{fig:Rabi_T2}b. We separate the contributions from the STT and FLT driven regimes in Fig.~\ref{fig:Rabi_T2}b, as these evolve quite differently under varying DC bias. The behavior of $T_2$ is relatively straightforward to understand based on the previous discussion. In the low-bias CB regime, the current through the impurity is not a full one-electron process. Since scattering between the transport electrons and the impurity is the primary source of decoherence in this model, the coherence time is maximized. Therefore, the CB regime effectively protects the impurity spin from the decoherence induced by the polarized current. As we approach the charging thresholds $\varepsilon$ and $\varepsilon+U$ exiting the CB regime, $T_2$ rapidly decreases. Due to the single orbital nature of our model, $T_2$ then reaches a constant and very small value once the DC bias $V_\mathrm{DC}$ exceeds these thresholds. This demonstrates that the choice of $V_\mathrm{DC}$ is critical to achieve favorable conditions for quantum control.

Fig.~\ref{fig:Rabi_T2}b illustrates the behavior of the Rabi rate in both the STT and FLT driven regimes. In the CB regime, the FLT Rabi rate dominates over the STT Rabi rates, following a logarithmic behavior, as indicated by the imaginary part of Eq.~\eqref{Rabi_eq_analytical}. For low biases, $\Omega_{\mathrm{FLT}}$ exhibits a linear dependence that does not vanish at zero DC bias, which explains the presence of an ESR signal at $V_\mathrm{DC}=0$ meV, even though the signal intensity is low due to the minimal current passing through the impurity. As the bias approaches the CB regime threshold, $\Omega_{\mathrm{FLT}}$ becomes highly nonlinear, reaching a maximum (minimum) at $\varepsilon + U$ ($\varepsilon$). According to Eq.~\eqref{Rabi_eq_analytical}, $\Omega_{\mathrm{FLT}}$ would theoretically diverge at $\varepsilon + U$ and $\varepsilon$ for $\hbar/\tau_c \ll \varepsilon + U, |\varepsilon|$, but finite temperature will act as a natural cut-off if considered, preventing this divergence. Above the energy thresholds, the STT Rabi rate emerges strongly, leading to an incoherent ESR signal, as shown in Fig. \ref{fig:SEQ_out_of_plane} and \ref{fig:SEQ_in_plane}. The sign of the Rabi rates in each driven regime determines the rotation direction: a negative sign induces clockwise rotation in the Bloch sphere, while a positive sign results in counterclockwise rotation.

At the electron-hole symmetry point, located at $eV_\mathrm{DC} = \varepsilon + U / 2 = 25$ meV, the additional energy required to add or remove an electron is equal, balancing the electron and hole processes and causing the exchange field, $\Omega_{\mathrm{FLT}}$, to vanish, resulting in zero ESR signal. Notably, at this symmetry point, the coherence time $T_2$ reaches its maximum value, indicating that the impurity spin is most effectively shielded from the influence of the decoherent polarized current when the electron and hole processes cancel out each other. It is important to note that the electron-hole symmetry point may be absent if we do not apply the wide-band limit approximation in the density of states of the electrodes.

Figure~\ref{fig:Rabi_T2}c shows the figure of merit $\Omega T_2$, indicating a maximum of roughly 6 Rabi cycles before the steady state is reached for the FLT Rabi rate, as clearly depicted in Fig.~\ref{fig:CB_out_of_plane}. This underscores the importance of optimizing the DC bias to operate in the CB regime to achieve the best possible conditions for quantum control. Contrary, the $\Omega_{\mathrm{STT}}$ behavior only emerges outside the CB regime when the spin accumulation dominates the dynamics, as Fig.~\ref{fig:Rabi_T2}b shows. Therefore, the product $\Omega_{\mathrm{STT}} T_2$ does not have a large value for any DC bias, as we inferred from Eq.~\eqref{RabiT2_SEQ}. Therefore, the STT-driven regime is inadequate for quantum control, but it can be utilized to polarize the QI to initialize the spin, as we already showed in Fig.~\ref{fig:SEQ_out_of_plane}. 

\subsection{Full DC voltage mapping and energy dressing}

\begin{figure}
    \centering
\includegraphics[width=1\linewidth]{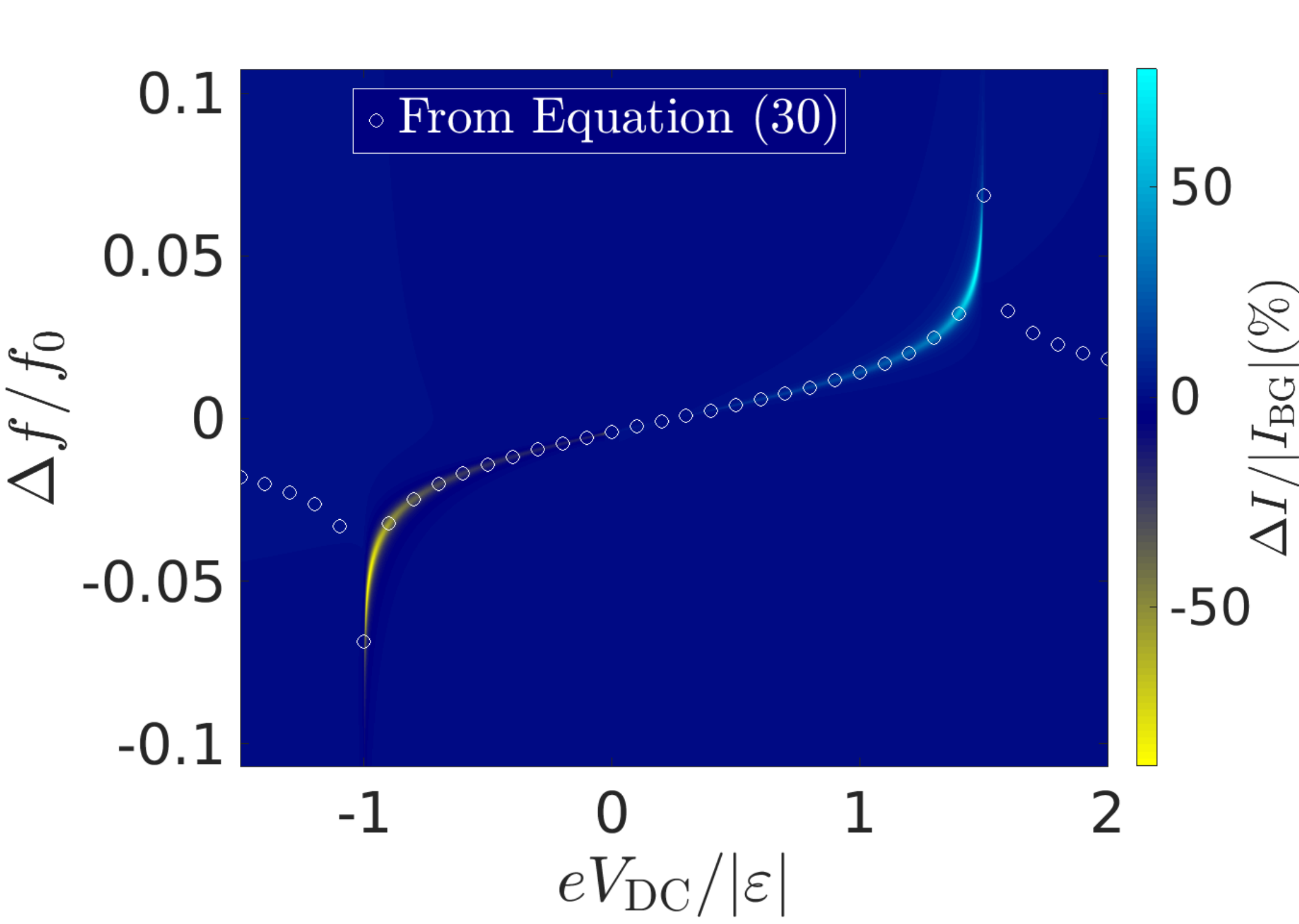}
\caption{\textbf{Shifting of the ESR resonance position with DC bias}. The ESR signal intensity, normalized to the background current, is shown as a color map, revealing peaks at positive DC biases and dips at negative DC biases. As the voltage approaches the thresholds $\varepsilon$ and $\varepsilon + U$, the ESR efficiency decreases substantially. The observed frequency shift agrees with the prediction of Eq.\eqref{energy_dress}, {\corr we have computed $\delta f/f_0=(f_0'-f_0)/f_0$}, which deviates by up to approximately 8\% (about 1.3 GHz) from the bare resonance frequency $f_0 = g \mu_B B/h = 16.8$ {\corr GHz. $\Delta f = f - f_0$ and $\Delta I=I_L-I_\mathrm{BG}.$}}
\label{fig:freq_renormalization}
\end{figure}

In this section, we examine the ESR signal response under varying DC bias for a fixed polar angle of $\theta = \pi/18$. The results are presented in Fig.~\ref{fig:freq_renormalization}, where a 2D color map highlights the normalized ESR signal relative to the background (baseline off-resonance) current, preserving the characteristic peak and dip behaviors. The efficiency of the ESR signal closely follows the figure of merit $\Omega_{\mathrm{FLT}} T_2$ shown in Fig.~\ref{fig:Rabi_T2}b, demonstrating low efficiency outside the CB regime beyond the energy thresholds $\varepsilon$ and $\varepsilon + U$ {\corr for the parameter used here}. 

As the DC bias approaches zero, the background current diminishes, albeit at a different rate compared to $\Delta I$. This results in a non-zero ESR signal at zero bias, a phenomenon reported previously~\cite{kot2023electric,J_Cuevas_C_Ast_ESR_theory_2024}. This observation aligns with the earlier discussion that $\Omega_{\mathrm{FLT}}$ does not vanish at zero bias but rather at the electron-hole symmetry point, $eV_\mathrm{DC}/|\varepsilon|=1+U/2|\varepsilon| = 0.25$. Indeed, there the signal completely disappears, as predicted by the exchange field formula, Eq.~\eqref{Rabi_simple}, and observed in Fig.~\ref{fig:Rabi_T2}. {\corr It is worth mentioning that if we boost the STT-driven ESR in the CB regime by forcing a larger $\hbar/\tau_c$, the signal at the symmetry point can become finite since it will be STT-mediated. This outcome of the model is not strictly physical, since increasing the broadening in this way is not fully consistent within a sequential tunneling picture; however, a similar effect could occur if cotunneling processes were incorporated into the model.}

An intriguing phenomenon observed in Fig.~\ref{fig:freq_renormalization} is the shifting of the ESR resonance position. We emphasize that in these simulations, the tip-sample distance is kept constant; therefore, the couplings remain unchanged while sweeping the DC bias. This dressing of the resonance energy arises from the projection of the exchange field onto the {\corr eigenbasis} axis of the spin and only vanishes at the electron-hole symmetry point or when the spin is initialized exactly perpendicular to the exchange field. In these cases, the resonance occurs at the bare resonance frequency $f_0= g\mu_B B/h = 16.8$ GHz.

This frequency shift has been consistently reported in the literature \expSTMESR, with various explanations proposed. Recently, it was attributed to a change in the $g$-factor induced by the electric field of the DC bias~\cite{kot2023electric}. However, experimental results~\cite{zhang2024electricfieldcontrolexchange}, supported by DFT simulations, show that the net effect of the DC bias on the magnetic moment is orders of magnitude too small, ruling out $g$-factor variation as the leading cause. Ref.~\cite{Wang2023AnPlatform} further excludes a $g$-factor contribution: a $g$-modulation would preclude the use of Fe atoms to drive the single-spin $1/2$ Ti atom, contradicting experimental observations, a point discussed in detail by some of us in Refs.~\cite{reinagalvez2024efficientdrivingspinqubitusing,zhang2024electricfieldcontrolexchange}. This leaves the exchange field as the most plausible mechanism for the phenomenon, though minor contributions from other sources cannot be entirely ruled out.

{\corr In fact, recent measurements provide supporting evidence for this mechanism~\cite{zhang2024electricfieldcontrolexchange,greule2025spinelectriccontrolindividualmolecules}, with the latter work clearly illustrating the nonlinear region in Fig.~\ref{fig:freq_renormalization}. Additionally,} the observable resonance frequency shift has been employed as a direct experimental method to extract the impurity parameters $\varepsilon$ and $\varepsilon + U$~\cite{zhang2024electricfieldcontrolexchange,greule2025spinelectriccontrolindividualmolecules}, in good agreement with DFT simulations~\cite{Corina_TMPc_paper_2025}. Since the shift depicted in Fig.~\ref{fig:freq_renormalization} is proportional to $\Omega_{\mathrm{FLT}}$ shown in Fig.~\ref{fig:Rabi_T2}a, see Eq.~\eqref{dress_rabi}, measuring the resonance shift provides a practical way to determine the Rabi rate. This prediction could be tested using alternative Rabi-rate estimation methods, such as that presented in~\cite{Bui2024All-electricalSpinb}.

\subsection{Vector field dependence of the Rabi rate and ESR signal: Homodyne detection}

The polar angle $\theta$ formed between the external magnetic field and spin polarization of the tip appears naturally in all quantities of interest and is closely related to measuring an efficient ESR signal, as we observed by comparing Figs.~\ref{fig:CB_out_of_plane}c and~\ref{fig:CB_in_plane}c. As such, it is important to understand the behavior of our model with respect to $\theta$. Furthermore, since the static magnetic field angle is a parameter easily adjusted in some experiments, confirming the agreement we have between model and experiment will provide validity of the driving field emerging from the modulation of the exchange magnetic field and the hopping.

\begin{figure}
    \centering
    \includegraphics[width=1\linewidth]{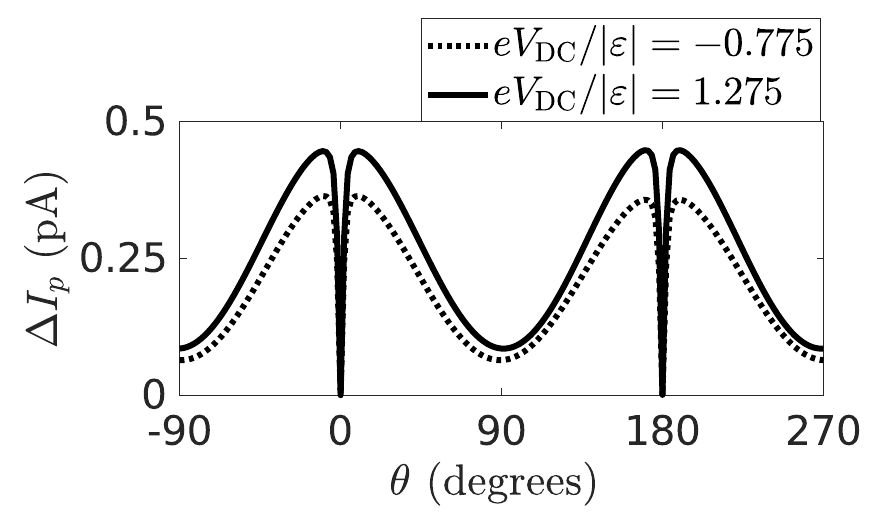}
    \caption{Example of the ESR signal behavior \textit{on resonance}, calculating the total amplitude $\Delta I_p=\sqrt{\Delta I_s^2 + \Delta I_a^2}$, at negative (dotted line) and positive (solid line) DC bias as a function of the polar angle $\theta$ formed between the magnetic field and the tip spin polarization. The maximum ESR signal occurs at a predominantly out-of-plane angle, as a result of homodyne detection. A sizable signal for an in-plane angle arises due to the spin accumulation. 
    }
    \label{fig:ESR_vs_angle}
\end{figure}

As an example of the angle behavior, we have Fig.~\ref{fig:ESR_vs_angle} showing the total ESR amplitude $\Delta I_p=\sqrt{\Delta I_s^2 + \Delta I_a^2}$ for the optimal DC bias values that maximize the quality factor $\Omega_\mathrm{FLT} T_2$, see Fig.~\ref{fig:Rabi_T2}c. $\Delta I_s$ denotes the symmetric part of the ESR signal, while $\Delta I_a$ provides the asymmetric part. Both are extracted from fitting the raw ESR signal to a symmetric and an antisymmetric Lorentzian center on the resonance frequency{\corr~following Ref.~\cite{kovarik_spin_2024},
\begin{equation}
\Delta I=I_L - I_\mathrm{BG}= \Delta I_s \frac{1}{1+\delta^2} - \Delta I_a \frac{2\delta}{1+\delta^2}, \label{asy_sym_current}
\end{equation}
where $\delta = 2(f-f_0')/W= 2(f-f_0-\delta f)/W$, with $W$ the full width at half maximum.}

\begin{figure}
\includegraphics[width=1.\linewidth]{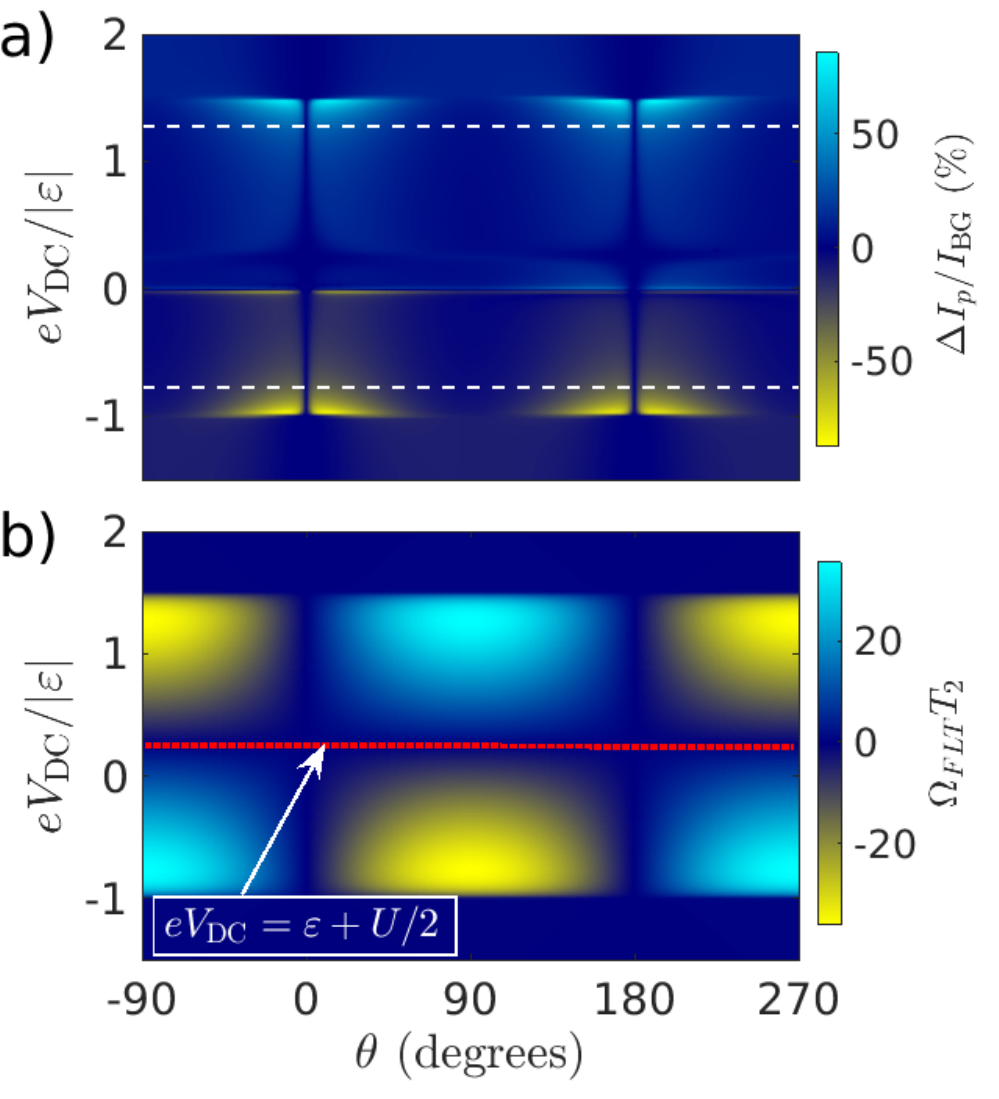}
    \caption{\textbf{Full DC and vector field dependence (Homodyne detection).} 
a) 2D color map of the ESR total amplitude $\Delta I_p=\sqrt{\Delta I_s^2+\Delta I_a^2}$, normalized to the background current, keeping its sign, as a function of the polar angle and DC bias, highlighting regions of high ESR efficiency. $\Delta I_s$ and $\Delta I_a$ denote symmetric and asymmetric component, see Eq.~\eqref{asy_sym_current}. At $\theta = 0, \pi$ and along the electron-hole symmetry lines, no ESR signal is observed. White dashed lines indicate the cuts shown in Fig.~\ref{fig:ESR_vs_angle}.
b) Figure of merit $\Omega_{\mathrm{FLT}} T_2$ as a function of the polar angle $\theta$ and DC bias, showing a sinusoidal dependence within the Coulomb
blockade regime proportional to $\sin\theta$, and a rapid suppression outside of it. The red dotted line marks the electron-hole symmetry line at $eV_\mathrm{DC}/|\varepsilon|=1+U/2|\varepsilon|=0.25$.}
    \label{fig:ESR_RabiT2_vs_angle}
\end{figure}

At $\theta = 0$ or $\pi$, the ESR signal vanishes completely because the driving field and the spin are parallel. Experimentally, this is not observed, likely because the tip forms an angle with the plane containing the spin i.e. in experiments the magnetization direction of the tip is generally not exactly aligned with $z$ as in the model. This can be incorporated in the simulations by tilting the plane in which the magnetic field lies by an angle $\gamma$ relative to the spin polarization along $z$, {\corr so that the spin and tip directions do not lie in the same plane, see Appendix~\ref{app:Homodyne_tilted_B}.}

{\corr Outside $\theta = 0, \pi$, the ESR amplitude at positive and negative DC bias follows the homodyne detection pattern, with a maximum out-of-plane} ($\theta\sim 10$ degrees), which matches the experimental evidence~\cite{JK_vector_field_2021,Fe_driving_SH_2023}. However, at 90 degrees, we encounter a local minimum, which goes against the conventional ESR-STM measurements, since the homodyne detection {\corr predicts zero amplitude for in-plane angles} as the ESR signal is proportional to the scalar product of the spin and the magnetization direction of the tip. As we already mentioned, we identify this discrepancy as {\corr a~consequence of spin accumulation, which is less pronounced in the experiments, but produces a small} current contribution along $z$ axis in the model, {\corr enabling visualization of the ESR response. To suppress this feature we can lower the DC bias to move deeper into the CB regime; however, the measured current then becomes very small, Fig.~\ref{fig:Sacc_no_driving}c. In this situation, the model demands including higher-order tunneling processes (cotunneling) to produce a measurable current. } It is worth noting that, since we are plotting $\Delta I_p$, we do not visualize the flipping of the {\corr symmetric} signal happening when the angle is sweeping {\corr from $0$ to $90$ degrees}, going through a pure Fano profile, {\corr see Appendix~\ref{app:Homodyne_tilted_B}.} 

Turning our sight to Fig.~\ref{fig:ESR_RabiT2_vs_angle}a, we observe the ESR amplitude dependence versus both the DC bias {\corr and $\theta$. The voltage dependence is
similar to} in Fig.~\ref{fig:freq_renormalization} and independent of the angle, but resolving the resonance shifting since we are plotting $\Delta I_p$ on resonance. If we focus on the angle dependence, we see again the ESR signal vanishing in the case of parallel ($\theta=0$ degrees) and anti-parallel ($\theta=180$ degrees) orientations, with a local minimum for a perpendicular alignment ($\theta=\pm90$ degrees) influence by the spin accumulation. As we mentioned earlier, reducing the DC bias reduces the ESR amplitude at the perpendicular alignment, even though the quality factor $\Omega_\mathrm{FLT} T_2$ maximizes at $\theta=90$ degrees, see Fig.~\ref{fig:ESR_RabiT2_vs_angle}b. 

Additionally, Fig.~\ref{fig:ESR_RabiT2_vs_angle}b demonstrates the sinusoidal dependence of the Rabi rate on $\theta$, as experimentally reported in~\cite{Fe_driving_SH_2023}, since $T_2$ is angle-independent. The electron-hole symmetry line, $eV_\mathrm{DC} = \varepsilon + U/2$, represented by a red dotted line in Fig.~\ref{fig:ESR_RabiT2_vs_angle}b, provides vanishing Rabi rate and ESR signal, consistent with Fig.~\ref{fig:ESR_RabiT2_vs_angle}a at $eV_\mathrm{DC}/|\varepsilon| = 0.25$.
Outside the CB regime, the quality factor drops to extremely low values, following the decrease in $T_2$ observed in Fig.~\ref{fig:Rabi_T2}a. Near zero DC bias, the ESR amplitude relative to the background exhibits a local maximum or minimum, depending on whether $\theta \in (-90^\circ,90^\circ)$ or falls within the range $[90^\circ, 270^\circ]$, respectively. This ESR signal behavior aligns with the fact that the Rabi rate remains nonzero at zero bias.

\section{Conclusions}

In this work, we have derived an intuitive physical picture of ESR in an STM, building on the solid mathematical aspects of non-equilibrium transport through a spin-polarized electrode under harmonic modulation of the bias. {\corr Using a single-orbital Anderson impurity model (SAIM), which is fairly general and can describe a wide variety of localized spin defects, we capture} the emergence of ESR driven purely by electric fields, which is experimentally realized in ESR-STM. Our model \textit{accounts for the driving amplitude} in both the FLT-driven and the STT-driven regimes. However, it does not fully incorporate the complete transport picture, as cotunneling is not considered, leaving the door open for future extensions of this work.

Two transport regimes emerge. Within the charging window ($\varepsilon, \varepsilon+U$),  the transport occurs in the Coulomb blockade regime and the impurity is singly occupied, yielding ESR driven by {\corr the perpendicular component of} the exchange field $B_{\mathrm{exch}}$ that arises as a consequence of the spin polarization of the electrode. {\corr This produces FLT-driven ESR, following results that align with conventional ESR, presenting homodyne detection and Rabi oscillations. On the other hand, the exchange field component parallel} to the {\corr eigenbasis} axis leads to a renormalization of the resonance frequency $f_0$, which is logarithmically dependent on the applied bias relative to the energy thresholds of the SAIM, $\varepsilon, \varepsilon+U$. This shift can be used to spectroscopically probe the energy thresholds and the Rabi rate. We further show that the quality factor $\Omega_\mathrm{FLT} T_2$ peaks near the charging energy thresholds. In general, in the Coulomb blockade regime, the impurity spin is an ideal candidate for quantum-coherent control.

Outside the Coulomb blockade, transport allows electron hopping in and out of the quantum impurity from the electrodes. Consequently, the current through the impurity increases rapidly and $T_2$ decreases drastically. ESR is now STT-driven, dominated by the spin-polarized current, which torques the spin via the {\corr time-dependent} spin accumulation, preventing coherent control ($\Omega_\mathrm{STT} T_2 \sim 1/3$). Despite this, the quantum impurity can be strongly polarized along the spin's {\corr eigenbasis axis, where the Hamiltonian is diagonal}, provided it is not initialized perpendicular to the spin polarization. This offers an alternative method for initializing the spin in a preferred orientation, independently of temperature. Previously, STT was analyzed semiclassically using the Landau–Lifshitz–Gilbert–Slonczewski equation.  {\corr Later, a quantum STT treatment explained the ESR signal in the cotunneling regime~\cite{Ribeiro2}. In contrast, our model captures quantum spin-transfer torque in sequential tunneling, including both coherent and incoherent ESR, highlighting the role of FLT-driven regime in spin control, a feature particularly relevant to ESR-STM.}

Finally, our model is able to reproduce a range of experimental observations in both qualitative and quantitative agreement, such as homodyne detection and resonance shifts. This solidifies the validity of our driving-mechanism approach and provides a guide for the numerical optimization of on-surface electron spin systems for quantum simulations, quantum sensing, and quantum information processing. Moreover, it constitutes a significant step toward a unified framework that treats both sequential and cotunneling transport regimes, {\corr while explicitly accounting for first-order coherent effects such as the exchange field or the Lamb shift.}

\begin{acknowledgments}
We acknowledge fruitful discussions with A. Ardavan, A. J. Heinrich,  C. Ast, C. P. Lutz, D.-J. Choi, D. Ruckert, D. Tomaszewski, F. Delgado, F. Donati, G. Burkard, G. Czap, H. T. Bui, J. Barnaś, J. König, J. C. Cuevas, L. Hoang-Anh, M. Kelai, M. Ternes, P. Gambardella,  P. Kot,  P. Busz, S.-H. Phark, S. Baumann, S. Loth, S. Stepanow, S. Kovarik, R. Broekhoven, and R. Schlitz during the process of writing this manuscript.
\newline
This work was supported by the Institute for Basic Science (IBS-R027-D1) and the German Research Foundation [Deutsche Forschungsgemeinschaft (DFG)] under Projects No. 450396347. Projects PID2021-127917NB-I00 funded by MCIN/AEI/10.13039/501100011033, QUAN-000021-01 funded by the Gipuzkoa Provincial Council, IT-1527-22 funded by the Basque Government, and ESiM project 101046364 supported by the European Union are gratefully acknowledged. Support from the National Science Centre of Poland (Grant No. 2020/36/C/ST3/00539) is also appreciated. Funded by the European Union. Views and opinions expressed are, however, those of the author(s) only and do not necessarily reflect those of the European Union. Neither the European Union nor the granting authority can be held responsible for them.
\end{acknowledgments}

\appendix

\section{Expectation value \texorpdfstring{$\langle s_i\rangle $}{}}
\label{app:spin_expectation}

In the main text, we introduced the general expression for the expectation values $\langle s_i\rangle = \mathrm{Tr}(\hat{\rho}\hat{s}_i)$ with $i=x,y,z$. Here, we derive these expressions for a general magnetic field $\mathbf{B}=B(\sin\phi \sin\theta,\cos\phi \sin\theta,\cos\theta)$, where $\theta$ is the polar angle and $\phi$ the azimuthal angle.

A key consideration when computing the trace is ensuring that both operators are represented in the same basis. Otherwise, the expectation value would be incorrect. This is particularly relevant when considering an external magnetic field, as the spin aligns with the magnetic field for an isotropic $g$-factor {\corr and the QI Hamiltonian will be diagonal in the axis parallel to $\mathbf{B}$}. Consequently, the density matrix and the Pauli matrices do not necessarily share the same basis, since the former is defined in the eigenstate basis, which is tilted by the magnetic field.

To circumvent this problem, one could change the basis set of either operator into the other one. In this work, we chose to perform a basis transformation of the standard Pauli matrices to the eigenvector basis set:
\begin{equation*}
\langle l |\hat{s}_i |j\rangle =\sum_{\sigma,\sigma'}  \langle l| \sigma\rangle\langle \sigma|\hat{s}_i|\sigma'\rangle  \langle \sigma'| j\rangle ,
\end{equation*}
where $l,j$ are singly occupied eigenstates of the spin-1/2, while $\sigma,\sigma'$ correspond to the spin-up and spin-down states and $s_i$ are the Pauli matrices. We focus exclusively on the singly occupied states, as the empty and doubly occupied states provide zero spin to spin-1/2 QI. For a general magnetic field, {\corr the singly ground and excited eigenstates in the up/down basis set} have the following expression:
\begin{equation}
    |g\rangle =\left(\sin\frac{\theta}{2},-e^{i\phi}\cos\frac{\theta}{2}\right),\ 
    |e\rangle =\left(\cos\frac{\theta}{2},e^{i\phi}\sin\frac{\theta}{2}\right),
    \label{eigenvectors_eg}
\end{equation}
where $g$ and $e$ denote the ground and excite states of the QI respectively. Meanwhile, the standard up and down spin states of the Pauli matrices are: $|\mkern-5mu \downarrow\rangle= (0,1),\ |\mkern-5mu \uparrow\rangle=(1,0).$
The density matrix is always expressed with the ground state as the first matrix element:
\begin{equation*}
\hat{\rho}=\begin{pmatrix}
\rho_{gg} & \rho_{eg} \\
\rho_{ge} & \rho_{ee}
\end{pmatrix}
\end{equation*}
Thus, the first matrix element of the spin operator must correspond to the ground state. Computing every matrix element for the $z$-component leads to: 
\begin{eqnarray*}
    \langle g |\hat{s}_z |g\rangle &=&  \frac{\hbar}{2}\left(\langle g|\mkern-5mu \uparrow\rangle \langle  \uparrow\mkern-5mu | g\rangle - \langle g|\mkern-5mu \downarrow\rangle \langle  \downarrow\mkern-5mu| g\rangle\right) \\
    &=& \frac{\hbar}{2} \left(\sin^2\left(\frac{\theta}{2}\right) - \cos^2\left(\frac{\theta}{2}\right)\right) =- \frac{\hbar\cos\theta}{2}\\
    \langle g |\hat{s}_z |e\rangle &=& \frac{\hbar\sin\theta}{2}\ \ ; \ \
    \langle e |\hat{s}_z |g\rangle = \frac{\hbar\sin\theta}{2}\\
    \langle e |\hat{s}_z |e\rangle &=& \frac{\hbar\cos\theta}{2}\\
\end{eqnarray*}
Thus we write
\begin{equation*}
\hat{s}_z=\frac{\hbar}{2}\begin{pmatrix}
-\cos\theta & \sin\theta \\
\sin\theta & \cos\theta
\end{pmatrix},
\end{equation*}
which is the general expression of the spin operator on z. This implies that:
\begin{equation*}
\hat{\rho}\hat{s}_z =\frac{\hbar}{2}\begin{pmatrix}
-\rho_{gg}\cos\theta + \rho_{ge}\sin\theta & -\rho_{eg}\cos\theta + \rho_{ee}\sin\theta \\
\rho_{gg}\sin\theta + \rho_{ge}\cos\theta & \rho_{eg}\sin\theta + \rho_{ee}\cos\theta
\end{pmatrix},
\end{equation*}
arriving at
\begin{eqnarray*}
    \langle s_z \rangle &=&\frac{\hbar}{2} (\rho_{ee}-\rho_{gg}) \cos\theta + \frac{\hbar}{2} (\rho_{ge}+\rho_{eg}) \sin \theta  \\
& \underbrace{\approx}_{\theta \ll 1 }& \frac{\hbar}{2}
(\rho_{\uparrow\uparrow}-\rho_{\downarrow\downarrow}),
\end{eqnarray*}
which is the equation used in the main text. Only when $\theta$ is sufficiently small we recover the up and down notation, indicating that the spin is in the vertical axis of the Bloch sphere.
Turning our attention to the $x$-component of the spin we obtain:
\begin{eqnarray*}
    \langle g |\hat{s}_x |g\rangle &=& \frac{\hbar}{2}\left(\langle g|\mkern-5mu \uparrow\rangle \langle  \downarrow\mkern-5mu| g\rangle + \langle g|\mkern-5mu \downarrow\rangle \langle  \uparrow\mkern-5mu| g\rangle\right) \\
    &=& -\frac{\hbar}{2}\left(e^{i\phi} + e^{-i\phi} \right)\sin\frac{\theta}{2} \cos\frac{\theta}{2} \\
    &=& -\frac{\hbar}{2}\cos\phi \sin\theta  
    \end{eqnarray*}
    \begin{eqnarray*}
    \langle g |\hat{s}_x |e\rangle &=& -\frac{\hbar}{2}( \cos\theta\cos\phi -i\sin\phi) \\
     \langle e |\hat{s}_x |g\rangle &=& -\frac{\hbar}{2}( \cos\theta\cos\phi +i\sin\phi)\\
    \langle e |\hat{s}_x |e\rangle &=& \frac{\hbar}{2}\cos\phi \sin\theta\\
\end{eqnarray*}
Hence the transformed operator is
\begin{equation*}
    \hat{s}_x =-\frac{\hbar}{2}\begin{pmatrix}
\cos\phi \sin\theta &\cos\theta\cos\phi +i\sin\phi \\
\cos\theta\cos\phi -i\sin\phi & -\cos\phi \sin\theta
\end{pmatrix},
\end{equation*}
which leads to the expectation value
\begin{eqnarray*}
    \langle{s_x}\rangle &=&\frac{\hbar}{2}
(\rho_{ee}-\rho_{gg})\sin\theta \cos\phi \\ &-&
\frac{\hbar}{2}(\rho_{ge}  +  \rho_{eg})\cos\theta\cos\phi -i\frac{\hbar}{2}(\rho_{ge}-\rho_{eg})\sin\phi.
\end{eqnarray*}
Finally, for the $y$-component we have
\begin{eqnarray*}
    \langle g |\hat{s}_y |g\rangle &=& \frac{\hbar}{2}\left(-i\langle g|\mkern-5mu \uparrow\rangle \langle  \downarrow\mkern-5mu| g\rangle + i\langle g|\mkern-5mu \downarrow\rangle \langle  \uparrow\mkern-5mu| g\rangle\right) \\
    &=& i\frac{\hbar}{2} \left( e^{i\phi}\sin\frac{\theta}{2} \cos\frac{\theta}{2} - e^{-i\phi}\sin\frac{\theta}{2} \cos\frac{\theta}{2}\right) \\
    &=&  -\frac{\hbar}{2} \sin\phi \sin\theta \\
    \langle g |\hat{s}_y |e\rangle &=& -\frac{\hbar}{2} (i\cos\phi + \cos\theta \sin\phi) \\
    \langle e |\hat{s}_y |g\rangle &=&  \frac{\hbar}{2}(i\cos\phi - \cos\theta \sin\phi)\\
    \langle e |\hat{s}_y |e\rangle &=& \frac{\hbar}{2} \sin\phi \sin\theta
\end{eqnarray*}
Therefore,
\begin{equation*}
    \hat{s}_y =-\frac{\hbar}{2}\begin{pmatrix}
 \sin\phi \sin\theta &  -i\cos\phi + \cos\theta \sin\phi \\
i\cos\phi + \cos\theta \sin\phi & -\sin\phi \sin\theta
\end{pmatrix},
\end{equation*}
and we obtain 
\begin{eqnarray*}
    \langle {s_y}\rangle &=&\frac{\hbar}{2}
(\rho_{ee}-\rho_{gg})\sin\theta \sin\phi \\ &-&
\frac{\hbar}{2}(\rho_{ge}  +  \rho_{eg})\cos\theta\sin\phi +i\frac{\hbar}{2}(\rho_{ge}-\rho_{eg})\cos\phi.
\end{eqnarray*}
For simplicity, we have been considering the case of $\phi=0$ for the derivation of Eqs.~\eqref{spin_equations} and~\eqref{spin_equations_b} and throughout this work, arriving at the expressions for the spin of x and y used in the main text. From the spin components, we can obtain the population difference as well as the real and imaginary parts of the coherences, which are defined along the {\corr eigenbasis} axis:
{\small
\begin{eqnarray}
    \frac{\rho_{ee} - \rho_{gg}}{2} &=& \frac{ \langle s_z \rangle \cos\theta + \langle s_x \rangle \sin\theta \cos\phi + \langle s_y \rangle \sin\theta \sin\phi }{\hbar} \nonumber \\ 
    \frac{\rho_{ge} + \rho_{eg}}{2} &=& \frac{ \langle s_z \rangle \sin\theta - \langle s_x \rangle \cos\theta \cos\phi - \langle s_y \rangle \cos\theta \sin\phi }{\hbar} \nonumber \\ 
    \frac{\rho_{ge} - \rho_{eg}}{2} &=& \frac{i}{\hbar} \left( \langle s_x \rangle \sin\phi - \langle s_y \rangle \cos\phi \right). 
    \label{pop_spin_general}
\end{eqnarray}}

\section{Coefficients \texorpdfstring{$\sum_\sigma \lambda_{lj,\sigma}\mu_{uv,\sigma}(1+2\sigma P_\alpha))$}{}}\label{app:lambda_coef}

These coefficients appear in the rates, and knowing how to evaluate them is essential for computing certain equations in this work. For the general eigenstates of a spin-1/2, Eqs.~\eqref{eigenvectors_eg}, 
\begin{eqnarray*}
   |g\rangle &=&\left(\sin\frac{\theta}{2},-e^{i\phi}\cos\frac{\theta}{2}\right) \nonumber \\ 
   |e\rangle &=&\left(\cos\frac{\theta}{2},e^{i\phi}\sin\frac{\theta}{2}\right),
   \label{eigenvectors_eg_1}
\end{eqnarray*}
the non-zero matrix elements between ground/excited states and the transient ones are:
\begin{eqnarray*}
    &&\lambda_{0g, \uparrow}=\langle 0 |0\rangle \langle  \uparrow\mkern-5mu|g\rangle=\sin\frac{\theta}{2} \\ 
    &&\lambda_{0e, \uparrow}=\langle 0 |0\rangle \langle  \uparrow\mkern-5mu|e\rangle = \cos\frac{\theta}{2} 
    \\
    &&\lambda_{0g, \downarrow}= \langle 0 |0\rangle \langle  \downarrow\mkern-5mu|g\rangle= -e^{i\phi}\cos\frac{\theta}{2}\\ 
    &&\lambda_{0e, \downarrow}= \langle 0 |0\rangle \langle  \downarrow\mkern-5mu|e\rangle= e^{i\phi}\sin\frac{\theta}{2} \\
    &&\lambda_{g2, \uparrow}= \langle g  |\mkern-5mu \downarrow\rangle \langle 2|2\rangle = -e^{-i\phi}\cos\frac{\theta}{2} \\
    &&\lambda_{e2, \uparrow}= \langle e  |\mkern-5mu \downarrow\rangle \langle 2|2\rangle = e^{-i\phi}\sin\frac{\theta}{2} \\
    &&\lambda_{g2, \downarrow}= \langle g  |\mkern-5mu \uparrow\rangle \langle 2|2\rangle = \sin\frac{\theta}{2} \\
    &&\lambda_{e2, \downarrow}= \langle e  |\mkern-5mu \uparrow\rangle \langle 2|2\rangle =\cos\frac{\theta}{2}
\end{eqnarray*}
Taking into account that $\gamma_{\alpha\sigma} ^0=\gamma_{\alpha}^0(1/2+\sigma P_\alpha)$, we can compute all the relevant coefficients:
\begin{eqnarray*}
    &&\sum_\sigma\lambda_{0g,\sigma}\mu_{g0,\sigma}(1+2\sigma P_\alpha)\\ 
    &=& 
    (1+P_\alpha)\sin^2\frac{\theta}{2} + (1-P_\alpha)\cos^2\frac{\theta}{2}=1 - P_\alpha\cos\theta\\ 
    &&\sum_\sigma\lambda_{0e,\sigma}\mu_{e0,\sigma}(1+2\sigma P_\alpha)\\
    &=&(1+P_\alpha)\cos^2\frac{\theta}{2} + (1-P_\alpha)\sin^2\frac{\theta}{2}=1 + P_\alpha\cos\theta \\
    &&\sum_\sigma\lambda_{e2,\sigma}\mu_{2e,\sigma}(1+2\sigma P_\alpha)=1 - P_\alpha\cos\theta\\
    &&\sum_\sigma\lambda_{g2,\sigma}\mu_{2g,\sigma}(1+2\sigma P_\alpha)=1 + P_\alpha\cos\theta\\
    &&\sum_\sigma\lambda_{0g,\sigma}\mu_{e0,\sigma}(1+2\sigma P_\alpha)=(1+P_\alpha)\sin\frac{\theta}{2}\cos\frac{\theta}{2} \\
    &-& (1-P_\alpha)\cos\frac{\theta}{2}\sin\frac{\theta}{2}=P_\alpha\sin\theta\\
    &&\sum_\sigma\lambda_{g2,\sigma}\mu_{2e,\sigma}(1+2\sigma P_\alpha)=-P_\alpha\sin\theta.
\end{eqnarray*}

\section{Details on current expression for a spin 1/2}\label{app:homodyne}

We will refer to results from the previous two appendices~\ref{app:spin_expectation} and~\ref{app:lambda_coef} for $\phi=0$, as in the main text. The current leaving the left lead is
\begin{eqnarray}
    I_L(t)&=&\frac{2e}{\hbar} \Bigg[\rho_{ee}(t) \mbox{Re} \left\{\Gamma_{e 2,2 e,L}^{+}(t)-\Gamma_{e 0,0e,L}^{-}(t) \right\} \nonumber \\
    &+&\left. \rho_{gg}(t) \mbox{Re} \left\{\Gamma_{g 2,2 g,L}^{+}(t) - \Gamma_{g 0,0 g,L}^{-}(t) \right\} \right. \nonumber \\
    &+& \mbox{Re}\{\rho_{ge}(t)\} \mbox{Re} \left\{\Gamma_{g 2,2 e,L}^{+}(t) - \Gamma_{g 0,0 e,L}^{-}(t)\right. \nonumber \\
    &+&\left. \Gamma_{e 2,2 g,L}^{+}(t) - \Gamma_{e 0,0 g,L}^{-}(t)\right\} \nonumber\\    
&+&\hspace{-4.5pt}\sum_{j=e,g}\left(\rho_0(t)\mbox{Re}\left\{\Gamma_{0 jj0,L}^{+} \right\} -\rho_2(t)\mbox{Re}\left\{ \Gamma_{2jj2,L}^{-}\right\}\right) \Bigg]\nonumber\\
    &=&  I_L^2(t)+I_L^0(t)=-I_R(t),
    \label{current_app}
\end{eqnarray}
where we have separated hole and electron contributions to the current and canceled out the ones that are zero due to having matrix elements $\lambda_{lj,\sigma}$ null. The major difference between the hole and electron rates comes from the first one containing $\varepsilon+U$, while the second one $\varepsilon$, but the spin dependence is similar and encoded in terms such as $\sum_{\sigma} \lambda_{l v\sigma}\lambda_{v j\sigma}^*(1+2\sigma P_L)$, which we have already computed in Appendix~\ref{app:lambda_coef}. These expressions, taking into account that the magnetic field is in the GHz regime, so the integrals ${\cal I}_L^\pm(\Delta_{v j})$ will not depend on the Zeeman energy but only on $\varepsilon$, $U$, and $\mu_\alpha$, allow us to write:
\begin{eqnarray}
    &&\Gamma_{e v,v e,L}^{\pm}(t) =\frac{1}{2}(1\mp P_L\cos\theta)
       \gamma_{L}(t) {\cal I}_L^\pm(\Delta_{vg}) \nonumber \\
     &&  \Gamma_{g v,v g,L}^{\pm}(t) = \frac{1}{2}(1\pm P_L\cos\theta)
       \gamma_{L}(t){\cal I}_L^\pm(\Delta_{vg}) \nonumber \\
       &&\Gamma_{e v,v g,L}^{\pm}(t) =\mp \frac{1}{2}P_L\gamma_{L}(t)
        {\cal I}_L^\pm(\Delta_{vg})\sin\theta \nonumber \\
    &&\sum_{j=e,g}\Gamma_{0 j,j 0,L}^{+}(t) = 
       \gamma_{L}(t) {\cal I}_L^+(\Delta_{g0}) \nonumber \\
    &&\sum_{j=e,g}\Gamma_{2 j,j 2,L}^{-}(t) = 
       \gamma_{L}(t) {\cal I}_L^-(\Delta_{g 2}) \nonumber \\ \nonumber 
\end{eqnarray}
where, on the first three equations, the ones where the first subscript in the rates start with a singly occupied state, the $+$ sign signifies the electron-like transport, implying that $v=0$, while the $-$ sign applies to the hole-like process, where $v=2$. Additionally, we introduce the time-dependent non-polarized coupling
$\gamma_{L}(t)=\left[1+A_{L}\cos (\omega t)\right]^2 \gamma_{L}^0$. Substituting these rates into Eq.~\eqref{current_app} yields
\begin{eqnarray}
    I_L^0(t)&=&-\frac{e}{\hbar} \gamma_L^0\left(1+A_{L}\cos (\omega t) \right)^2\left( \mbox{Re} \left\{ {\cal I}_L^-(\Delta_{0e}) \right\} \right. \nonumber \\
    &\times&\left[\rho_{ee}(t) + \rho_{gg}(t) +P_L\cos\theta (\rho_{ee}(t) - \rho_{gg}(t))\right. \nonumber \\
    &+& \left. 2\mbox{Re}\{\rho_{ge}(t)\} P_L\sin\theta  \right] -\left. 2\rho_0  \mbox{Re} \left\{ {\cal I}_L^+(\Delta_{e0})\right\}\right). \nonumber
    \end{eqnarray}
Introducing Eq.~\eqref{General_Sz} leads to
\begin{eqnarray}
    I_L^0(t)&=& -\frac{e}{\hbar} \gamma_L(t) \left[(\rho_1+ 2P_L \langle s_z \rangle )\mbox{Re} \left\{ {\cal I}_L^-(\Delta_{0e}) \right\} \right.\nonumber \\ 
    &-&\left. 2\rho_0 \mbox{Re} \left\{ {\cal I}_L^+(\Delta_{e0})\right\} \right],\nonumber
\end{eqnarray}
and similarly 
\begin{eqnarray}
    I_L^2(t)&=& \frac{e}{\hbar} \gamma_L(t) \left[(\rho_1 - 2P_L \langle s_z \rangle ) \mbox{Re} \left\{ {\cal I}_L^+(\Delta_{2e}) \right\}\right. \nonumber \\ 
    &-&\left. 2\rho_2 \mbox{Re} \left\{ {\cal I}_L^-(\Delta_{e 2})\right\}\right], \nonumber 
\end{eqnarray}
which together gives Eqs.~\eqref{homo_current} and~\eqref{homo_current_final}, the latter by introducing Eq.~\eqref{General_Sz} and when the population of the transient states, $\rho_0$ and $\rho_2$, are negligible and $\rho_{ee}(t) + \rho_{gg}(t) = \rho_1 = 1 - \rho_2 - \rho_0 \approx 1$. 
{\corr For $E_c\gg |\varepsilon|,\varepsilon+U\gg \hbar/\tau_c$, ${\cal I}_L^+(\Delta_{e0})={\cal I}_L^-(\Delta_{e2})\rightarrow 0$ outside of the CB regime while $\rho_0=\rho_2=0$ inside of it, as Fig.~\ref{fig:Sacc_no_driving}c shows. This simplifies the equation of the current, obtaining the following analytical form at low temperatures:
\begin{eqnarray}
  &&  I_L(t)= \frac{e}{h} \gamma_L(t) \nonumber\\ 
    &\times& \left[ (\rho_1 - 2P_L \langle s_z \rangle ) \left\{\frac{\pi}{2}+\arctan\left(\frac{eV_\mathrm{DC}-\varepsilon-U}{\hbar/\tau_c}\right) \right\} \right.\nonumber \\
    &-&\left.(\rho_1+ 2P_L \langle s_z \rangle )\left\{\frac{\pi}{2}-\arctan\left(\frac{eV_\mathrm{DC}-\varepsilon}{\hbar/\tau_c}\right) \right\} \right]
    \label{eq:current_analytical}
\end{eqnarray}
Expression that gives zero if $\hbar/\tau_c=0$ and we are inside the CB regime, and $$I_L(t)=\pi e\  \mathrm{sign}(V_\mathrm{DC}) \gamma_L(t)(\rho_1 -2\ \mathrm{sign}(V_\mathrm{DC}) P_L \langle s_z \rangle )/h,$$ for DC bias values far above the energy thresholds, i.e. $|eV_\mathrm{DC}|\gg |\varepsilon|,\varepsilon+U$.}

\section{\corr Homodyne detection for a tilted magnetic-field plane}
\label{app:Homodyne_tilted_B}

\begin{figure}
    \centering
\includegraphics[width=0.99\linewidth]{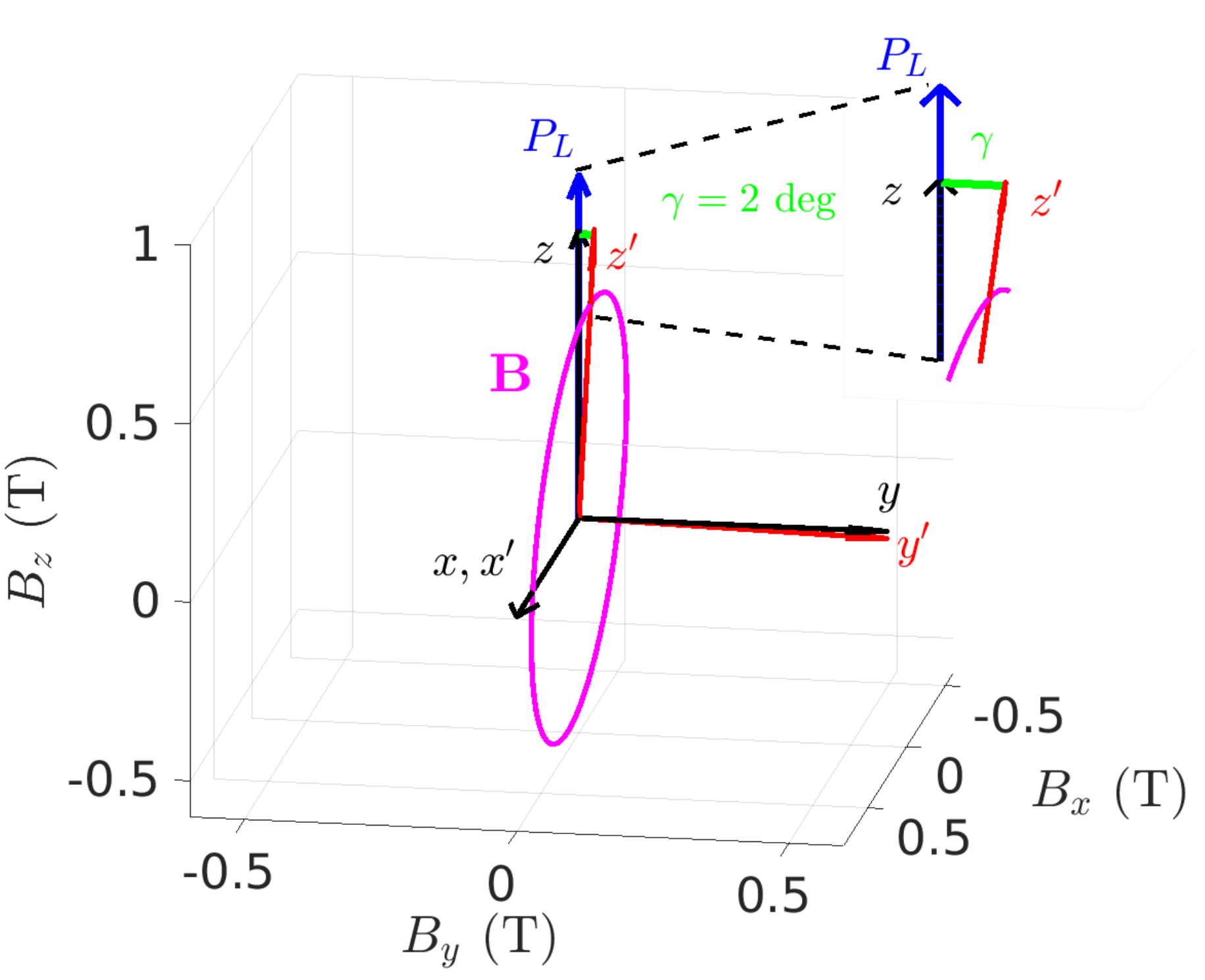}
\caption{\textbf{Schematic of the magnetic-field plane tilted by} $\boldsymbol{\gamma}=2$ degrees \corr (green), as shown in the zoom, formed between the tip's magnetization direction, along $z$, and the new $z'$ axis (in red). The magnetic field lies on the magenta curve with $\mathbf{B}=0.6$ T.}
\label{fig:tilted_magnetic_field}

    \centering
\includegraphics[width=0.99\linewidth]{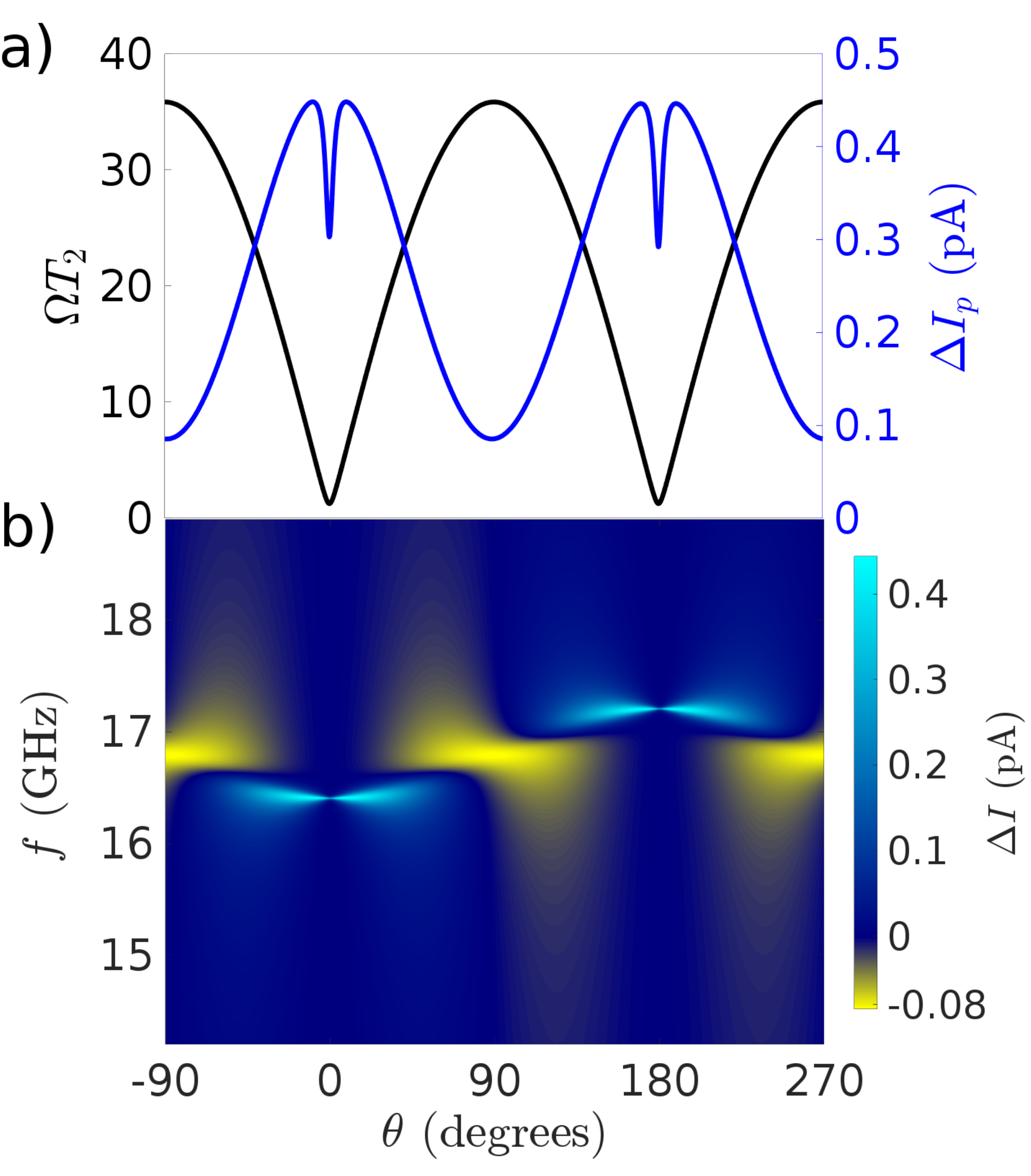}
\caption{\textbf{Homodyne detection for the tilted magnetic-field plane of Fig.~\ref{fig:tilted_magnetic_field}.} \corr 
(a) $\theta$ dependence of the figure of merit $\Omega_\mathrm{FLT}T_2$ (black, left) and the total ESR intensity $\Delta I_{p}$ (blue, right). The overall behavior mirrors that of Fig.~\ref{fig:ESR_RabiT2_vs_angle}b and Fig.~\ref{fig:ESR_vs_angle}, but with a finite signal now observed at $\theta = 0$ degrees and $180$ degrees, consistent with experiments. Parameters are identical to those of the main text at positive DC bias.  
(b) ESR intensity map, Eq.~\eqref{asy_sym_current}, versus homodyne angle and frequency, showing the ESR maximum near $\theta = 10$ degrees, the flipping of the symmetric signal (peak-to-dip transition), and the evolution of the asymmetric component.}
\label{fig:homodyne_detection_tilted_magnetic_field}
\end{figure}

{\corr In this appendix we examine the dependence of the homodyne angle $\theta$ when the external magnetic field lies in a plane not containing the tip magnetization, as illustrated in Fig.~\ref{fig:tilted_magnetic_field}. The plane is tilted by $\gamma = 2$ degrees with respect to the spin-polarization direction. Increasing $\gamma$ gradually reduces the ESR signal until $\theta = 90$ degrees, at which point homodyne detection vanishes and reemerges beyond this angle. This behavior arises because the ESR signal for in-plane orientations relative to the spin-polarization direction is STT-driven. Importantly, here $\theta$ denotes the angle between the tilted axis $z'$ and the magnetic field, not between the tip magnetization (along $z$) and the field.}

{\corr Figure~\ref{fig:homodyne_detection_tilted_magnetic_field}a (blue) shows $\Delta I_p=\sqrt{\Delta I_s^2 + \Delta I_a^2}$, where the small tilt of the magnetic field yields finite ESR signals at $\theta = 0$ degrees and $180$ degrees, absent in the main-text results. This occurs because the magnetization direction of the tip can no longer be perfectly parallel to the magnetic field, a more realistic configuration in closer agreement with experiments~\cite{JK_vector_field_2021}. For further discussion see Ref.~\cite{Fe_driving_SH_2023}, Fig.~S15, noting that in experiments the tip is typically more in-plane, shifting the ESR maximum closer to $\theta=90$ degrees. For comparison, we also show $\Omega_\mathrm{FLT}T_2$ in Fig.~\ref{fig:homodyne_detection_tilted_magnetic_field}a (in black), illustrating that its maximum does not coincide with the largest ESR signal.}

{\corr Overall, the agreement between our model and experiment is remarkably good, again highlighting the importance of properly including the exchange field in homodyne detection. Without it, the ESR signal would resemble an STT-driven detection mechanism with a maximum at $\theta = 90$ degrees, where the torque is strongest, which contradicts the experimental evidence, an issue we have already identified in other works~\cite{Ribeiro2,PRL_Ye_ESR_theory_2024}.}

{\corr Figure~\ref{fig:homodyne_detection_tilted_magnetic_field}b further presents a colormap of the raw ESR signal, Eq.~\eqref{asy_sym_current}, versus frequency and $\theta$. The plot reproduces the homodyne detection and reveals a sign reversal of the symmetric signal as $\theta$ approaches $90$ degrees from below, consistent with Figs.~\ref{fig:CB_out_of_plane}c and \ref{fig:CB_in_plane}c. This flipping marks the crossover from FLT- to STT-driven detection, since the FLT contribution vanishes at $\theta=90$ degrees under homodyne detection, as discussed in the main text. The asymmetric component also evolves with angle, increasing from the out-of-plane configuration, peaking near $\theta = 42$ degrees, and vanishing again for in-plane configurations, where the signal is STT-mediated. The exchange-field-induced frequency shift following Eq.~\eqref{energy_dress} is also visible in Fig.~\ref{fig:homodyne_detection_tilted_magnetic_field}b.}

\section{Insights on mapping the QME onto a spin problem} \label{app:mapping_onto_spin}

Here we clarify the steps for mapping the spin-1/2 version of Eqs.~\eqref{rho_master_eq_6_0} into a spin system, leading to Eqs.~\eqref{spin_equations} and~\eqref{spin_equations_b}. We derive equations depending only on the spin components and transient QI populations. First, we write the full set for the populations and coherences for the simplest quantum impurity (spin 1/2):
\begin{eqnarray*}
    \hbar\dot{\rho}_{gg}&=&-2\sum_{v=0,2}\mbox{Re}(\Gamma_{gv,vg}(t)) \rho_{gg} - \sum_{v=0,2}\Gamma_{e v,vg}(t)\rho_{ge} \\
    &-&\sum_{v=0,2}\Gamma_{e v,vg}^*(t)\rho_{eg} + 2\sum_{v=0,2} \mbox{Re}(\Gamma_{v g,g v}(t))\rho_v \\  
\hbar\dot{\rho}_{ee} &=&- 2\sum_{v=0,2}\mbox{Re}(\Gamma_{ev,ve}(t))\rho_{ee} -\sum_{v=0,2}\Gamma_{e v,vg}^*(t)\rho_{ge} \\ 
&-& \sum_{v=0,2}\Gamma_{e v,vg}(t)\rho_{eg} +2\sum_{v=0,2}\mbox{Re}\Gamma_{ve,e v}(t)\rho_v \\
\hbar\dot{\rho}_0 &=& 2\mbox{Re}(\Gamma_{e0,0e}(t))\rho_{ee}+2\mbox{Re}(\Gamma_{g0,0g}(t))\rho_{gg} \\[0.1cm] 
&+& 2\mbox{Re}(\Gamma_{e 0,0g}(t)) (\rho_{ge}+\rho_{eg}) \\[0.1cm] 
&-& 2\mbox{Re}(\Gamma_{0g,g 0}(t)+\Gamma_{0e,e 0}(t))\rho_0 \\[0.1cm] 
\hbar\dot{\rho}_2 &=& 2\mbox{Re}(\Gamma_{e2,2e}(t))\rho_{ee}+2\mbox{Re}(\Gamma_{g2,2g}(t))\rho_{gg} \\[0.1cm] 
&+& 2\mbox{Re}(\Gamma_{e 2,2g}(t)) (\rho_{ge}+\rho_{eg}) \\[0.1cm] 
&-& 2\mbox{Re}(\Gamma_{2g,g 2}(t)+\Gamma_{2e,e 2}(t))\rho_2
\end{eqnarray*}
\onecolumngrid
{\small \begin{eqnarray*}
\hbar \dot{\rho}_{ge} &=& -\sum_{v=0,2}\Gamma_{e v,vg}(t)\rho_{gg} + \left[i\Delta_{ge}-\sum_{v=0,2}\left(\Gamma_{ev,ve}(t) +\Gamma_{gv,vg}^*(t)\right)\right]\rho_{ge} - \sum_{v=0,2}\Gamma_{e v,vg}^*(t) \rho_{ee} + 2\sum_{v=0,2}\mbox{Re}(\Gamma_{vg,e v}(t))\rho_v  \\ 
\hbar\dot{\rho}_{eg}&=& -\sum_{v=0,2}\Gamma_{e v,vg}^*(t)\rho_{gg} - \sum_{v=0,2}\Gamma_{e v,vg}(t)\rho_{ee} - \left[i\Delta_{ge}+\sum_{v=0,2}\left(\Gamma_{gv,vg}(t) +\Gamma_{ev,ve}^*(t)\right)\right]\rho_{eg} + 2\sum_{v=0,2}\mbox{Re}(\Gamma_{vg,e v}(t))\rho_v. 
\end{eqnarray*}}

We need to make use of several relations for the rates that apply when both the magnetic field and the frequency of the time-dependent field are small compared to the energy thresholds $\varepsilon$ and $\varepsilon+U$. Choosing $\pm$ depending on $v=0,2$ respectively, we have: 
\begin{eqnarray*}
 \mathrm{Re}(\Gamma_{ev,ve})+\mathrm{Re}(\Gamma_{gv,vg}) &=& 2\,\mathrm{Re}(\Gamma_{ev,ve}^{P_\alpha=0}), \ \ \mathrm{Re}(\Gamma_{ev,ve})-\mathrm{Re}(\Gamma_{gv,vg}) = \pm 2 \sum_\alpha P_\alpha \,\mathrm{Re}(\Gamma_{gv,vg,\alpha}^{P_\alpha=0})\cos\theta , \\
\mathrm{Im}(\Gamma_{ev,ve})+\mathrm{Im}(\Gamma_{gv,vg}) &=& 2\,\mathrm{Im}(\Gamma_{gv,vg}^{P_\alpha=0}), \ \ 
\mathrm{Im}(\Gamma_{ev,ve})-\mathrm{Im}(\Gamma_{gv,vg}) = \pm 2 \sum_\alpha P_\alpha \,\mathrm{Im}(\Gamma_{ev,ve,\alpha}^{P_\alpha=0})\cos\theta . 
\end{eqnarray*}  
and $\Gamma_{ev,vg} = \Gamma_{gv,ve}$. In fact, we can write $\Gamma_{ev,ve} = \sum_\alpha \Gamma_{ev,ve,\alpha}^{P_\alpha=0} (1 \pm P_\alpha \cos\theta), \ \Gamma_{gv,vg} = \sum_\alpha \Gamma_{gv,vg,\alpha}^{P_\alpha=0} (1 \mp P_\alpha \cos\theta)$. With these relations and Eqs.~\eqref{General_Sz},~\eqref{Sx_ave_eq}, and~\eqref{Sy_ave_eq} we write the spin system of equations:
{\small
\begin{eqnarray*}
     \langle \dot{s}_z \rangle &=&{\cos\theta}\left[ -\sum_{v=0,2}\mbox{Re}(\Gamma_{gv,vg}^{P_\alpha=0})(\rho_{ee}-\rho_{gg}) + i\sum_{v=0,2}\mbox{Im}(\Gamma_{e v,vg}) \rho_{ge} - i\sum_{v=0,2}\mbox{Im}(\Gamma_{e v,vg}) \rho_{ge}^* \right. \\ 
     &+&\left. \sum_\alpha P_\alpha\cos\theta\left(-\mbox{Re}(\Gamma_{g0,0g,\alpha}^{P_\alpha=0})(\rho_{ee}+\rho_{gg})+ \mbox{Re}(\Gamma_{g2,2g,\alpha}^{P_\alpha=0})(\rho_{ee}+\rho_{gg}) +2\mbox{Re}(\Gamma_{0g,g 0,\alpha}^{P_\alpha=0}) \rho_0 -2\mbox{Re}(\Gamma_{2g,g 2}^{P_\alpha=0}) \rho_2\right)\right]  \\ 
     &+& {\sin \theta}\left[ -\sum_{v=0,2}\mbox{Re}(\Gamma_{e v,vg}) (\rho_{gg}+ \rho_{ee}) + 2\sum_{v=0,2}\mbox{Re}(\Gamma_{vg,e v})\rho_v \right.  \\
     &-&\left. (\Delta_{ge} - 2\sum_\alpha P_\alpha\cos\theta(\mbox{Im}(\Gamma_{g0,0g,\alpha}^{P_\alpha=0})-\mbox{Im}(\Gamma_{g2,2g,\alpha}^{P_\alpha=0})) \mbox{Im}(\rho_{ge}) -  2\sum_{v=0,2}\mbox{Re}(\Gamma_{gv,vg}^{P_\alpha=0}) \mbox{Re}(\rho_{ge})\right] ,
     \end{eqnarray*}
     \begin{eqnarray*}
     \langle \dot{s}_x \rangle &=&{\sin\theta}\left[ -\sum_{v=0,2}\mbox{Re}(\Gamma_{gv,vg}^{P_\alpha=0})(\rho_{ee}-\rho_{gg}) + i\sum_{v=0,2}\mbox{Im}(\Gamma_{e v,vg}) \rho_{ge} - i\sum_{v=0,2}\mbox{Im}(\Gamma_{e v,vg}) \rho_{ge}^* \right. \\
     &+&\left. \sum_\alpha P_\alpha\cos\theta\left(-\mbox{Re}(\Gamma_{g0,0g,\alpha}^{P_\alpha=0})(\rho_{ee}+\rho_{gg})+ \mbox{Re}(\Gamma_{g2,2g,\alpha}^{P_\alpha=0})(\rho_{ee}+\rho_{gg}) +2\mbox{Re}(\Gamma_{0g,g 0,\alpha}^{P_\alpha=0}) \rho_0 -2\mbox{Re}(\Gamma_{2g,g 2,\alpha}^{P_\alpha=0}) \rho_2\right)\right]  \\
     &-& {\cos \theta}\left[ -\sum_{v=0,2}\mbox{Re}(\Gamma_{e v,vg}) (\rho_{gg}+ \rho_{ee}) + 2\sum_{v=0,2}\mbox{Re}(\Gamma_{vg,e v})\rho_v \right.  \\
     &-&\left. (\Delta_{ge} - 2\sum_\alpha P_\alpha(\mbox{Im}(\Gamma_{g0,0g,\alpha}^{P_\alpha=0})-\mbox{Im}(\Gamma_{g2,2g,\alpha}^{P_\alpha=0})) \mbox{Im}(\rho_{ge})\cos\theta -  2\sum_{v=0,2}\mbox{Re}(\Gamma_{gv,vg}^{P_\alpha=0}) \mbox{Re}(\rho_{ge})\right] ,
     \\
          \langle \dot{s}_y \rangle &=& 
       \sum_{v=0,2}\left[\mbox{Im}(\Gamma_{e v,vg}) (\rho_{ee}-\rho_{gg})-2\mbox{Re}(\Gamma_{gv,vg}^{P_\alpha=0}) \mbox{Im}(\rho_{ge}) \right]+(\Delta_{ge} - 2\sum_\alpha P_\alpha(\mbox{Im}(\Gamma_{g0,0g,\alpha}^{P_\alpha=0})-\mbox{Im}(\Gamma_{g2,2g,\alpha}^{P_\alpha=0})) \mbox{Re}(\rho_{ge})\cos\theta, 
\end{eqnarray*}}
\hspace*{-0.375cm} which can be expressed independently of the singly occupied state populations and coherences by introducing $\hbar\mbox{Re}(\rho_{ge})=\langle s_z\rangle \sin\theta -\langle s_x\rangle \cos\theta$, $\langle s_y \rangle = -\hbar\mbox{Im}(\rho_{ge})$, and $\rho_{ee}-\rho_{gg}=2 (\langle s_z\rangle \cos\theta +\langle s_x\rangle \sin\theta)/\hbar$, which follow from Eqs.~\eqref{pop_spin_general} when $\phi=0$. This leads to:
{\small
\begin{eqnarray*}
     \langle \dot{s}_z \rangle &=&\left[ \sum_{v=0,2}2{\cos\theta}\, \mbox{Im}(\Gamma_{e v,vg}) +\sin\theta (\Delta_{ge} - 2\sum_\alpha P_\alpha\cos\theta(\mbox{Im}(\Gamma_{g0,0g,\alpha}^{P_\alpha=0})-\mbox{Im}(\Gamma_{g2,2g,\alpha}^{P_\alpha=0})) \right]\langle s_y\rangle/\hbar -2\sum_{v=0,2}\mbox{Re}(\Gamma_{gv,vg}^{P_\alpha=0})\langle {s}_z \rangle/\hbar \\
     &+&\left. \sum_\alpha P_\alpha\cos^2\theta\left([\mbox{Re}(\Gamma_{g2,2g,\alpha}^{P_\alpha=0})-\mbox{Re}(\Gamma_{g0,0g,\alpha}^{P_\alpha=0})](1-\rho_0-\rho_2)+ 2\mbox{Re}(\Gamma_{0g,g 0,\alpha}^{P_\alpha=0}) \rho_0 -2\mbox{Re}(\Gamma_{2g,g 2,\alpha}^{P_\alpha=0}) \rho_2\right)\right.  \\ 
     &+& {\sin \theta}\left[ 2\mbox{Re}(\Gamma_{g0,0g,e 0})\rho_0 +2\mbox{Re}(\Gamma_{2g,e 2})\rho_2 -(\mbox{Re}(\Gamma_{e 0,0g})+\mbox{Re}(\Gamma_{e 2,2g})) (1-\rho_0-\rho_2) \right] ,   \\
      \langle \dot{s}_x \rangle &=&\left[ \sum_{v=0,2}2{\sin\theta}\, \mbox{Im}(\Gamma_{e v,vg}) -\cos\theta (\Delta_{ge} - 2\sum_\alpha P_\alpha\cos\theta(\mbox{Im}(\Gamma_{g0,0g,\alpha}^{P_\alpha=0})-\mbox{Im}(\Gamma_{g2,2g,\alpha}^{P_\alpha=0})) \right]\langle s_y\rangle/\hbar -2\sum_{v=0,2}\mbox{Re}(\Gamma_{gv,vg}^{P_\alpha=0})\langle {s}_x \rangle/\hbar \\ 
     &+&\left. \sum_\alpha P_\alpha\cos\theta\sin \theta\left([\mbox{Re}(\Gamma_{g2,2g,\alpha}^{P_\alpha=0})-\mbox{Re}(\Gamma_{g0,0g,\alpha}^{P_\alpha=0})](1-\rho_0-\rho_2)+ 2\mbox{Re}(\Gamma_{0g,g 0,\alpha}^{P_\alpha=0}) \rho_0 -2\mbox{Re}(\Gamma_{2g,g 2,\alpha}^{P_\alpha=0}) \rho_2\right)\right.  \\ 
     &-& {\cos \theta}\left[ 2\mbox{Re}(\Gamma_{g0,0g,e 0})\rho_0 +2\mbox{Re}(\Gamma_{2g,e 2})\rho_2 -(\mbox{Re}(\Gamma_{e 0,0g})+\mbox{Re}(\Gamma_{e 2,2g})) (1-\rho_0-\rho_2) \right], \\[0.25cm] 
         \langle \dot{s}_y \rangle &=& 
     -\sum_{v=0,2}2\mbox{Im}(\Gamma_{e v,vg}) (\langle s_z\rangle \cos\theta +\langle s_x\rangle \sin\theta)/\hbar 
     - 2\sum_{v=0,2}\mbox{Re}(\Gamma_{gv,vg}^{P_\alpha=0}) \langle s_y \rangle/\hbar  \\
      &-&\left(\Delta_{ge} - 2\sum_\alpha P_\alpha\cos\theta(\mbox{Im}(\Gamma_{g0,0g,\alpha}^{P_\alpha=0})-\mbox{Im}(\Gamma_{g2,2g,\alpha}^{P_\alpha=0})\right) \left(\langle s_z\rangle \sin\theta -\langle s_x\rangle \cos\theta \right)/\hbar.
\end{eqnarray*}}
Doing the same for the equations of the empty and doubly state yield:
\begin{eqnarray*}
    \hbar\dot{\rho}_0 &=& 2\mbox{Re}(\Gamma_{g0,0g}^{P_\alpha=0})(1-\rho_0-\rho_2) +4\sum_\alpha \mbox{Re}(\Gamma_{g0,0g,\alpha}^{P_\alpha=0})P_\alpha\cos\theta ( \langle s_z\rangle \cos\theta +\langle s_x\rangle \sin\theta)/\hbar \\ 
    &+& 4\mbox{Re}(\Gamma_{e 0,0g}) (\langle s_z\rangle \sin\theta -\langle s_x\rangle\cos\theta)/\hbar -4\mbox{Re}(\Gamma_{0g,g 0}^{P_\alpha=0})\rho_0 ,\\
       \hbar\dot{\rho}_2 &=& 2\mbox{Re}(\Gamma_{g2,2g}^{P_\alpha=0})(1-\rho_0-\rho_2) -4\sum_\alpha \mbox{Re}(\Gamma_{g2,2g,\alpha}^{P_\alpha=0})P_\alpha\cos\theta ( \langle s_z\rangle \cos\theta +\langle s_x\rangle \sin\theta)/\hbar\\ 
    &+& 4\mbox{Re}(\Gamma_{e 2,2g} )(\langle s_z\rangle \sin\theta -\langle s_x\rangle\cos\theta)/\hbar -4\mbox{Re}(\Gamma_{2g,g 2}^{P_\alpha=0})\rho_2.
\end{eqnarray*}
By considering that $\sum_\alpha P_\alpha \left[\mbox{Im}(\Gamma_{ g 0,0 g,\alpha}^{P_\alpha=0}- \Gamma_{ g 2,2 g,\alpha}^{P_\alpha=0}) \right] \sin\theta = \sum_{v=0,2} \mbox{Im} (\Gamma_{ e v,v g}(t))$, one sees that the second and third lines in the $s_x$ equation, both proportional to the polarization, cancel out, while in the $s_z$ equation, we obtain the non-equilibrium spin accumulation. At the same time, this relation further simplifies the equations for the transient state by merging the second and third terms in both equations, ultimately leading to Eqs.~\eqref{spin_equations} and~\eqref{spin_equations_b}.

\twocolumngrid


%


\end{document}